\documentclass[12pt,a4,sans]{article}
\pdfoutput=1
\usepackage[latin1]{inputenc}
\usepackage{amssymb}
\usepackage{amsmath}
\usepackage{amsthm}
\usepackage{amsfonts}
\usepackage[pdftex]{graphicx}
\usepackage{geometry}
\usepackage{multirow}
\usepackage{braket}
\usepackage{enumerate}

\allowdisplaybreaks

\usepackage[table,usenames,dvipsnames]{xcolor}

\usepackage{setspace}
\usepackage[backgroundcolor=white,linecolor=black]{todonotes}
\usepackage[in]{fullpage}
\usepackage{hyperref}
\usepackage{array}
\usepackage{cancel}
\usepackage{wrapfig}
\usepackage[font=small,labelfont=bf]{caption}
\usepackage{pifont}

\usepackage[dvipsnames]{xcolor}
\usepackage{mathtools}
\usepackage{verbatim}
\usepackage{color}

\numberwithin{equation}{section}

\hypersetup{
	colorlinks=true,
	linktoc=page,
	citecolor=Blue,
	linkcolor=Blue,
	urlcolor=Blue} 

\makeatletter 
\renewcommand{\maketitle} 
{ \begingroup \begin{center} \large {\bf \@title}
		\vskip 5pt \large \@author \\ \vskip 5pt \@date \end{center}
	\vskip 5pt \endgroup \setcounter{footnote}{0} }
\makeatother 

\newcommand{\pic}[2]{\vcenter{\hbox{\includegraphics[scale=#1]{#2}}}}

\newcommand{\comments}[1]{}
\newcommand{\la}{\langle}
\newcommand{\ra}{\rangle}

\newcommand{\A}{\mathcal{A}}

\newcommand{\N}{\mathcal{N}}
\newcommand{\Op}{\mathcal{O}}
\newcommand{\Tr}{\text{Tr}}

\newcommand{\OC}{\mathcal{O}_{\mathcal{C}}}
\newcommand{\OS}{\mathcal{O}_{\mathcal{S}}}

\renewcommand{\b}[1]{\braket{#1}}

\renewcommand{\O}{\mathcal{O}}

\newcommand{\be}{\begin{equation}}
\newcommand{\ee}{\end{equation}}

\newcommand{\hf}{\frac{1}{2}}

\def\beqa{\begin{eqnarray}}
\def\eeqa{\end{eqnarray}}
\def\beq{\begin{equation}}
\def\eeq{\end{equation}}

\def\Tr{{\rm Tr}}
\def\one{\mbox{1 \kern-.59em {\rm l}}}

 \def\C_B{{\cal B}} \def\cC{{\cal C}}
  \def\C_F{{\cal F}}
  
  \def\cL{{\cal L}}
\def\cM{{\cal M}} \def\cN{{\cal N}} \def\cO{{\cal O}}
  \def\cR{{\cal R}}
\def\cS{{\cal S}}  
  
 \def\cZ{{\cal Z}}

\def\uno{\mbox{1 \kern-.59em {\rm l}}}

\def\lan{\langle}
\def\ran{\rangle}

\def\one{1\!\!1\,\,}

\def\bcomment#1{}

\def\eps{\epsilon}

\usepackage{slashed}
\usepackage{caption}

\usepackage[nottoc,notlot,notlof]{tocbibind}
\usepackage[nosort]{cite}
\usepackage{color}

\usepackage{parskip}

\graphicspath{{./Images/}}

\long\def\symbolfootnote[#1]#2{\begingroup%
	\def\thefootnote{\fnsymbol{footnote}}\footnote[#1]{#2}\endgroup}

\begin{document}
	
	\begin{flushright}
		QMUL-PH-18-01\\
		CERN-TH-2018-063
	\end{flushright}
	
	\vspace{15pt} 

	\begin{center}
		
		{\Large \bf $\Tr(F^3)$ supersymmetric form factors  }\\ 
		\vspace{0.3 cm} {\Large \bf   
		and maximal transcendentality}\\ 
		\vspace{0.3 cm}
		{\Large \bf Part I: $\cN\!=\!4$ super Yang-Mills}

		%
		\vspace{25pt}


		{\mbox {\bf Andreas~Brandhuber$^{a,\S}$, 
				Martyna~Kostaci\'{n}ska$^{a,\S}$,}} \\ \vspace{0.2cm}
				{\mbox{\bf
				Brenda~Penante$^{b,\star}$ and %
				Gabriele~Travaglini$^{a,\S}$}}%

		\vspace{0.5cm}
		
		\begin{quote}
			{\small \em
				\begin{itemize}
					\item[\ \ \ \ \ \ $^a$]
					\begin{flushleft}
						Centre for Research in String Theory\\
						School of Physics and Astronomy\\
						Queen Mary University of London\\
						Mile End Road, London E1 4NS, United Kingdom
					\end{flushleft}

					\item[\ \ \ \ \ \ $^b$]CERN Theory Division, 1211 Geneva 23, Switzerland

				\end{itemize}
			}
		\end{quote}


		\vspace{15pt} 

{\bf Abstract}
	\end{center}
	
	\vspace{0.3cm} 
	
\noindent

\noindent

In the large  top-mass limit, Higgs plus multi-gluon amplitudes in QCD can be computed using  an effective field theory. This approach turns the computation of such amplitudes into that of form factors of operators of increasing classical dimension. In this paper we focus on the first finite top-mass  correction, arising from the operator ${\rm Tr} (F^3)$, up to two loops and three gluons.  Setting up the calculation in the maximally supersymmetric theory requires  identification of an appropriate supersymmetric completion of ${\rm Tr} (F^3)$, which we recognise  as a descendant of the Konishi operator. We provide detailed computations for both this operator and the component operator ${\rm Tr} (F^3)$, preparing the ground for the calculation in $\cN\!<\!4$, to be detailed in a companion paper. 
Our results for both operators are  expressed in terms of a few universal functions of transcendental degree four and below, some of which have appeared in other contexts, hinting at universality of such quantities. An important  feature  of the result is  a delicate cancellation of unphysical poles appearing in soft/collinear limits of the remainders which links terms of different transcendentality. Our calculation provides another example of the principle of maximal transcendentality for observables with non-trivial kinematic dependence. 
	
	\vfill
	\hrulefill
	\newline
\vspace{-1cm}
$^{\S}$~\!\!{\tt\footnotesize\{a.brandhuber, m.m.kostacinska, g.travaglini\}@qmul.ac.uk}, \ $^{\star}$~\!\!{\tt\footnotesize b.penante@cern.ch}

	\setcounter{page}{0}
	\thispagestyle{empty}
	\newpage


	\setcounter{tocdepth}{4}
	\hrule height 0.75pt
	\tableofcontents
	\vspace{0.8cm}
	\hrule height 0.75pt
	\vspace{1cm}
	
	\setcounter{tocdepth}{2}


\section{Introduction}\label{sec:Introduction}

Form factors of local gauge-invariant operators appear ubiquitously in gauge theories and compute quantities of great phenomenological interest. For a certain  operator $\O(x)$, we  define the form factor between the vacuum and  and an $n$-particle state as  
\begin{align}
\label{eq:Fourier-FF}
\begin{split}
F_\O (1,\dots,n;q)\coloneqq\, &\int d^4x\, e^{-iq\cdot x}\la 1\ldots n|\mathcal{O}(x)|0\rangle\,=\,(2\pi)^4 \ \delta^{(4)}\Big(q-\sum_{i=1}^n p_i\Big)\la 1\ldots n|\mathcal{O}(0)|0\rangle\ , 
\end{split}
\end{align}
where the momentum conserving $\delta$-function follows from translational invariance. 
Noteworthy examples of such quantities include  form factors of the hadronic electromagnetic current with  external hadronic states, which  are the building blocks of the  $e^+ e^{-}\!\to\!$ hadrons and deep inelastic scattering matrix elements; and the form factor of the electromagnetic current, which computes the (electron) $g\!-\!2$.

An important class of form factors, which will be the focus of this paper and its companion, makes its appearance in  the study of  amplitudes involving the Higgs boson and many gluons in QCD. 
  At one loop, the Higgs couples to the gluons through a loop of quarks, with the top quark loop giving the largest contribution to the gluon fusion process. These amplitudes  can then  be treated in an effective Lagrangian description, where the quark loop is effectively replaced by a set of local interactions of increasing classical dimension. 
  
In the limit where the mass of the Higgs $m_H$ is much smaller  than the mass of the top quark $m_t$, the leading interaction   is a dimension-five operator of the form \cite{Wilczek:1977zn,Shifman:1979eb,Dawson:1990zj}
\beq
\cL_5 \sim H \, {\rm Tr} (F^2)
\ , 
\eeq
 where $H$ represents the Higgs boson and $F$ is the gluon field strength. Hence the scattering amplitude of a Higgs and a gluonic state $\langle g\ldots g |$   in the infinite top-mass  limit, is nothing but a  form factor of  the dimension-four operator ${\rm Tr} \big(F^2 (0)\big)$, {i.e.} 
$\langle g\ldots g | \, {\rm Tr} \big(F^2 (0)\big) \, | 0 \rangle$. 
Subleading interactions (in  $1/m_t$)  will appear at dimension seven and include terms of the type \cite{Buchmuller:1985jz, Neill:2009tn, Neill:2009mz, Harlander:2013oja, Dawson:2014ora}
\beq
\cL_7 \sim H\, {\rm Tr}  (F^3)\, , \qquad  \cL_{7,i}^{\prime}  \,  \sim\, H \, {\rm Tr} (DF DF)
\ , 
\eeq
where $i$ schematically labels the three possible index contractions. In pure Yang-Mills, only   one  of the three possible operators among $\cL_{7,i}^{\prime}$ is  independent  due to the equations of motion \cite{Dawson:2014ora,Gracey:2002he}, and we  pick 
\beq
\cL_{7}^{\prime}  \,  \sim\, H \, {\rm Tr} (D_\mu F_{\nu \rho} D^\mu F^{\nu \rho})
\ . 
\eeq
We also mention an additional source of interests in such quantities -- at zero momentum transfer ($q=0$ in \eqref{eq:Fourier-FF}), a form factor of an operator $\cO$ represents a potential correction  to a certain Standard Model scattering amplitude due to the inclusion in the theory of a new local interaction proportional to $\cO(x)$. For instance, the operator $F^3$, a close relative of which will be the primary focus of this paper,  arises as the first correction in the low-energy effective action of bosonic strings. Interestingly,  it is also the  only gauge-invariant modification to the three-gluon vertex  which is non-vanishing at three points \cite{Benincasa:2011pg}, see for instance \cite{Dixon:1993xd,Dixon:2004za, Cohen:2010mi, Broedel:2012rc} for examples of such effective amplitudes. 

While it is  clearly of great phenomenological importance to study such quantities directly in QCD, experience shows that many interesting properties and underlying structures  may better be highlighted by focusing on  simpler models  such as supersymmetric theories, \mbox{$\cN\!=\!4$} supersymmetric Yang-Mills (SYM) being the prime example of such a model. 
When making comparisons between form factors in different theories, however, one must face the issue that operators with the same classical dimensions and quantum numbers mix under renormalisation. Furthermore, in different theories the set of operators involved in the mixing will generically be different. As a case in point, 
in pure Yang-Mills $\Tr(F^2)$ does not mix with any other operator \cite{Dawson:2014ora}
while in $\mathcal{N}\!=\!4$ SYM there is a large number of operators that can potentially mix with it. 
Therefore, the question arises as to which form factors are we to compare in the two theories if we wish to gain some deeper understanding of the secret structures of such quantities.

Focusing initially on the  operator  $\Tr(F^2)$, in the case of $\mathcal{N}\!=\!4$ SYM  the answer to this question is suggested  by  supersymmetry, since
$\Tr(F^2)$ appears in the so-called on-shell Lagrangian, which has the schematic form 
\beq
\label{1.3}
\mathcal{L}_\text{on-shell} \sim \Tr(F^2)
+ g\, \Tr(\psi \psi \phi) + g^2 \,\Tr( [\phi,\phi]^2)
\,. 
\eeq
This operator is obtained as a supersymmetric descendant of the protected operator $\Tr(\phi^2)$, where $\phi$ is any given scalar in $\N\!=\!4$ SYM, 
by acting with four supersymmetry charges.  Both ${\rm Tr} (\phi^2)$ and $\mathcal{L}_\text{on-shell} $  are components of the chiral part of the stress-tensor multiplet 
$\mathcal{T}_2$ \cite{Eden:2011yp}. Their supersymmetric form factors have been studied and formulated in superspace in \cite{Brandhuber:2011tv}, which puts them on a similar footing as superamplitudes \cite{Witten:2003nn}.

The extra length-three and four terms in \eqref{1.3}  ensure that the operator $\mathcal{L}_\text{on-shell}$ is protected (half-BPS)  and does not mix with other operators, in contradistinction to $\Tr(F^2)$. Given the special status of $\mathcal{L}_\text{on-shell}$,  
it is therefore natural to compare form factors of $\Tr(F^2)$ in QCD with form factors of $\mathcal{L}_\text{on-shell}$ in $\mathcal{N}\!=\!4$ SYM. Furthermore, supersymmetric Ward identities   can be used to relate form factors of $\mathcal{L}_\text{on-shell}$ to those of $\Tr (\phi^2)$ with different external states, as was done in \cite{Brandhuber:2012vm}. Supersymmetry also allows to package form factors of the stress-tensor multiplet operator $\mathcal{T}_2$ into supersymmetric form factors \cite{Brandhuber:2011tv}.

Before addressing the story for  $\Tr(F^3)$, we should first ask ourselves  what lesson  we can learn by computing form factors of half-BPS operators in $\cN\!=\!4$ SYM, when comparing them to form factors of $\Tr(F^2)$ in, say, pure Yang-Mills. 
A surprising answer to this question was found in \cite{Brandhuber:2012vm} where, following earlier studies in \cite{Brandhuber:2010ad}, the form factors $\langle \phi \, \phi \, g^+ | {\rm Tr} (\phi^2) | 0 \rangle $ of the lowest-weight operator ${\rm Tr} (\phi^2)$ in the stress-tensor multiplet were studied at two loops, with the particular state containing two scalars and one gluon $g^+$. 
Comparing this quantity to the result for  $\langle  g^+g^+g^{\pm} | {\rm Tr} (F^2) | 0 \rangle $  at two loops 
\cite{Gehrmann:2011aa} showed that, remarkably -- and for reasons currently not explainable via symmetries  -- the maximally transcendental part of these form factors is identical to the result for $\langle \phi \, \phi\,  g^+ \, | {\rm Tr} (\phi^2) | 0 \rangle $ (which by itself contains only terms of maximal transcendentality -- four, at two loops).

For $\Tr(F^3)$, the main subject of this paper, the situation is more involved since this operator 
mixes with a variety of operators both in QCD/pure Yang-Mills and in $\mathcal{N}\!=\!4$ SYM. In pure Yang-Mills, it can mix with $\Tr(D_\mu F_{\nu\rho} D^\mu F^{\nu \rho})$; in QCD with three additional dimension-six operators; while in $\mathcal{N}\!=\!4$ SYM, mixing can potentially occur with a large number of operators formed by elementary fermion and scalar  fields.
What is the  appropriate translation  of the operator ${\rm Tr} (F^3)$ to the $\cN\!=\!4$ theory? 

A first thought might indicate that the form factor $\langle \phi \, \phi \, \phi \, | {\rm Tr} (\phi^3) | 0\rangle$, studied in \cite{Penante:2014sza} and \cite{Brandhuber:2014ica},  might be the correct translation of 
$\langle g^+ g^+ g^{+} | {\rm Tr} (F^3) | 0\rangle$, however one quickly realises that ${\rm Tr}(\phi^k)$ is half BPS for any $k$, while ${\rm Tr} (F^3)$ is not protected. 
One may however note that at one loop, 
 ${\rm Tr} (F^3)$   has the same anomalous dimension as the  Konishi operator.  
 An obvious candidate is therefore the Konishi descendant obtained by acting with eight $\bar{Q}$-supersymmetries on the Konishi operator $\epsilon^{ABCD}\, {\rm Tr} (\phi_{AB} \phi_{CD})$, which is proportional  to  ${\rm Tr} (F^3)$ plus 
appropriate  additional terms generated  by  supersymmetry.%
\footnote{$\phi_{AB}$  are the scalar fields of the theory, and  $A, \ldots, D = 1, \ldots , 4$ are  fundamental  indices of $SU(4)$.} 
This descendant is obtained by acting with tree-level supersymmetry generators,  and therefore mixing   is deferred to one  loop. 
Supersymmetric form factors of the full Konishi multiplet were recently studied in  \cite{Koster:2016loo,Chicherin:2016qsf}, allowing for the efficient use of supersums in our calculations.  This also allows for an immediate generalisation to $\cN\!<\!4$, which will be discussed in \cite{Part2}.

In this paper we outline in detail the calculation in $\cN\!=\!4$ SYM  of the two-loop form factors of  two operators: ${\rm Tr} (F^3)$ and the appropriate translation given by the Konishi descendant mentioned above, with an external  state  of  three positive-helicity gluons. This expands the results and observations of \cite{Brandhuber:2017bkg} and sets the stage for the calculations in $\cN\!<\!4$ which will be discussed in \cite{Part2}.

The most interesting observation,  already made in  \cite{Brandhuber:2017bkg},  is the 
remarkable  similarity  of the QCD  and the SYM results, regardless of the amount of supersymmetry. 
First of all,  there is a  {\it universality} of the maximally transcendental part of the  results across all theories, {\it including   pure Yang-Mills}  \cite{Brandhuber:2017bkg}. Furthermore, this maximally transcendental part is the same as the complete result for the minimal form factor of the  half-BPS operator ${\rm Tr} (\phi^3)$, which was computed in \cite{Brandhuber:2014ica}. Hence this is another illustration of the fact  that half-BPS operators in $\cN\!=\!4$ SYM play a surprising role in theories with less or no supersymmetry including QCD \cite{Brandhuber:2012vm,Brandhuber:2017bkg}. 
It is also a beautiful  appearance  of the 
 principle of maximal transcendentality \cite{Kotikov:2002ab,Kotikov:2004er} which,  in its original formulation,  relates the  anomalous dimensions of twist-two operators in $\cN\!=\!4$ SYM to  those  calculated  in QCD \cite{Moch:2004pa,Vogt:2004mw}
	by simply deleting all terms of   transcendentality degree less than  maximal  (or $2L\!-\!1$ at $L$ loops, in Mellin moment space).
	In our framework we see another incarnation of this principle across different theories, however for complicated, kinematic-dependent quantities.  
	This is even more surprising since 
	scattering amplitudes in general do not  have this property, {\it e.g.} one-loop MHV  amplitudes  in pure Yang-Mills  contain additional pieces  that have maximal transcendental degree    \cite{Bern:1994cg, Bern:1993mq,Bedford:2004nh}.  
We also note a different type of universality across form factors of different operators in $\cN\!=\!4$ SYM 
  namely for the scalar Konishi and the three closed  $SU(2)$, $SU(2|3)$ and $SL(2)$ sectors in the $\cN\!=\!4$ theory, respectively   
  \cite{Banerjee:2016kri,Loebbert:2015ova,Brandhuber:2016fni,Loebbert:2016xkw}. 
Further recent manifestations  of the principle of maximal transcendentality   include configurations of semi-infinite Wilson lines  \cite{Li:2014afw,Li:2016ctv} and the four-loop collinear anomalous dimension \cite{Dixon:2017nat}.
	
Second, our form factors (or more precisely their  remainders) contain terms of transcendentality ranging from  four  to zero. In an earlier paper \cite{Brandhuber:2016fni} we considered the  simpler scalar   descendant of the Konishi operator 
\beq
\cO_{\mathcal{K}} = \mathcal{O}_B \, - \, {gN\over  8 \pi^2}\, \mathcal{O}_F\, , 
\eeq
where 
$\mathcal{O}_B := {\rm Tr} (X[Y , Z])$ and $\mathcal{O}_F := (1/2) {\rm Tr} (\psi \psi)$, 
 with $X:=\phi_{12}$, $Y:=\phi_{23}$, $Z:=\phi_{31}$ and  $\psi_\alpha := \psi_{123, \alpha}$. This operator is part of the $SU(2|3)$ closed subsector of the $\cN\!=\!4$ theory \cite{Beisert:2003ys}. In that paper we considered the two-loop minimal form factor of $\cO_{\mathcal{K}}$ which also contains terms with transcendentality ranging from four to zero. While, as mentioned earlier, the maximally transcendental part is universal, 
we find that the transcendentality three and two terms are also {\it universal building blocks} of the two-loop form factors considered here, as already shown in  \cite{Brandhuber:2017bkg}, and to be expanded upon in the companion paper \cite{Part2}. 
 For the two operators considered in this paper -- ${\rm Tr} (F^3)$ and the particular Konishi descendant described earlier -- 
a new feature appears: 
the result of their minimal form factor remainders at two loops also contains polylogarithmic functions  multiplied by ratios of kinematic invariants. 
Only few universal functions are needed which, interestingly, also appeared in  \cite{Brandhuber:2016fni} as well as in related spin-chain Hamiltonian computations  in 
 \cite{Loebbert:2015ova,Loebbert:2016xkw}. What is more, we find that the rational factors we find are precisely needed to cancel potential unphysical simple and double poles. This requires unexpected, delicate inter-transcendental cancellations.

 Third, and even more remarkably, the computations  in   $\cN\!<\!4$ SYM  to appear  in the companion paper \cite{Part2}  will reveal further striking  similarities with $\cN\!=\!4$ SYM.%
\footnote{These results were anticipated at the 2017 IFT Christmas workshop and the 2018 Bethe forum  \cite{talk12}. We thank the organisers of these events for their invitations.}
In particular we will make an  important observation on the terms subleading in transcendentality: 
  the difference between the result  in  different theories with any amount (or no) supersymmetry and the result in  $\cN\!=\!4$ SYM is confined to a tiny class of terms, mostly simple $\zeta_n$ terms and coefficients of simple logarithms. 
 This can be explained by the fact that, for the operator ${\rm Tr}\, (F^3)$, the matter content of the different theories only enters through one-loop sub-diagrams, hence allowing effectively for a supersymmetric decomposition of the computation similar to that for one-loop amplitudes~\cite{Bern:1994cg}.%
 \footnote{However note that for the supersymmetric completion of this bosonic operator, called $\mathcal{O}_\mathcal{S}$ throughout this paper and introduced in Section \ref{section:mixing-first}, there would be  additional two-loop topologies not of this type, and  this statement would not apply.}
 This diagrammatic explanation also implies that the  form factor of ${\rm Tr} (F^3)$ in QCD differs from the corresponding calculation in  $\cN\!=\!4$ SYM only by certain single-scale integrals  of sub-maximal transcendentality which only bring about logarithms or constant terms. 
The consequence of this observation, already made in \cite{Brandhuber:2017bkg},  is that in the three-gluon case, $\cN\!=\!4$ SYM captures not only the  
 maximally transcendental  part of the leading-order (in $1/ m_t$)  Higgs plus three-gluon amplitudes \cite{Brandhuber:2012vm}, but also of the subleading corrections  from ${\rm Tr} (F^3)$.
The universal building blocks observed in \cite{Brandhuber:2016fni}  also make another appearance in the context of   $\cN\!<\!4$  SYM \cite{Part2}.

 A final comment is in order here. Throughout this paper we have made use of the four-dimensional helicity scheme and four-dimensional cuts to compute our two-loop form factors. At present there is no proof that the so-called $\mu^2$-terms, potentially arising from $D$-dimensional cuts, would not affect the final result for remainder functions. However, there are a number of examples where it has explicitly been proved that  four-dimensional cuts are sufficient for calculational purposes, namely the two-loop computations  of the four- \cite{0309040} and five-point \cite{Bern:2006vw} MHV amplitudes in $\cN\!=\!4$ SYM and the remainder of the six-point MHV amplitude  \cite{Bern:2008ap}. The latter case is particularly  interesting since  there is a remarkable cancellation between such $\mu^2$-terms  in the two-loop amplitude, and terms vanishing strictly in four dimensions in the one-loop amplitude, which also enters the definition of the remainder and contribute when  multiplied by $1/\epsilon$ poles in the one-loop amplitude.%
\footnote{In our case, we note that the one-loop form factor, which  enters the form factor remainder, computed using four-dimensional cuts is valid in $D$ dimensions \cite{Neill:2009mz}.}
We mention that our result passes a number of important consistency checks, including reproducing the correct infrared and ultraviolet divergences (and hence anomalous dimensions), and soft/collinear factorisation at two loops. Also note that issues encountered with dimensional regularisation in the case of the Konishi operator in \cite{Nandan:2014oga} do not arise in the present work since the operator definition does not involve state sums.

The rest of the paper is organised as follows. In Section \ref{Sec:2} we discuss the various operators considered in the paper and their tree-level form factors. In Section \ref{Sec:3} we describe the calculation of the one-loop form factors of these operators, finding their one-loop anomalous dimensions. In Section \ref{Sec:4} we move on to the two-loop form factor calculations and provide the details of the computations of results presented in \cite{Brandhuber:2017bkg}. In Section \ref{Sec:5} we solve the operator mixing, finding  an appropriate operator that diagonalises the dilatation operator, and then compute the BDS remainder function of  renormalised  operators in $\N\!=\!4$ SYM. Finally, in Section \ref{Sec:Discussion} we conclude by discussing  the results of our paper. 
 
\section{Operators and tree-level form factors}
\label{Sec:2}

\subsection{Form factors of ${\Tr(F^3)}$}

We begin our investigation by considering form factors of the operator $\Tr(F^3)$. In four dimensions it can be rewritten as a sum of selfdual and anti-selfdual terms
\begin{align}
\label{split}
\Tr(F^3) = \Tr(F^3_{\mathrm{ASD}}) + \Tr(F^3_{\mathrm{SD}}) \, \propto \, \O_\mathcal{C} + \overline{\O}_\mathcal{C} \ ,
\end{align}
where the subscript $\mathcal{C}$ stands for $\mathcal{C}$omponent. The precise normalisation involved in the definition of $\O_\mathcal{C}\propto \Tr(F^3_{\mathrm{ASD}}) $ and $\overline{\O}_\mathcal{C}$ is conveniently fixed in such a way that 
 the minimal tree-level form factor
of $\O_\mathcal{C}$ with three positive helicity gluons as external states is given by
\begin{align}
\label{eq:tree-level-F3}
F^{(0)}_{\O_{\mathcal C}}(1^+,2^+,3^+; q) = -[12][23][31] \, ,
\end{align}
and hence the minimal form factor for $\overline{\O}_\mathcal{C}\propto \Tr(F^3_{\mathrm{SD}})$ is
\begin{align}
\label{eq:tree-level-F3bar}
F^{(0)}_{\overline{\O}_{\mathcal C}}(1^-,2^-,3^-; q) = \langle 12 \rangle \langle 23 \rangle \langle 31 \rangle \, .
\end{align}
Examples of non-minimal form factors of $\O_{\mathcal C}$ at tree level that will be needed later on include%
\footnote{In the expressions for the $n$-particle form factors of $\cO_{\cS, \cC}$ ($\cO_{\cM}$) in this and the coming sections we omit a factor of $g^{n-3}$ ($g^{n-2}$) to make the formulae more transparent.}
\begin{align}
	\label{eq:treeFFs-n4}
	\begin{split}
	&F_{\O_{\mathcal C}}^{(0)}(1^+,2^+,3^+,4^-;q)\,=\, \frac{([12][23][31])^2}{[12][23][34][41]}\, , 
	\\[10pt]
	&F_{\O_{\mathcal C}}^{(0)}(1^+,2^+,3^+,4^+;q)\,=\,\frac{[12][23][34][41]}{s_{12}}\left(1+\frac{[31][4|q|3\ra}{s_{23}[41]}\right)+\text{cyclic}(1,2,3,4)\, ,
	\end{split}
\end{align}
where the first line of \eqref{eq:treeFFs-n4} can be obtained from \eqref{eq:tree-level-F3} multiplying by the soft factor $-\frac{[31]}{[34][41]}$, while the second line has been calculated using Feynman diagrams and MHV diagrams in \cite{Dawson:2014ora} (and confirmed now by an independent calculation). The first line of \eqref{eq:treeFFs-n4} is a member of an infinite family of $\overline{\mathrm{MHV}}$ form factors with three positive helicity gluons and an arbitrary number of negative helicity gluons:
\begin{align}
F_{\O_{\mathcal C}}^{(0)}(1^-,\ldots,i^+,\ldots, j^+, \ldots, k^+, \ldots, n^-;q)\,=\, (-1)^n \frac{([ij][jk][ki])^2}{[12][23]\cdots[n1]}\, .
\end{align}
Note that form factors belonging to this family but with different number of negative helicity gluons are related by soft factors $-\frac{[s-1,\,s+1]}{[s-1,\, s^-][s^-,\, s+1]}$. We also mention that the expression of these form factors at $q\!=\!0$ was known already for four and five points  in \cite{Dixon:1993xd}, and later extended to a generic number of particles in \cite{Dixon:2004za}.

\subsection{Supersymmetric form factors and mixing}
\label{section:mixing-first}

The operator $\O_\cC$ can mix with other operators under renormalisation, and hence we need to address mixing before embarking on concrete calculations. An important observation is that in $\N\!=\! 4$ SYM $\O_\cC$ is contained within a certain descendant of the Konishi operator generated by acting with tree-level supercharges%
\footnote{As opposed to the free supersymmetry generators which are implicit in the Nair superspace formalism used to define the states.} 
$Q_{A\alpha}$ and $\overline{Q}^A_{\dot\alpha}$ on the 
 lowest-dimensional operator
 \begin{align}
 \label{eq:Konishi}
\O_{\mathcal K}\, \sim \,\eps_{ABCD}\,\Tr (\phi^{AB}\phi^{CD})\,.
\end{align}
Here we denote $A=1,\dots,4 $ the $R$-symmetry index and $\alpha,\dot\alpha=1,2$ the Lorentz spinor indices.
Importantly, acting with eight tree-level supercharges $\overline{Q}^A_{\dot \alpha}$ on $\O_{\mathcal K}$ we generate an operator $\O_{\mathcal S}$ such that 
\begin{align}
\label{OS}
\O_{\mathcal S}\, = \, \O_{\mathcal C}+\O(g)\,, 
\end{align}
where the subscript $\mathcal{S}$ stands for $\mathcal{S}$upersymmetric and the additional $\O(g)$ terms are of length four or more in fields.

To be more concrete we give the schematic structure of $\O_{\mathcal S}$,  up to terms with four fields, 
\begin{align}
\begin{split}\
\label{extra}
\O_{\mathcal S} & \sim  \Tr(F_{\mathrm{ASD}}^3) \, +\,  g\,  \Tr(F_{\mathrm{ASD}}^2 \phi \bar{\phi}) 
\, + \, g\,    \Tr(F_{\mathrm{ASD}} \phi F_{\mathrm{ASD}} \bar{\phi})   \\
&+g \, \Tr(F_{\mathrm{ASD}} \psi \psi \phi) \, + \, g \, \Tr(F_{\mathrm{ASD}} \psi \phi \psi) + g\,  \Tr( \psi \psi \psi \psi) \ ,
\end{split}
\end{align}
where we also assume that all Lorentz or $R$-symmetry indices are contracted to form an invariant.

These correction terms appear multiplied by powers of the Yang-Mills coupling $g$, and not the 't Hooft coupling.%
\footnote{A simpler situation was addressed in \cite{Brandhuber:2016fni} in the $SU(2|3)$ sector, where it is known that two operators mix at dimension three, see Section 7 of that paper for a detailed discussion.}
Furthermore, they only
affect tree-level non-minimal form factors with more than three external lines. 
In particular, in Section \ref{sec:treeFFs} we illustrate in detail the effects of these terms on four-point tree-level form factors where they give rise to extra contact term interactions.
At loop level, this mixing can affect also minimal form factors. Importantly, $\O_{\mathcal S}$ solves the mixing problem at one loop, thus any further corrections to $\O_{\mathcal S}$ due to mixing can only be detected in a calculation at two loops or higher -- see Section \ref{section:mixing} for the resolution of the mixing at two loops. 

Luckily the explicit expression for the supersymmetric completion terms are not required
for our computations.
Indeed, the tree-level MHV form factors of the full Konishi multiplet in $\cN \!=\! 4$ SYM have been constructed and expressed in a compact formula in \cite{Chicherin:2016qsf},
\begin{align}\label{eq:FF-Konishi}
\begin{split}
\hspace{-0.15cm}&\langle 1, 2, \ldots , n | {\mathcal K} (\theta, \bar\theta) | 0 \rangle_{\rm MHV}^{(0)}= 
{ {e}^{\sum_{l=1}^n [ l |\bar\theta\theta |l\rangle + \eta_l \langle \theta l \rangle } \over \langle 1 2\rangle \cdots \langle n 1 \rangle }\sum_{i\leq j < k \leq l}\!\!
( 2 \!-\! \delta_{ij}) ( 2 \!-\! \delta_{kl})
\epsilon^{ABCD} \hat\eta_{iA}\hat\eta_{jB}\hat\eta_{kC}\hat\eta_{lD}\langle jk\rangle \langle li \rangle \,, 
\end{split}
\end{align}
where $\hat{\eta}_A\coloneqq \eta_A + 2 [ \tilde\lambda \, \bar\theta_A]$ and $\eta_A$ are the usual on-shell superspace coordinates labelling the external on-shell states  \cite{Nair:1988bq}, with $A=1, \ldots , 4$.  The $\theta^A_\alpha$ and $\bar{\theta}_{A \dot\alpha}$ label the components of the Konishi super-multiplet.

MHV form factors of $\O_{\mathcal K}$ are obtained by setting $\theta = \bar{\theta}=0$, while the form factors of $\overline{\O}_{\mathcal S}$ are obtained by setting $\bar{\theta}=0$
and extracting the $\theta^8$-term:
\begin{align}\label{eq:FF-OSbar}
\begin{split}
F^{(0)}_{\overline{\O}_{\mathcal S}, \mathrm{MHV}} (1, 2, \ldots , n ;q) = 
\frac{1}{144}{ \delta^{(8)}(\sum_{i=1}^n \eta_i \lambda_i ) \over \langle 1 2\rangle \cdots \langle n 1 \rangle }\sum_{i\leq j < k \leq l}\!\!
( 2 \!-\! \delta_{ij}) ( 2 \!-\! \delta_{kl})
\epsilon^{ABCD} \eta_{iA}\eta_{jB}\eta_{kC}\eta_{lD}\langle jk\rangle \langle li \rangle \, .
\end{split}
\end{align}
We notice that for this particular component operator we recover the on-shell supermomentum conservation $\delta$-function for the external on-shell particles, which simplifies calculations of supersymmetric unitarity cuts such as the ones we employ below in Section \ref{Sec:4}.

In this paper we perform two-loop computations of form factors with an external state of three positive-helicity gluons. Taking into account these constraints, there are several further gluonic operators which will appear in the mixing at two loops and need to be considered, namely 
 ${\rm Tr}(D^\mu F^{\nu \rho}D_\mu F_{\nu\rho})$ and two further operators with different Lorentz contractions. The equations of motion relate these to $\O_\cC$, the operator $q^2\, {\rm Tr}(F^2)$, and further operators containing fermions and scalars%
\footnote{See \cite{Dawson:2014ora} for a discussion of operator bases in QCD.}, 
which are irrelevant for the present discussion given the gluonic external state. The effect of this for the two-loop mixing problem is that the only other operator we expect to enter in the two-loop mixing is 
\beq
\O_\cM\!\propto\!q^2\, {\rm Tr}(F^2)
\ . 
\eeq
We choose its specific normalisation in such a way that
\begin{align}
\label{eq:operator2}
F_{\O_\cM}^{(0)}(1^+,2^+,3^+;q) \, = \, \frac{q^6}{\b{12}\b{23}\b{31}}\, = \,  \frac{F_{\O_\cC}^{(0)}(1^+,2^+,3^+;q)}{u v w},
\end{align}
where $u\coloneqq s_{12}/q^2$, $v\coloneqq s_{23}/q^2$, and $w\coloneqq s_{31}/q^2$. 

\subsection{Further tree-level form factors}
\label{sec:treeFFs}

To conclude this section we present further examples of tree-level MHV form factors of ${\O}_{\mathcal S}$ up to four external legs and contrast them with those of ${\O}_{\mathcal C}$. We will make use of these results in our explicit two-loop calculations in Section \ref{Sec:4}.  They also illustrate the effects of the $\mathcal{O}(g)$ terms of ${\O}_{\mathcal S}$ presented in \eqref{extra}. 

Firstly, from \eqref{eq:FF-OSbar} and its appropriately chosen prefactor, we find that the minimal form factors are independent of the choice of operator:
\begin{align}
\label{FF-OSbar-n3}
F^{(0)}_{\overline{\O}_{\mathcal S},\overline{\O}_{\mathcal C}}(1^-,2^-,3^-;q)\, = \, \langle 12 \rangle \langle 23 \rangle \langle 31 \rangle \, ,
\end{align}
and correspondingly 
\begin{align}
\label{FF-OS-n3}
F^{(0)}_{\O_{\mathcal S},\O_{\mathcal C}}(1^+,2^+,3^+;q)\, = \, -[12][23][31]\, .
\end{align}
The situation for four external particles is more involved, and the results depend in general on which of the two operators is chosen. However, for purely gluonic external lines there is no difference and from \eqref{eq:FF-OSbar} we recover 
\begin{align}	
F_{\O_{\mathcal S},\O_{\mathcal C}}^{(0)}(1^+,2^+,3^+,4^-;q)\,=\, \frac{[12][23][31]^2}{[34][41]} \, ,
\end{align}
in agreement with \eqref{eq:treeFFs-n4}. Similarly, if there are two fermions on the external lines the result does not depend on the operator, and only if the fermions are adjacent the result is non-vanishing: 
\begin{align}	
F_{\O_{\mathcal S},\O_{\mathcal C}}^{(0)}(1^+,2^+,3^{\psi^4},4^{\bar\psi^{123}};q)\,=\, \frac{[12][23][31]}{[34]} \, ,\qquad F_{\O_{\mathcal S},\O_{\mathcal C}}^{(0)}(1^+,2^+,3^{\bar\psi^{123}},4^{\psi^4};q)\,=\, -\frac{[12][24][41]}{[34]} \, , 
\end{align}
where we have explicitly indicated the $R$-symmetry indices. If at least one scalar is included in the external states, then we need to distinguish the two cases, 
{\it e.g.}
\begin{align}
\label{FF-OC-2scalars}
F_{\O_{\mathcal C}}^{(0)}(1^+,2^+,3^{\phi^{12}},4^{\phi^{34}};q)\,=\, -\frac{1}{2}\frac{[12]}{[34]}([13][24]+[14][23]) \, , 
\end{align}
while
\begin{align}
\label{FF-OS-2scalars}
F_{\O_{\mathcal S}}^{(0)}(1^+,2^+,3^{\phi^{12}},4^{\phi^{34}};q)\,=\, F_{\O_{\mathcal C}}^{(0)}(1^+,2^+,3^{\phi^{12}},4^{\phi^{34}};q)+\frac{1}{6} [12]^2 \, ,
\end{align}
where the extra term arises due to a correction of the form, schematically,  $\Tr(F^2 \phi\bar\phi)$ in $\O_{\mathcal S}$.
On the other hand if the two scalars are not adjacent we find
\begin{align}
\label{FF-OCS-2scalars}	
F_{\O_{\mathcal C}}^{(0)}(1^+,2^{\phi^{12}},3^+,4^{\phi^{34}};q)\,=\, 0 \, , \qquad F_{\O_{\mathcal S}}^{(0)}(1^+,2^{\phi^{12}},3^+,4^{\phi^{34}};q)\,=\, -\frac{1}{3}\, [13]^2 \, .
\end{align}
Finally we present a few examples involving fermions in the external states which have vanishing form factor for the operator $\O_{\mathcal C}$.
\begin{align}
\label{FF-OS-2fermions}
\begin{split}
F_{\O_{\mathcal S}}^{(0)}(1^+,2^{\psi^4},3^{\phi^{23}},4^{\psi^1};q)\,& =\, -\frac{2}{3} [12][14] \, , \\ 
F_{\O_{\mathcal S}}^{(0)}(1^+,2^{\psi^4},3^{\psi^1},4^{\phi^{23}};q)\,&=\, \frac{1}{3} [12][13] \, , \\
F_{\O_{\mathcal S}}^{(0)}(1^{\psi^4},2^{\psi^3},3^{\psi^2},4^{\psi^1};q)\,&=\, \frac{1}{3} \left(
[12][34]-[14][23] \right) \, .
\end{split}
\end{align}
The examples in \eqref{FF-OS-2fermions} and \eqref{FF-OCS-2scalars} have no kinematic poles and are produced by the contact terms inside $\O_{\mathcal S}$.

A final comment is in order. One could equivalently consider form factors of the parity-conjugate operator $\overline{\O}_{\mathcal C}$, with all helicities of external particles flipped. These are obtained from the form factors of $\O_{\mathcal C}$ by the replacement $\langle a\, b \rangle \leftrightarrow -[a\, b]$. In terms of states, this corresponds to performing the transformation
\beq
\phi^{AB}\to {1\over 2} \epsilon_{ABCD} \phi^{CD} \coloneqq \phi_{AB} = (\phi^{AB})^{\ast}\, , \quad \psi^{ABC}\to \epsilon_{ABCD} \psi^{D} \, , \quad \psi^D \to {1\over 3!} \epsilon_{ABCD}\psi^{ABC}
\ . 
\eeq
Similarly, we also note that the $\overline{\mathrm{MHV}}$ form factors of $\O_{\mathcal S}$ are easily found
using the helicity-flip rule $\langle a\, b \rangle \leftrightarrow -[a\, b]$ on \eqref{eq:FF-OSbar}.

\section{One-loop minimal form factors}\label{Sec:3}

An important ingredient needed to compute two-loop form factors using generalised unitarity cuts is the one-loop correction to the minimal form factor of the operators $\O_{\mathcal S}$ and $\O_{\mathcal C}$. In both cases the only non-vanishing result is obtained for an external state of three positive-helicity gluons and we will shortly see that the form factors of operators $\O_{\mathcal S}$ and $\O_{\mathcal C}$ turn out to be identical at one loop. 

\begin{figure}[h]
\centering
\includegraphics[width=0.3\linewidth]{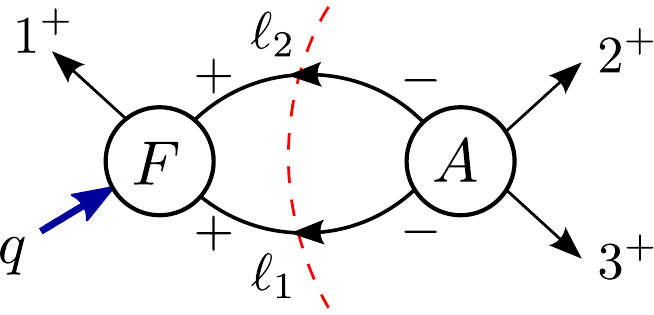}
\caption{\it A two-particle cut of the one-loop minimal form factor of $\O_{\mathcal{S}}$ or $\O_\mathcal{C}$.}
\label{fig:double-cut}
\end{figure}

The form factors of $\O_{\mathcal S}$ or $\O_{\mathcal C}$ are completely determined by the two-particle cut shown in Figure~\ref{fig:double-cut} together with its cyclic permutations.

\noindent 
The tree-level MHV gluon amplitude entering this cut is 
\begin{align}\label{eq:tree-MHV-amp}
A^{(0)}(\ell_1^{-},\ell_2^{-}, 2^+, 3^+)\,=\,i\,\frac{\b{\ell_1\ell_2}^3}{\b{\ell_2 2}\b{23}\b{3\ell_1}}\,,
\end{align}
whereas the tree-level form factor is given in \eqref{FF-OS-n3}.%
\footnote{Note that  in the pictorial notation we employ in this paper  each line represents a propagator stripped of the factor of $i$. Such factors of $i$ arising from (cut) propagators are collected separately.}
Denoting the $m$-particle cut of an $L$-loop form factor in a generic $P^2$-channel by
\begin{align}
F^{(L)}_{\Op}(\ldots;q)\Big|_{m,P^2}\,,
\end{align}
the two-particle cut of the one-loop form factor is given by%
\begin{align}
\label{eq:oneloop}
F^{(1)}_{\O_{\mathcal S},\O_{\mathcal C}}(1^+,2^+,3^+;q)\Big|_{2,s_{23}}\,=\,i\,[23]^2\,\frac{[1|\ell_1\,\ell_2|1]}{2(p_2\cdot\ell_1)}\,.
\end{align}
The cuts in the $s_{12}$- and $s_{13}$-channels are obtained by relabelling this expression. Putting everything together, manipulating the cut integrand and performing a Passarino-Veltman (PV) reduction, we arrive at an expression where the cut integrals can be lifted off shell unambiguously. Indeed, any  ambiguities would arise from the numerator of  
  \eqref{eq:oneloop} and would necessarily have the form  $[1| \ell_1 \ell_1 |1]=0$. 
We obtain%
\footnote{Expressions for the one-loop master integrals can be found in Appendix \ref{App:Integrals}.} 
\begin{align}\label{eq:one-loop-result}
F^{(1)}_{\O_{\mathcal S},\O_{\mathcal C}}(1^+,2^+,3^+;q)\,&=\,i\,F^{(0)}_{\O_{\mathcal S},\O_{\mathcal C}} \left(2\times\pic{0.4}{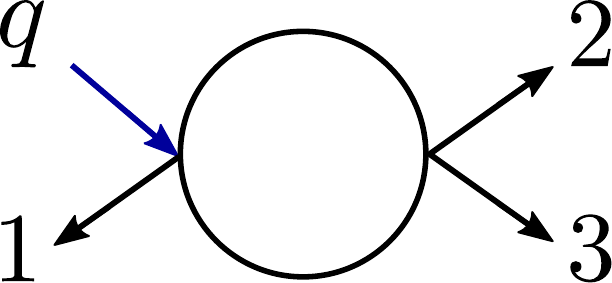}\,+\,s_{23}\times \pic{0.4}{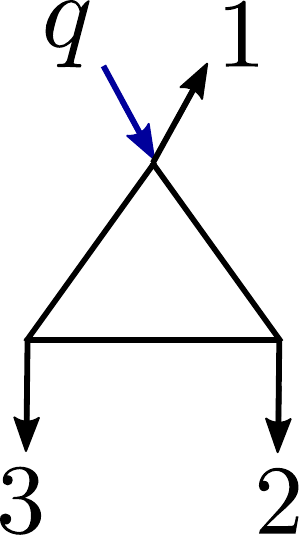}\,+\,(\text{cyclic}\,1,2,3)\right)\,.
\end{align}
Note that this formula should be multiplied by $g^2N$, which combines into a factor of 
the 't Hooft coupling 
\begin{align}\label{eq:tHooft}
a \, \coloneqq \, {g^2 N \over (4 \pi)^2}\,, 
\end{align}
after absorbing  a factor of $1/ ( 4 \pi)^2$ from the definition of the integral functions. 
Inspecting \eqref{eq:one-loop-result}, we can make the following observations:
\begin{itemize}
\item[\bf{1.}] Due to the normalisation of the tree-level form factor \eqref{eq:tree-level-F3} the one-loop correction is universal for both operators $\O_{\mathcal S}$ and $\O_{\mathcal C}$. It is moreover important for the results presented in \cite{Part2} to note that the one-loop form factor is theory-independent, {\it i.e.} the same whether computed in pure or supersymmetric Yang-Mills. Theory-dependence will manifest itself at two and more loops. 
\item[\bf{2.}] As mentioned in the Introduction, and crucially for  future investigations at higher loops, the result \eqref{eq:one-loop-result} has no additional rational terms even in pure Yang-Mills which could arise from the use of $D$-dimensional cuts as compared to four-dimensional cuts, see the discussion in \cite{Neill:2009mz}. \color{black}
\item[\bf{3.}]
Comparing \eqref{eq:one-loop-result} with the expression for the one-loop form factor of $\O_B=\Tr(X[Y,Z])$ obtained in \cite{Brandhuber:2016fni}\footnote{$X,\,Y$ and $Z$ are the three complex scalar fields of $\N\!=\!4$ SYM.}, we see that the one-loop form factors coincide, up to factoring out the corresponding tree-level form factor.
\item[\bf{4.}]
Using \eqref{eq:one-loop-result} we can extract the one-loop anomalous dimensions of $\O_{\mathcal S}$ and $\O_{\mathcal C}$ at one loop from the coefficient of the ultraviolet-divergent bubble integral. It turns out that at this order these operators are eigenstates of the dilatation operator with anomalous dimension
\begin{align}
\gamma_{\O_{\mathcal S},\O_{\mathcal C}}^{(1)} = 12\, a\,.
\end{align}
This is the same as the one-loop anomalous dimension of $\O_B$ found in \cite{Brandhuber:2016fni}.
\end{itemize}
The latter two observations, together with the fact that at zero coupling $\O_{B}$ and $\O_{\mathcal S}$ are related by supersymmetry transformations, was the original motivation for the study of the two-loop form factor of $\O_B$ presented in \cite{Brandhuber:2016fni} -- a stepping stone towards understanding the two-loop form factor of $\O_{\mathcal S}$.


\section{Two-loop minimal form factors in $\N\!=\!4$ SYM}
\label{Sec:4}

In this section we determine the two-loop form factors of the operators $\O_{\mathcal S}$ and $\O_{\mathcal C}$ 
introduced in Section \ref{Sec:2} using the following strategy. 

\begin{figure}[h]
\centering
\includegraphics[width=0.6\linewidth]{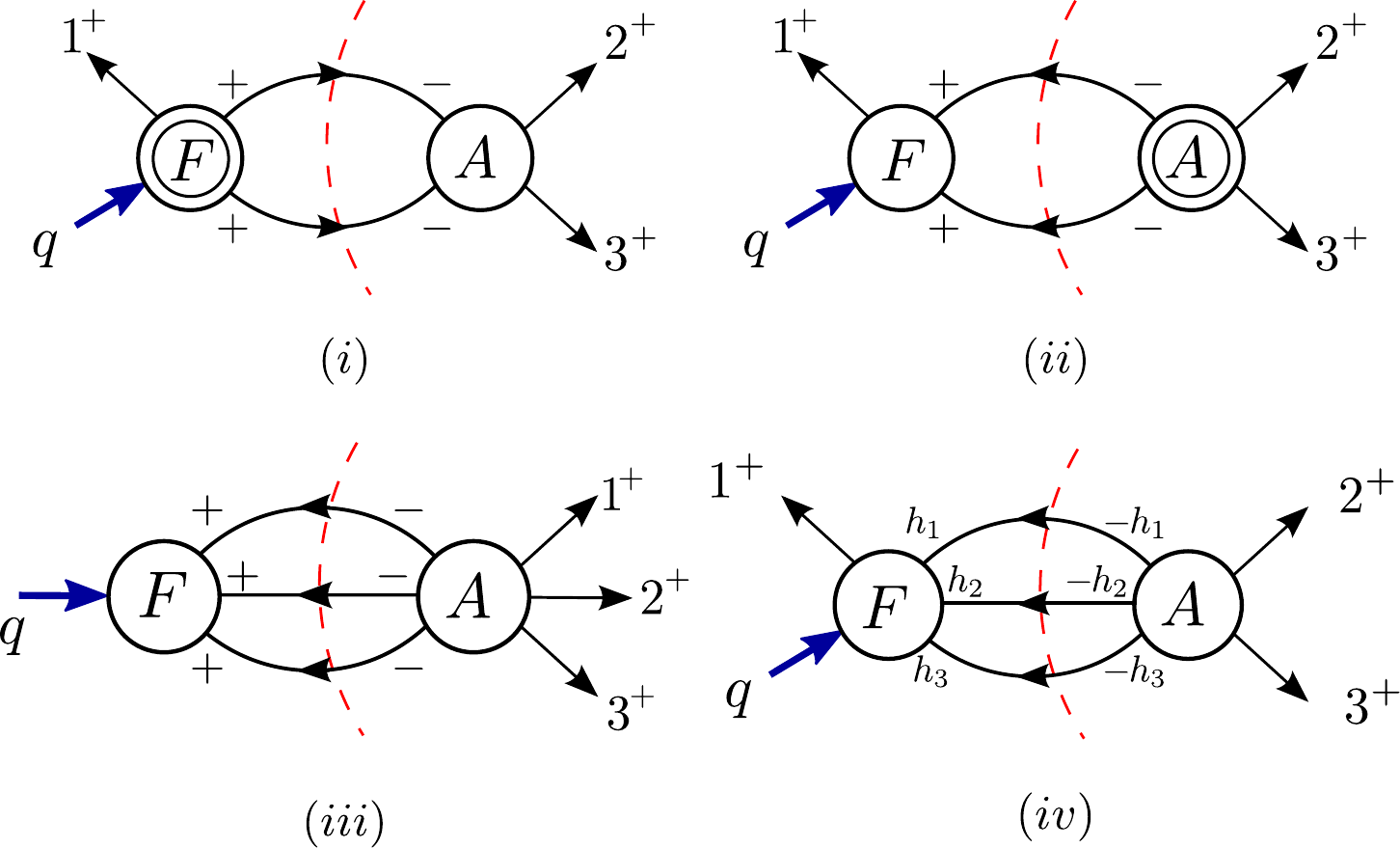}
\caption{\it Four different cuts of the two-loop form factors which will be used to construct the two-loop integrand.}
\label{fig:allcuts}
\end{figure}

\begin{itemize}
\item[{\bf 1.}]
First, we consider two-particle cuts in one of the possible kinematic channels, for example the $s_{23}$-channel. 
There are two cuts to consider, shown in Figure \ref{fig:allcuts}$(i)$ and $(ii)$. 
\item[{\bf 2.}]
We then move on to the three-particle cut in the $q^2$-channel, as in Figure \ref{fig:allcuts}$(iii)$, which we use to fix potential ``ambiguities" of the previous result and to detect integral topologies which do not have a two-particle cut. 
By ambiguity we mean here the fact that for two cut momenta, $p_i$ and $p_j$, it is impossible to distinguish between their Mandelstam invariant $(p_i+p_j)^2$ and their scalar product $2(p_i\cdot p_j)$. This is due to the fact that the cutting procedure puts the two momenta on shell, $p_{i,j}^2=0$. As a result, if a dot product involving these momenta features in the numerator of an integral detected by a cut involving $p_{i}$ and $p_j$ we must use further cuts, which do not involve simultaneously both momenta $p_{i}$ and $p_j$, in order to resolve the ambiguity. 
\item[{\bf 3.}]
Finally we turn to the more involved three-particle cut in the $s_{23}$-channel, presented in Figure \ref{fig:allcuts}$(iv)$, where we  fix all remaining ambiguities of the integrand. 
\item[{\bf 4.}]
By consistently merging the results of all the cuts, we construct the complete four-dimensional integrand at two loops.
\end{itemize}

\subsection{Two-particle cuts}

We begin by calculating the two-particle cuts of the two-loop form factor. These can only be considered in the $s_{23}$-channel as in the $q^2$-channel the two-particle cut would lead to a subminimal tree-level form factor, which does not exist at this loop order. 
We proceed to consider the following two different two-particle cuts in the $s_{23}$-channel: the case with $F^{(0)}\times A^{(1)}$ and that with $F^{(1)}\times A^{(0)}$.

\subsubsection{Tree-level form factor $\times$ one-loop amplitude}

\begin{figure}[h]
\centering
\includegraphics[width=0.30\linewidth]{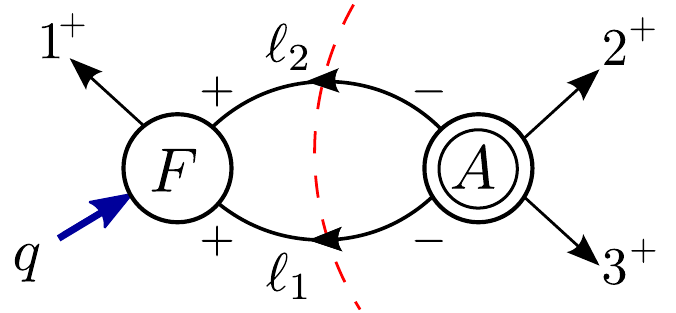}
\caption{\it A double cut of the two-loop minimal form factor of $\O_{\mathcal S}$, $\O_{\mathcal C}$: the case of a tree-level form factor joined to a one-loop amplitude.}
\label{fig:double-cut2}
\end{figure}

We consider the two-particle cut presented in Figure \ref{fig:double-cut2}, whose ingredients are a tree-level form factor and a one-loop amplitude. Similarly to the one-loop case, this cut is universal for the two operators, $\O_{\mathcal S}$ and $\O_{\mathcal C}$, due to the equality of the tree-level minimal form factors \eqref{FF-OS-n3}. 

\noindent The four-point one-loop amplitude in $\N\!=\!4$ SYM on the right-hand-side of the cut has a very simple form,
\begin{align}\label{eq:N=4-1-loop-amp}
A^{(1)}(\ell_1^{-}, \ell_2^{-},2^+, 3^+) = A^{(0)}(\ell_1^{-}, \ell_2^{-},2^+, 3^+)\left[-s_{23}s_{\ell_2 2}\times\pic{0.35}{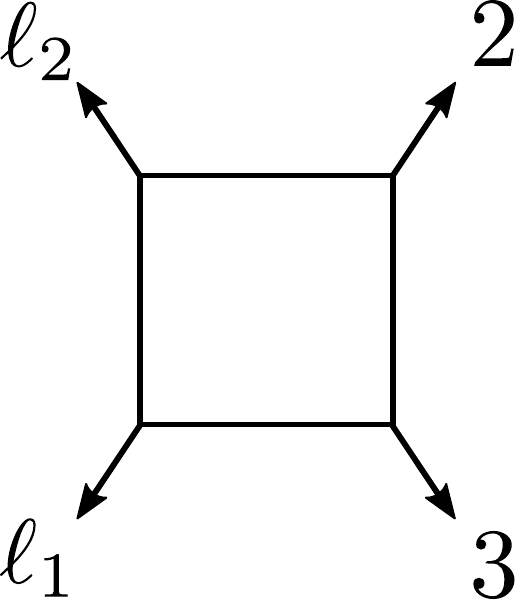}\,\right]\,,
\end{align}
Gluing the amplitude \eqref{eq:N=4-1-loop-amp} to the form factor \eqref{FF-OS-n3} and reinstating the cut propagators we arrive at the following result for this two-particle cut: 
\begin{align}
\begin{split}\label{eq:resultF0A1}
F^{(2)}_{\O_{\mathcal S}}(1^+,2^+,3^+;q)\Big|_{2,s_{23}} =\, \,F^{(0)}_{\O_{\mathcal S}}\,s_{23}^2\,\frac{[1|q \, \ell_1|1]}{[12]\la 23 \ra [31]}\times\pic{0.4}{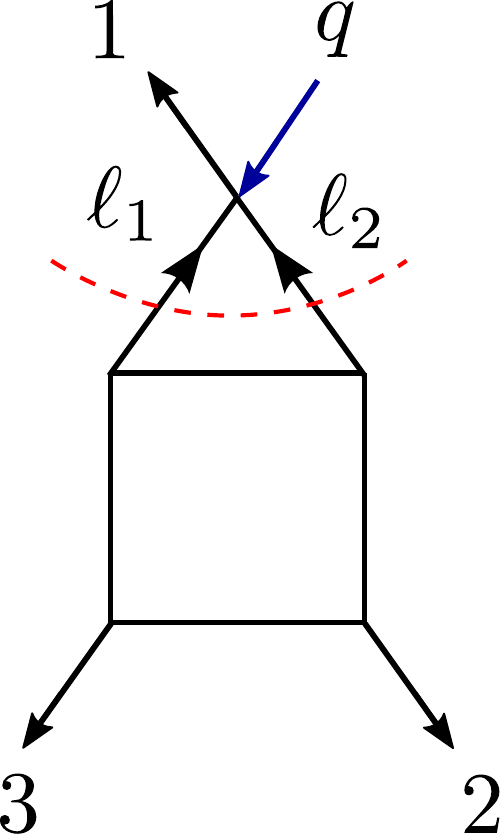}+\text{cyclic}(1,2,3)\,.
\end{split}
\end{align}

\subsubsection{One-loop form factor $\times$ tree-level amplitude}\label{subsec:ff1timesatree}

\begin{figure}[h]
\centering
\includegraphics[width=0.30\linewidth]{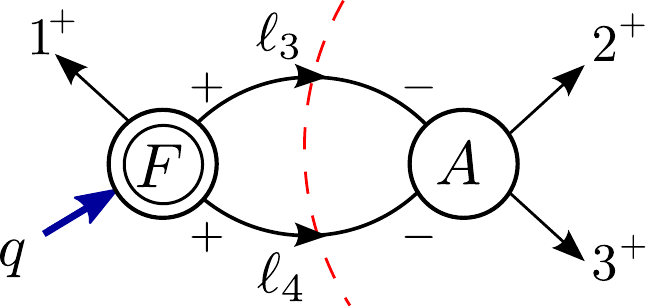}
\caption{\it A double cut of the two-loop minimal form factor of $\O_{\mathcal S},\,\O_{\mathcal C}$ -- the case of a one-loop form factor joined to a tree-level amplitude.}
\label{fig:double-cut3}
\end{figure}

Next we turn our attention to the second of the two-particle cuts, shown in Figure \ref{fig:double-cut3}, in which we glue a one-loop minimal form factor and a tree-level amplitude. As discussed in Section \ref{Sec:3} the one loop form factor is the same for $\O_{\mathcal{S}}$ and $\O_{\mathcal C}$ and as a result this entire cut is identical for the two operators. 

\begin{figure}[h]
\centering
\includegraphics[width=0.15\linewidth]{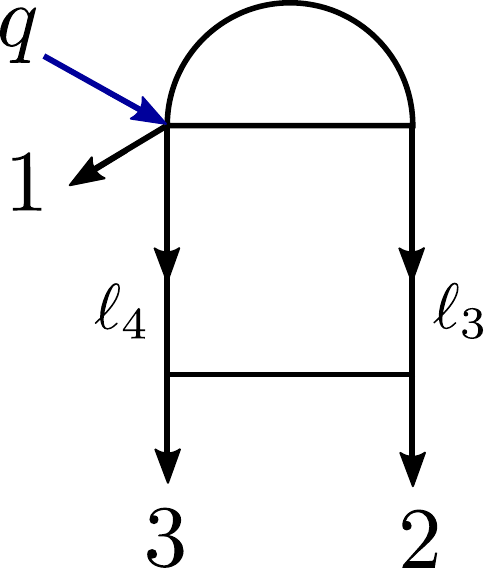}
\caption{\it Integral topology that   cannot be detected by the two-particle $s_{23}$-channel cut.}
\label{fig:peng-left}
\end{figure}

\noindent
In order to construct the integrand, it is important that we use the expression for the one-loop form factor \eqref{eq:one-loop-result} prior to PV reduction. One reason is that the reduction procedure discards certain integrals that vanish in dimensional regularisation, {\it e.g.}~scaleless bubbles.
Such an integral may appear as a sub-topology inside a two-loop integral, with the momentum flowing in the sub-bubble being now off shell (when lifted off the cut); this topology should therefore not be discarded. Thus, in order to obtain the complete result for this cut we use the expression for the one-loop form factor {\it before} the reduction, namely:
\begin{align}
F^{(1)}_{\O_{\mathcal S},\O_{\mathcal C}}(1^+,2^+,3^+;q)\,&=i\,\left(\frac{s_{23}}{\b{23}}\right)^2[1|q\,\ell|1]\times\pic{0.4}{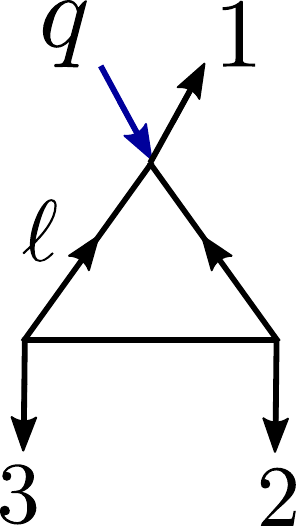}+\text{cyclic}(1,2,3)\,.
\end{align}
Using the tree-level amplitude in \eqref{eq:tree-MHV-amp} and conveniently rewriting it as
\begin{align}
A^{(0)}(\ell_4^{-},\ell_3^{-}, 2^+, 3^+)\,=\,
-\,i\,\frac{\la\ell_3\ell_4\ra^2}{\la23\ra^2}\frac{s_{23}}{2(p_2\cdot\ell_3)}=-i\,s_{23}\left(\frac{\la\ell_3\ell_4\ra}{\la23\ra}\right)^2\times\pic{0.4}{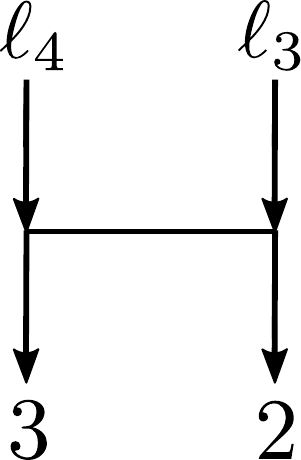}\,,
\end{align}
we arrive at the following expression for the two-particle cut:
\begin{align}
\label{eq:2p-cut-FF1loop}
\begin{split}
&F^{(2)}_{\O_{\mathcal S}}(1^+,2^+,3^+;q)\Big|_{2,s_{23}} \,=\, -s_{23}\left(\frac{\la\ell_3\ell_4\ra}{\la23\ra}\right)^2 \times \Big[
\left(\frac{s_{\ell_3\ell_4}}{\b{\ell_3\ell_4}}\right)^2 [1|q\,\ell|1]\times\pic{0.4}{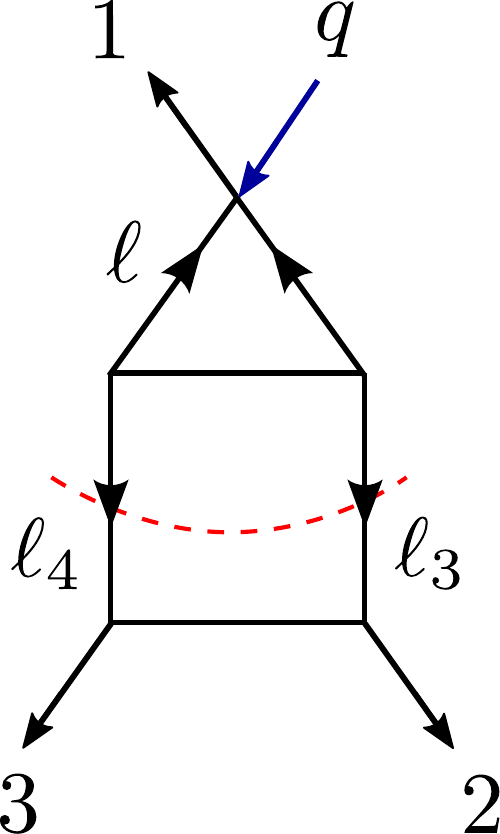}\\
& + \left(\frac{s_{\ell_4 1}}{\b{\ell_4 1}}\right)^2 [\ell_3|q\,\ell|\ell_3]\times\pic{0.4}{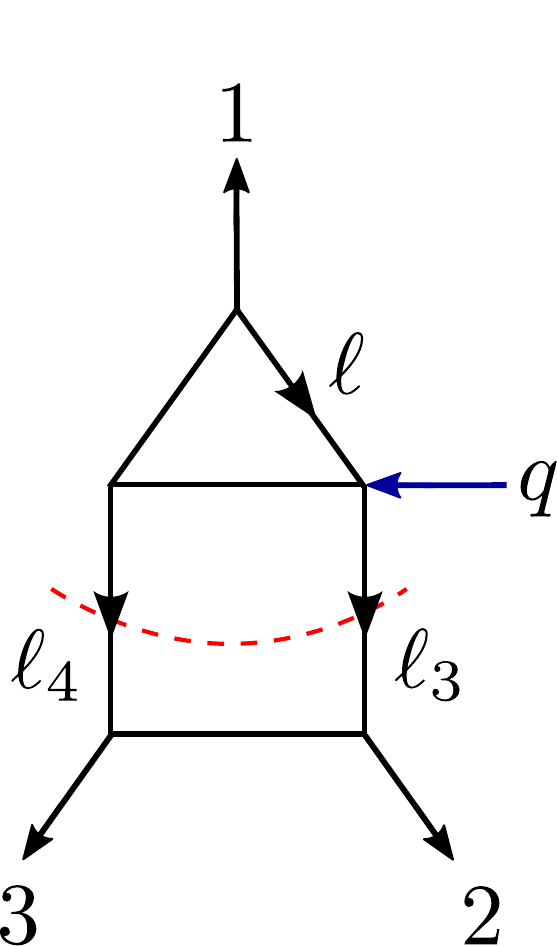}\ +\ \left(\frac{s_{1\ell_3}}{\b{1\ell_3}}\right)^2 [\ell_4|\ell\,q|\ell_4]\times\pic{0.4}{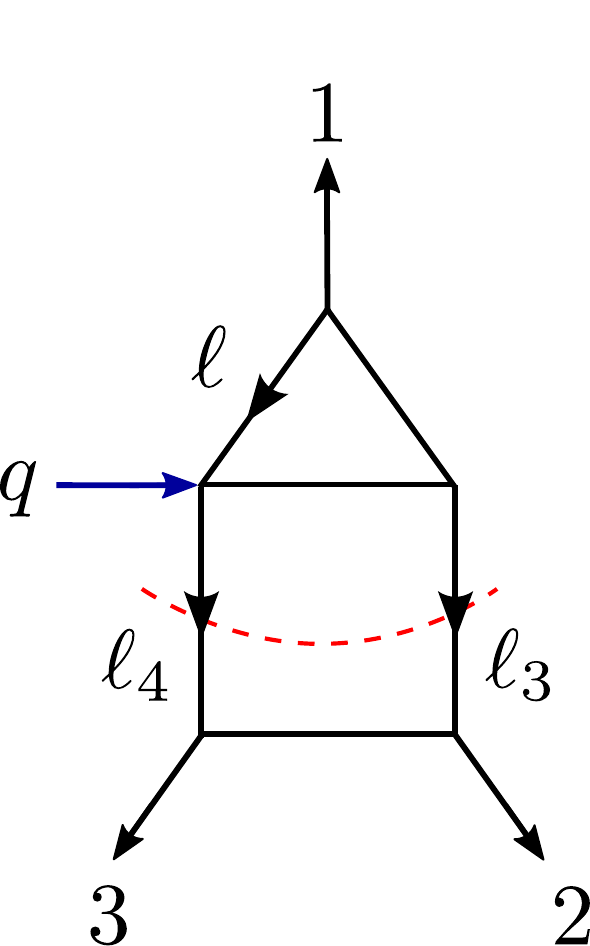}\Big]\ .
\end{split}
\end{align}
The first integral in \eqref{eq:2p-cut-FF1loop} with its numerator can be simplified to
\begin{align}
\begin{split}\label{eq:2p-cut-result}
-\frac{s_{23}^3}{\la23\ra^2}{[1|q\,\ell|1]}\times\pic{0.4}{SOTT-cut2}\,
=\,F^{(0)}_{\O_{\mathcal S}}\,s_{23}^2\,\frac{[1|q\, \ell|1]}{[12]\la23\ra[31]}\times\pic{0.4}{SOTT-cut2}\,.
\end{split}
\end{align}
We immediately see that this is identical to the result of the two-particle cut \eqref{eq:resultF0A1}, where we have computed the case of $F^{(0)}\times A^{(1)}$. This would lead to the conclusion that the correct answer is obtained by simply lifting \eqref{eq:2p-cut-result} off shell, however 
an important subtlety arises here. Indeed, any term proportional to $\ell^2$ (or $(\ell+p_2 + p_3)^2$) would cancel one of the propagators and generate the integral topology in Figure \ref{fig:peng-left} (or its mirror). 

When $\ell_3$ and $\ell_4$ are cut a (scale-free) bubble on the form factor side is isolated, which vanishes in dimensional regularisation. As a result, we cannot make any meaningful statement about the presence of this topology given the information provided only by this pair of two-particle cuts, and we must defer the verdict until three-particle cuts have been considered. This will be discussed in detail in Section \ref{Sec:merging}.

In order to perform an integral reduction using \texttt{LiteRed} \cite{Lee:2012cn, Lee:2013mka}, it is useful to rewrite the numerator of \eqref{eq:2p-cut-result} as
\begin{align}\label{eq:numerator-rewrite}
\begin{split}
s_{23}^2\frac{[1|q\, \ell|1]}{[12]\la23\ra[31]}\,=\,\,&\frac{s_{23}}{2s_{13}}\left(s_{23}s_{\ell1}-s_{\ell3}s_{12}+s_{13}s_{\ell2}\right)\,-\,\,\frac{s_{23}}{2s_{12}}\left(s_{23}s_{\ell1}-s_{\ell2}s_{13}+s_{12}s_{\ell3}\right)\,.
\end{split}
\end{align}
We now perform a PV reduction on the terms which contain the invariant $s_{1\ell}$ since any dependence on $p_1$ is unphysical (only the combination $q-p_1$ is relevant).
Following standard steps we find that
\begin{align}
s_{1\ell}\times \pic{0.4}{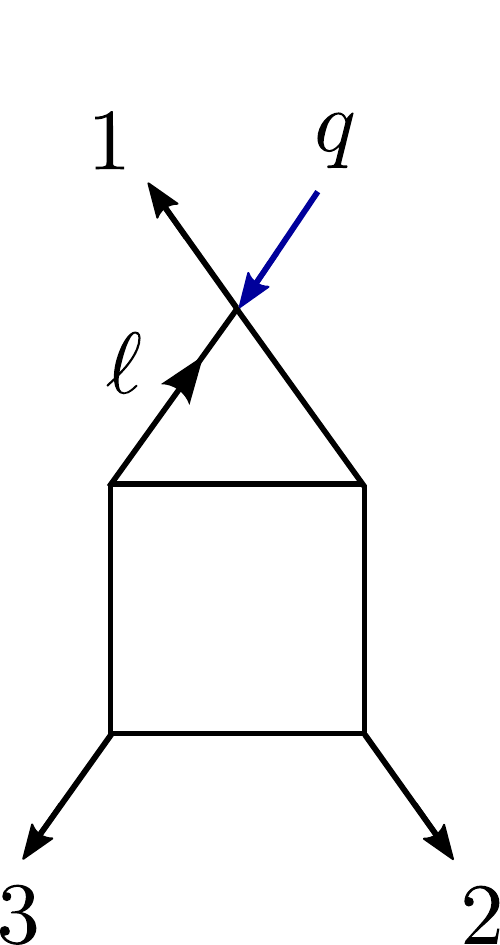}\,=\,\frac{1}{s_{23}}\Big[s_{12}s_{3\ell}+s_{13}s_{2\ell}\Big]\times \pic{0.4}{SOTT-nocut}\ .
\end{align}
Inserting this result into \eqref{eq:numerator-rewrite}, we find that \eqref{eq:2p-cut-result} becomes
\begin{align}\label{eq:SOTT-num-simplified}
-\frac{s_{23}^3}{\la23\ra^2}{[1|q\,\ell|1]}\times\pic{0.4}{SOTT-nocut}\,
=\,F^{(0)}_{\O_{\mathcal S}}(1^+,2^+,3^+;q)\,s_{23}\,(s_{2\ell}-s_{3\ell})\times \pic{0.4}{SOTT-nocut}\ . 
\end{align}
Note that $p_1$ no longer appears in the numerator, as desired.
Inspecting the result of the two-particle cut in \eqref{eq:SOTT-num-simplified} we see that, because of the numerator factor $(s_{2\ell}-s_{3\ell})$ it is impossible to say at this stage whether $s_{2\ell}$ and $s_{3\ell}$ stand for a full invariant or just a scalar product of two momenta -- the $\ell^2$-terms which would arise from the full invariants cancel in the difference. This is a manifestation of the ambiguity mentioned earlier, leading to topologies of the type depicted in Figure \ref{fig:peng-left}. This matter will be settled in Section~\ref{Sec:merging} by means of a three-particle cut. 

We now move to the second term of \eqref{eq:2p-cut-FF1loop}. After factoring out the tree-level form factor, it can be rewritten as
\begin{align}
\begin{split}\label{eq:num-SOTR}
\left(\frac{s_{\ell_4 1}}{\b{\ell_4 1}}\right)^2 [\ell_3|q\,\ell|\ell_3]\times\pic{0.4}{SOTR-cut2}=F^{(0)}_{\O_{\mathcal S}}(1^+,2^+,3^+;q)\ \frac{\Tr_+(1\,q\,\ell_3\, q\, \ell\,\ell_3\, q\,1\,3\,2)}{s_{12}s_{23}s_{13}}\,\times\pic{0.4}{SOTR-cut2}\,,	
\end{split}
\end{align}
while the numerator of the third integral of \eqref{eq:2p-cut-FF1loop} can be obtained from \eqref{eq:num-SOTR} upon relabelling $(\ell_3\leftrightarrow\ell_4\,, 2 \leftrightarrow 3)$
\begin{align}
\begin{split}\label{eq:num-SOTL}
F^{(0)}_{\O_{\mathcal S}}(1^+,2^+,3^+;q)\ \frac{\Tr_+(1\,q\,\ell_4\, q\, \ell\,\ell_4\, q\,1\,2\,3)}{s_{12}s_{23}s_{13}}\,\times\pic{0.4}{SOTL-cut2}\,.	
\end{split}
\end{align}

\subsubsection{Summary of results after two-particle cuts}

For the reader's convenience, we summarise in Table \ref{table:summary1} the results of the cuts we have performed so far. 
We have presented each distinct topology with the corresponding numerator we have detected. The result after the two particle cuts consists of the three topologies with their 
numerators and the two remaining cyclic shifts of the external momentum labels.
\begin{center}
\begin{table}
\centering
\begin{tabular}{|c|c|c|c|}
\hline
\shortstack[c]{\rule{0pt}{10pt}\bf Integral\\ \\ \bf topology}
&$\pic{0.4}{SOTT-nocut}$
&$\pic{0.4}{SOTR-cut2}$
&$\pic{0.4}{SOTL-cut2}$ 
\\ 
&&& \\ 
\hline 
&&& \\
{\bf Numerator} & $s_{23}\,(s_{2\ell}-s_{3\ell})$& $\frac{\Tr_+(1\,q\,\ell_3\, q\, \ell\,\ell_3\, q\,1\,3\,2)}{s_{12}s_{23}s_{13}}$&$\frac{\Tr_+(1\,q\,\ell_4\, q\, \ell\,\ell_4\, q\,1\,2\,3)}{s_{12}s_{23}s_{13}}$
\\ 
&&& 
\\ \hline 
&&& \\
{\bf Ambiguity} &$\ell$&$\ell_3$, $\ell_4$&$\ell_3$, $\ell_4$\\ 
&&&\\
\hline
\end{tabular}
\caption{\it Summary of the results of the two-particle cuts so far. All numerators have the tree-level form factor factored out. The propagators which appear cut are still ambiguous given the cuts performed so far.}
\label{table:summary1}
\end{table}
\end{center}

\subsection{Three-particle cut in $q^2$-channel}

\begin{figure}[!h]
\centering
\includegraphics[width=0.30\linewidth]{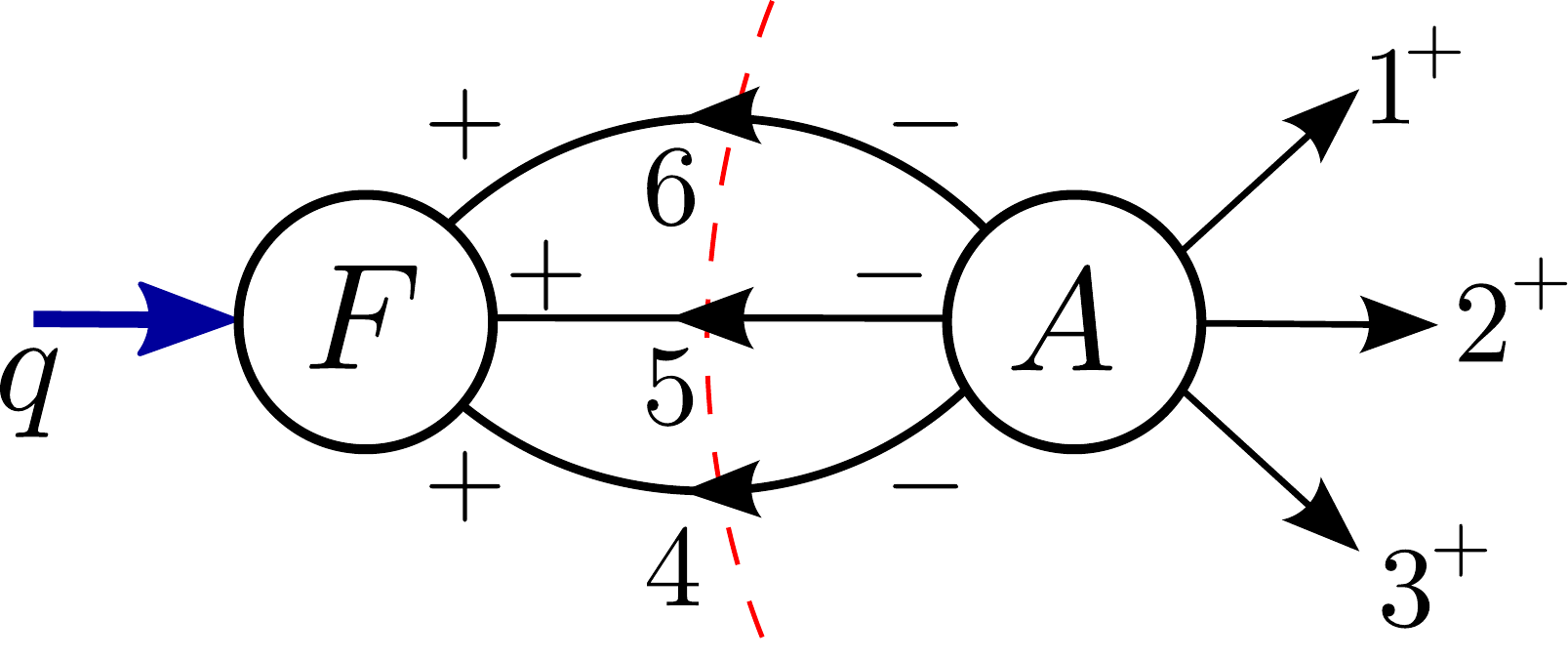}
\caption{\it Triple cut of the two-loop form factor in the $q^2$-channel. Only one possible helicity assignment exists.}
\label{fig:q2-triple-cut}
\end{figure}

In this section we consider the three-particle cut of the two-loop form factor in the $q^2$-channel, as presented in Figure \ref{fig:q2-triple-cut}. We note that for this channel there exists only one possible helicity assignment for the momenta running in the loop -- all gluons. 

\noindent For the six-point tree-level gluon amplitude, we use the expression of \cite{Mangano:1990by}, which reads
\begin{align}
\begin{split}\label{eq:mangano-amp}
A(1^+,2^+,3^+,4^-,5^-,6^-)\,&=\,i\,\Big[\overbrace{\frac{([23]\la56\ra[1|p_2\!+\!p_3|4\ra)^2}{s_{234}s_{23}s_{34}s_{56}s_{61}}}^{\beta^2}+\overbrace{\frac{([12]\la45\ra[3|p_1\!+\!p_2|6\ra)^2}{s_{345}s_{34}s_{45}s_{61}s_{12}}}^{\gamma^2}\\
\,&+\frac{\overbrace{s_{123}[23]\la56\ra[1|p_2\!+\!p_3|4\ra[12]\la45\ra[3|p_1\!+\!p_2|6\ra}^{\beta\gamma}}{s_{12}s_{23}s_{34}s_{45}s_{56}s_{61}}\Big]\,,
\end{split}
\end{align}
and for the tree-level form factor, as before, we use \eqref{FF-OS-n3}. We now consider the contribution of each term separately.

\noindent{\bf $\beta^2$-term:} The first term in \eqref{eq:mangano-amp} gives rise to a previously-detected topology, namely
\begin{align}\label{eq:SOTL-cut3}
F^{(0)}_{\O_{\mathcal S}}(1^+,2^+,3^+;q)\frac{\Tr_+(1\,q\,4\, 5\, 6\,4\, q\,1\,2\,3)}{s_{12}s_{23}s_{13}}\,\times \pic{0.4}{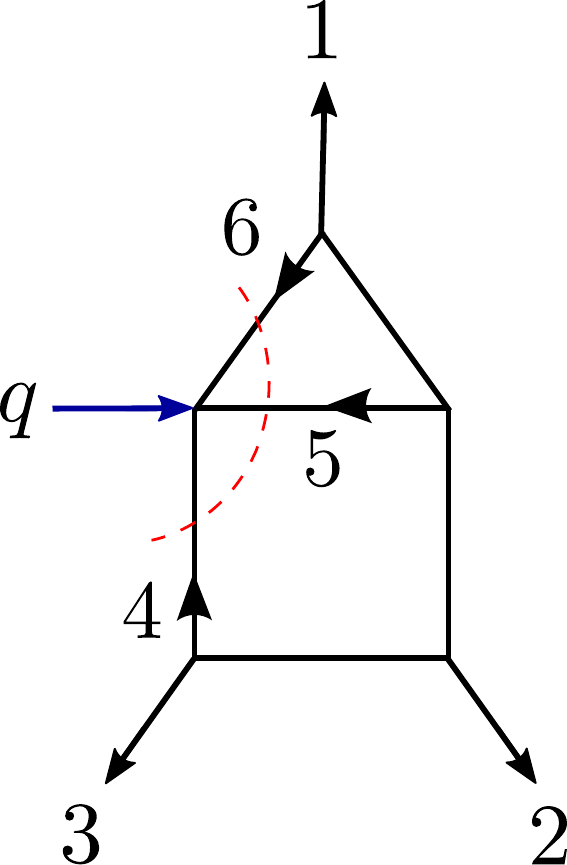}\,.
\end{align}
After an appropriate relabelling, it is easy to see that the numerator becomes identical to that of \eqref{eq:num-SOTL}, obtained from a two-particle cut. \\[10pt]
\noindent{\bf $\gamma^2$-term:} Considering the second term in \eqref{eq:mangano-amp} we detect a similarly familiar topology, namely
\begin{align}\label{eq:SOTR-cut3}
F^{(0)}_{\O_{\mathcal S}}(1^+,2^+,3^+;q)\frac{\Tr_+(3\,q\,6\, 5\, 4\,6\, q\,3\,2\,1)}{s_{12}s_{23}s_{13}}\,\times\pic{0.4}{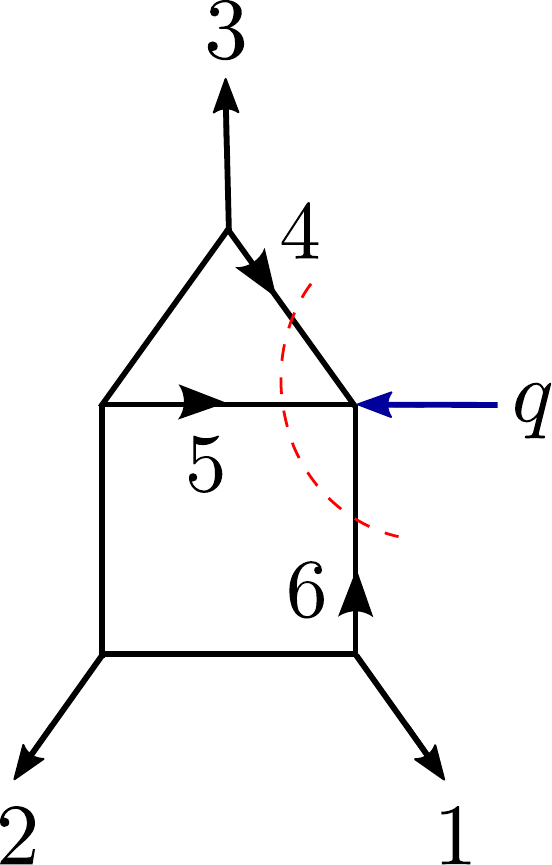}\,.
\end{align}
Once again, after an appropriate relabelling we observe that the numerator is the same as in \eqref{eq:num-SOTR}, showing that the results for this topology obtained from two- and three- particle cuts are mutually consistent.
\\[10pt]
\noindent{\bf $\beta\gamma$-term:} Finally, we consider the third term in \eqref{eq:mangano-amp}, for which we obtain
\begin{align}
\begin{split}\label{eq:num-rake}
F^{(0)}_{\O_{\mathcal S}}(1^+,2^+,3^+;q)\frac{s_{123}}{s_{12}s_{23}s_{13}}\Tr_+(1q46q3)\times\pic{0.4}{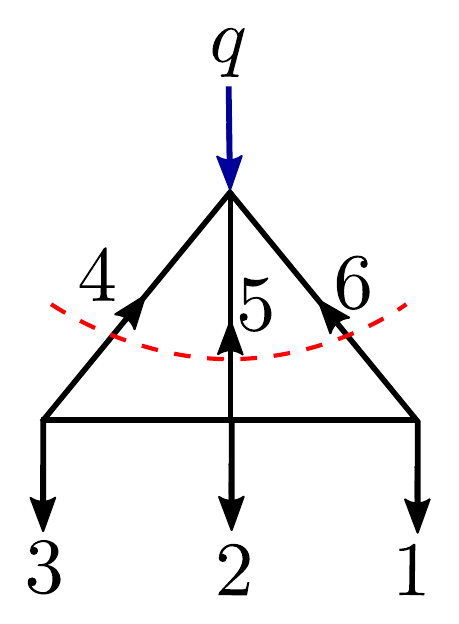}\,.
\end{split}
\end{align}
This is a new topology which could not have been detected by any of the two-particle cuts. As such, we add it to our result for the integrand. The numerator of this last integral will be confirmed by a different three-particle cut considered in the next section. Table~\ref{table:summary2} summarises the integrand as found by the cuts studied up to this point.

\begin{center}
\begin{table}[!ht]
\centering
\begin{tabular}{|c|c|c|c|c|}
\hline
\shortstack[c]{\rule{0pt}{20pt}\bf Integral\\ \\ \bf topology} 
&$\pic{0.4}{SOTT-nocut}$&$\pic{0.4}{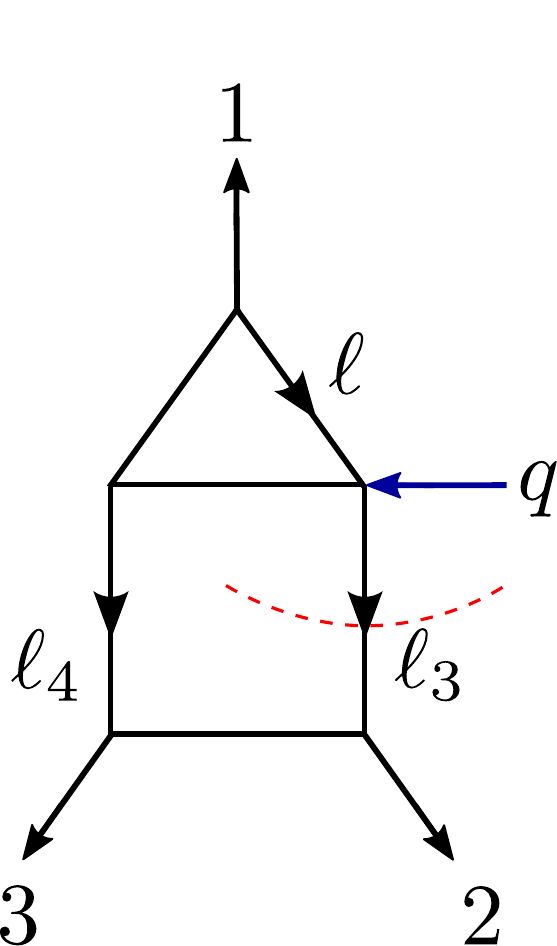}$&$\pic{0.4}{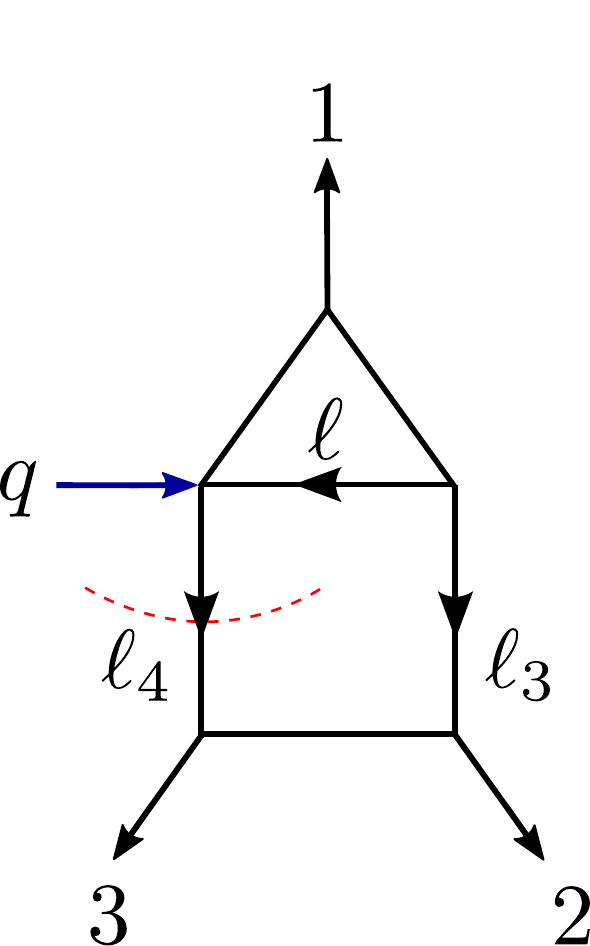}$&$\pic{0.4}{Rake3-3cut}$ \\ 
&&&& 
\\ \hline 
&&&& \\
{\bf Numerator} & $s_{23}\,(s_{2\ell}-s_{3\ell})$& $\frac{\Tr_+(1\,q\,\ell_3\, q\, \ell\,\ell_3\, q\,1\,3\,2)}{s_{12}s_{23}s_{13}}$&$\frac{\Tr_+(1\,q\,\ell_4\, q\, \ell\,\ell_4\, q\,1\,2\,3)}{s_{12}s_{23}s_{13}}$&$\frac{s_{123}}{s_{12}s_{23}s_{13}}\Tr_+(1q46q3)
$
\\ 
&&&& \\ 
\hline 
&&&& \\
{\bf Ambiguity} &$\ell$&$\ell_3$&$\ell_4$&$p_4$, $p_6$\\ 
&&&&\\
\hline
\end{tabular}
\caption{\it Summary of the result after the two-particle cuts and the three-particle cut in the $q^2$-channel. All numerators have the tree-level form factor factored out. The propagators which are cut are still ambiguous given the cuts performed so far.}
\label{table:summary2}
\end{table}
\end{center}

\subsection{Three-particle cut in $s_{23}$-channel}

\begin{figure}[h]
\centering
\includegraphics[width=0.30\textwidth]{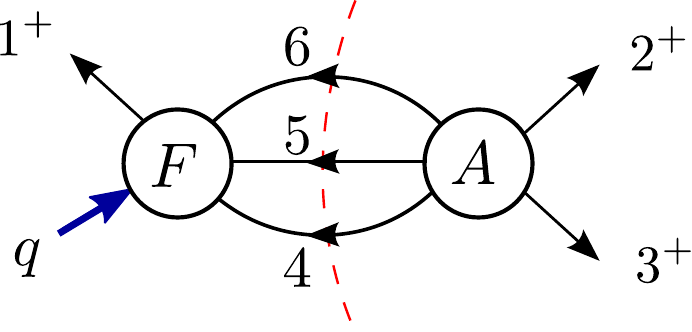}
\caption{\it Triple cut of the two-loop form factor in the $s_{23}$-channel.}
\label{fig:s23-generic}
\end{figure}

In this section we compute the last three-particle cut of the two-loop form factor we need to consider: the $s_{23}$-channel cut presented in Figure \ref{fig:s23-generic}. This is the most intricate cut, as it involves a non-minimal form factor, and we will see that it provides the necessary final constraints to fix the two-loop form factor integrand completely. The motivation to analyse this cut is two-fold: first, we would like to fix potential ambiguities in the numerators of the other previously detected topologies (shown in Table \ref{table:summary2}) since they all have a non-vanishing three-particle cut in the $s_{23}$-channel. Moreover, we expect to observe new integrals which have non-vanishing cuts only in this channel. 

This cut also carries important information that distinguishes the two-loop form factors of the operators $\O_\mathcal{S}$ and $\O_\mathcal{C}$, as well as the theory under study. Since it features a non-minimal tree-level form factor, fermions and scalars can run in the loops, unlike the case of the triple cut in the $q^2$-channel. As a result, the non-minimal form factor is sensitive to the choice of operator and number of supersymmetries, as confirmed by the expressions for tree-level form factors in Section \ref{sec:treeFFs}. In what follows, we will work first with the operator $\O_{\mathcal C}$, and then move on to consider the operator $\O_{\mathcal S}$. We begin by presenting the ingredients of the computation and subsequently discuss the methodology and results.
Form factors with reduced amount of supersymmetry are discussed in \cite{Part2}.

\subsubsection{Component calculation}

Working in components, the triple cut in the $s_{23}$-channel requires us to consider separately all possible configurations of gluons, fermions and scalars for the particles running in the loop. Below we discuss each case in turn.

\begin{figure}[h]
\centering
\includegraphics[width=0.30\textwidth]{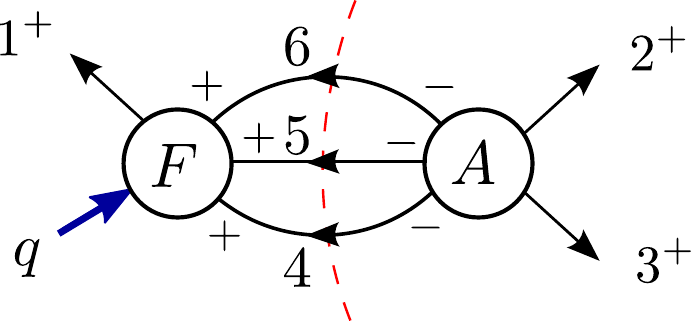}
\caption{\it Triple cut of the two-loop form factor in the $s_{23}$-channel with only gluons running in the loop involving an $\overline{\rm MHV}$ amplitude.}
\label{fig:s23-gluons-1}
\end{figure}

\noindent{\bf Gluons in the loop:}
First, we consider diagrams where only gluons are running in the loop. There are two possible cases, involving either an MHV or $\overline{\rm MHV}$ amplitude (and a corresponding $\overline{\rm MHV}$ or next-to-$\overline{\rm MHV}$ form factor respectively). The case with an $\overline{\rm MHV}$ amplitude is presented in Figure \ref{fig:s23-gluons-1}, and there is only one possible helicity configuration for the internal particles.

\begin{figure}[h]
\centering
\includegraphics[width=0.95\textwidth]{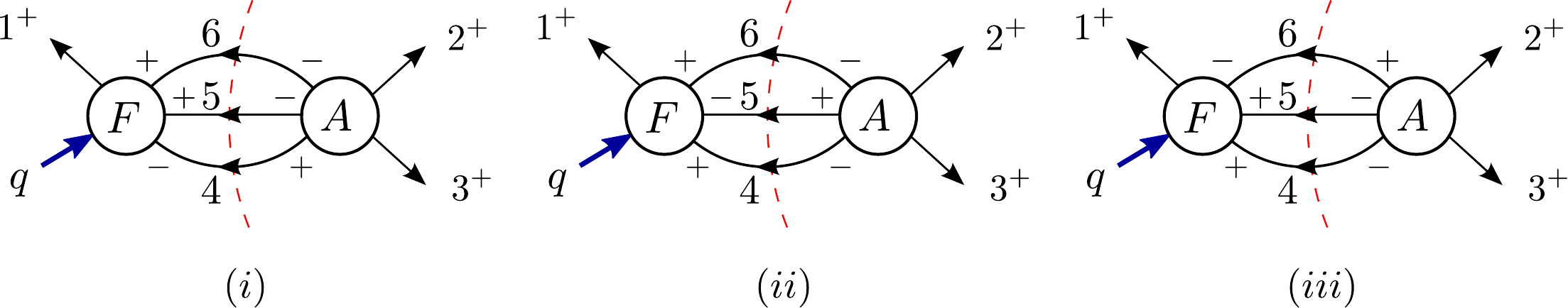}
\caption{\it Triple cut of the two-loop form factor in the $s_{23}$-channel with only gluons running in the loop: $F^{\overline{\rm MHV}}\times A^{\rm MHV}$.}
\label{fig:s23-gluons-2-3-4}
\end{figure}

We have computed the tree-level form factor entering the cut using MHV diagrams \cite{Cachazo:2004kj} applied to form factors \cite{Dixon:2004za, Brandhuber:2011tv, Dawson:2014ora}. The result was quoted in the second line of \eqref{eq:treeFFs-n4}, and we write here  for convenience: 
\begin{align}
\begin{split}
\label{eq:allplus}
F^{(0)}_{\O_{\mathcal C}}(1^+,-6^+,-5^+,-4^+;q)\,&=\,
-[16][65][54][41]\bigg[ 
{1\over s_{16}} \left(1-\frac{[51][4|q|5\ra}{s_{56}[41]}\right)\\[5pt]
\, -\, {1\over s_{56}} \left(1-\frac{[46][1|q|4\ra}{s_{45}[16]}\right)
\, &-\,{1\over s_{54}} \left(1-\frac{[15][6|q|1\ra}{s_{14}[65]}\right)
\, +\,{1\over s_{14}} \left(1+\frac{[64][5|q|6\ra}{s_{16}[54]}\right)
\bigg]\, ,
\end{split}
\end{align}
while the five-point tree-level $\overline{\rm MHV}$ amplitude is given by
\begin{align}\label{eq:5pt-amp}
A^{(0)}(2^+,3^+,4^-,5^-,6^-)\,=\,-i\,\frac{[23]^3}{[34][45][56][62]}\,.
\end{align}
The second possible internal helicity assignment involves an MHV amplitude. In this case, there are three configurations depending on the position of the internal positive-helicity gluon. These are indicated in Figure \ref{fig:s23-gluons-2-3-4}.
The form factors entering the cuts above are a part of an $\overline{\rm MHV}$ family whose expression is known for any number of legs \cite{Broedel:2012rc}, in particular 
\begin{align}
\label{eq:MHVFF}
\begin{split}
F^{(0)}_{\O_{\mathcal C}}(1^+,-6^+,-5^+,-4^-;q)\,&=\,\frac{[16][65][51]^2}{[54][41]}\,,\\
F^{(0)}_{\O_{\mathcal C}}(1^+,-6^+,-5^-,-4^+;q)\,&=\,\frac{[16][64]^2[41]}{[65][54]}\,,\\
F^{(0)}_{\O_{\mathcal C}}(1^+,-6^-,-5^+,-4^+;q)\,&=\,\frac{[15]^2[54][41]}{[16][65]}\,.
\end{split}
\end{align}
For the tree-level MHV amplitudes entering the cut we have
\begin{align}
\begin{split}
A^{(0)}(2^+,3^+,4^+,5^-,6^-)\,&=\,i\,\frac{\la56\ra^3}{\la23\ra\la34\ra\la45\ra\la62\ra}\,,\\
A^{(0)}(2^+,3^+,4^-,5^+,6^-)\,&=\,i\,\frac{\la46\ra^4}{\la23\ra\la34\ra\la45\ra\la56\ra\la62\ra}\,,\\
A^{(0)}(2^+,3^+,4^-,5^-,6^+)\,&=\,i\,\frac{\la45\ra^3}{\la23\ra\la34\ra\la56\ra\la62\ra}\,.
\end{split}
\end{align}\\

\begin{figure}[h]
\centering
\includegraphics[width=0.65\textwidth]{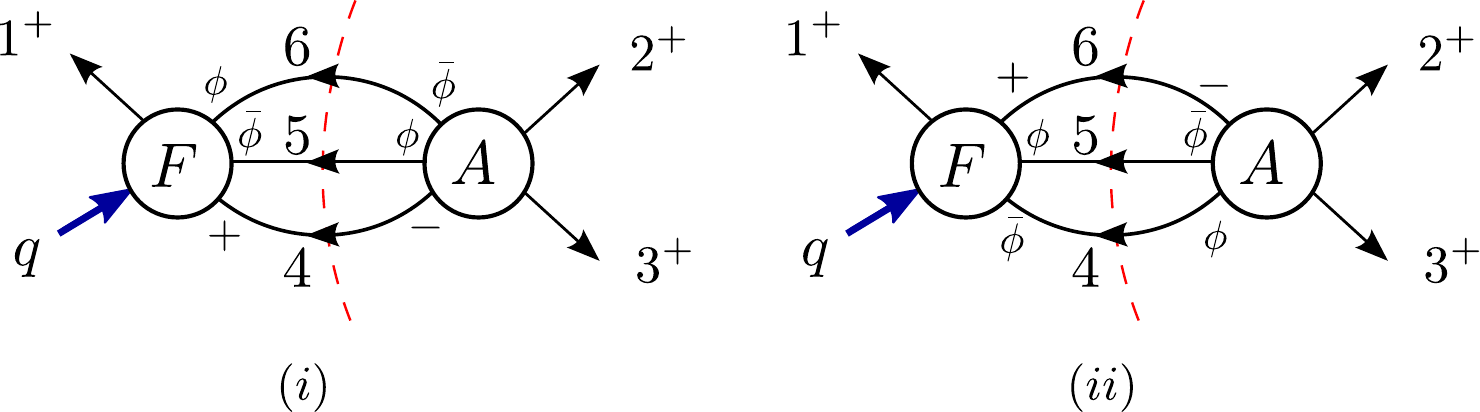}
\caption{\it Triple cut of the two-loop form factor in the $s_{23}$-channel with scalars and a gluon running in the loop.}
\label{fig:triple-cut-scalars}
\end{figure}

\noindent{\bf Scalars in the loop:} We now consider the case where we allow scalars to run in the loop in addition to gluons, as presented in Figure \ref{fig:triple-cut-scalars}.
The non-minimal tree-level form factor for the configuration in Figure~\ref{fig:triple-cut-scalars}~$(i)$ is
\begin{align}\label{eq:ff-scalars1}
F^{(0)}_{\O_{\mathcal C}}(1^{+},-6^{\phi},-5^{\bar\phi},-4^+;q)\,=\,
 -{1\over 2}
\frac{[14]}{[65]}\left([54][16]+[51][46]\right)\,,
\end{align}
while the tree-level amplitude is given by
\begin{align}
A^{(0)}(2^+,3^+,4^-,5^{\phi},6^{\bar\phi})\,=\,i\,\frac{\la45\ra\la46\ra^2}{\la23\ra\la34\ra\la56\ra\la62\ra}\,.
\end{align}
We note that the result of this diagram needs to be multiplied by a factor of 3 to account for the three distinct complex scalar/anti-scalar pairs arising from the splitting of the gluon in $\N\!=\!4$ SYM. 
One could also imagine diagrams where we assign the scalars in the opposite way, with $\bar\phi$ incoming into the form factor on leg $p_6$ and $\phi$ on leg $p_5$.
However, the form factor and amplitude turn out to be identical to those of the previous case, hence such diagram would lead to the same result as that in Figure~\ref{fig:triple-cut-scalars}~$(i)$. We multiply our result by a further factor of 2 to account for this. 

The second configuration of scalars we need to consider is presented in Figure~\ref{fig:triple-cut-scalars}~$(ii)$ (note that the two scalars can only be adjacent as they arise from the splitting of a gluon into a scalar/anti-scalar pair).
In this case, the tree-level form factor and amplitude read
\begin{align}
\begin{split}
F^{(0)}_{\O_{\mathcal C}}(1^+,-6^+,-5^{\phi},-4^{\bar\phi};q)\,&=\,-\hf\frac{[16]}{[54]}\left([46][51]+[41][56]\right)\,,\\
A^{(0)}(2^+,3^+,4^{\phi},5^{\bar\phi},6^{-})\,&=\,i\,\frac{\la56\ra\la46\ra^2}{\la23\ra\la34\ra\la45\ra\la62\ra}\,.
\end{split}
\end{align}
Similarly to the case discussed above, we need to multiply this result by $6$ in order to account for the helicity state sum and the opposite assignment of scalar/anti-scalar pair for the internal legs.

\begin{figure}[h]
\centering
\includegraphics[width=0.65\textwidth]{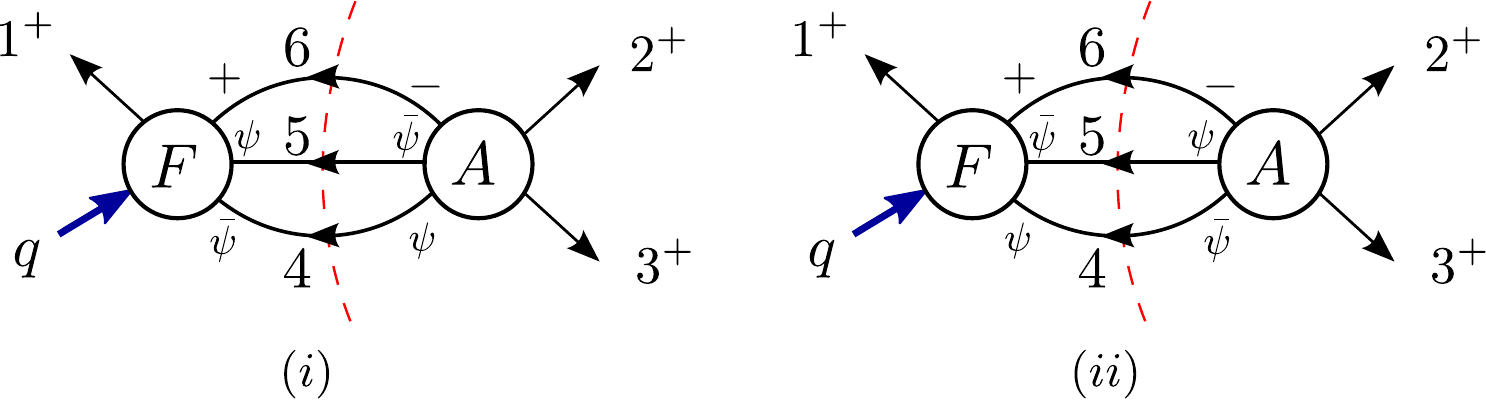}
\caption{\it Triple cut of the two-loop form factor in the $s_{23}$-channel -- fermions and a gluon running in the loop, the first possible configuration.}
\label{fig:s23-triplecut-cut-fermions-1}
\end{figure}

\begin{figure}[h]
\centering
\includegraphics[width=0.65\textwidth]{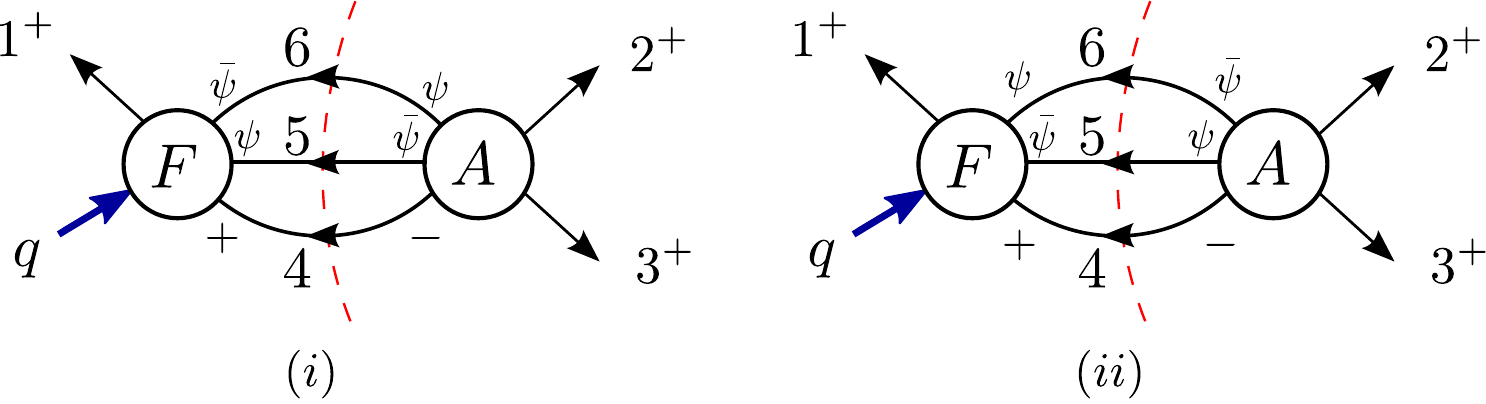}
\caption{\it Triple cut of the two-loop form factor in the $s_{23}$-channel - fermions and a gluon running in the loop, the second possible configuration.}
\label{fig:s23-triplecut-cut-fermions-2}
\end{figure}

\noindent{\bf Fermions in the loop:} Finally, we consider the case with fermions running in the loop, as shown in Figure \ref{fig:s23-triplecut-cut-fermions-1}.
The calculation of the non-minimal tree-level form factors gives
\begin{align}
\begin{split}
&F^{(0)}_{\O_{\mathcal C}}(1^{+},-6^{+},-5^{\psi},-4^{\bar\psi};q)\,=\,-\frac{[51][56][16]}{[54]}\,,\\
&F^{(0)}_{\O_{\mathcal C}}(1^{+},-6^{+},-5^{\bar\psi},-4^{\psi};q)\,=\,\frac{[41][46][16]}{[54]}\,,
\end{split}
\end{align}
while the tree-level amplitudes entering the cuts are
\begin{align}
\begin{split}
&A^{(0)}(2^{+},3^{+},4^{\psi},5^{\bar\psi},6^-)\,=\,i\,\frac{\la56\ra^2\la46\ra}{\la23\ra\la34\ra\la45\ra\la62\ra}\,,\\
&A^{(0)}(2^{+},3^{+},4^{\bar\psi},5^{\psi},6^-)\,=\,-i\,\frac{\la46\ra^3}{\la23\ra\la34\ra\la45\ra\la62\ra}\,.
\end{split}
\end{align}
The second possible helicity configuration is that presented in Figure \ref{fig:s23-triplecut-cut-fermions-2}.

\noindent In this case, the tree-level form factors are
\begin{align}
\begin{split}\label{eq:ff-fermion}
\quad&F^{(0)}_{\O_{\mathcal C}}(1^{+},-6^{\bar\psi},-5^{\psi},-4^{+};q)\,=\,\frac{[54][51][41]}{[65]}\,,\\
\quad&F^{(0)}_{\O_{\mathcal C}}(1^{+},-6^{\psi},-5^{\bar\psi},-4^{+};q)\,=\,-\frac{[64][61][41]}{[65]}\,,
\end{split}
\end{align}
and the tree-level amplitudes are
\begin{align}
\begin{split}\label{eq:amp-fermion}
\quad&A^{(0)}(2^{+},3^{+},4^{-},5^{\bar\psi},6^\psi)\,=\,-i\,\frac{\la45\ra^2\la46\ra}{\la23\ra\la34\ra\la56\ra\la62\ra}\,,\\
\quad&A^{(0)}(2^{+},3^{+},4^{-},5^{\psi},6^{\bar\psi})\,=\,i\,\frac{\la46\ra^3}{\la23\ra\la34\ra\la56\ra\la62\ra}\,.
\end{split}
\end{align}
We note that each of the results for the calculation of a cut involving fermions should be multiplied by a factor of $4$ in order to account for the possible $R$-symmetry index assignments.\\[10pt]
As mentioned earlier, this three-particle cut carries the most distinguishing information between the operators $\OC$ and $\OS$ and the theory. Having collected all of the ingredients necessary for the calculation of the two-loop form factor of the component operator $\O_{\mathcal C}$, we move on to do the same for the supersymmetric descendant of the Konishi, $\O_{\mathcal S}$. The methodology to derive this cut is the same for both operators and as such we defer the discussion of it to Section \ref{sec:strategy}.

\subsubsection{Supersymmetric calculation}\label{sec:calc-emery}
 
The operator $\O_{\mathcal S}$ introduced in Section \ref{Sec:2} is a tree-level descendant of the Konishi operator, whose MHV form-factors can be extracted from \eqref{eq:FF-Konishi} \cite{Chicherin:2016qsf}. Once an appropriate component of the super form factor (parity conjugate of \eqref{eq:FF-Konishi}) has been extracted, it captures all the helicity assignments discussed in the previous section, with the exception of the all-plus gluon case \eqref{eq:allplus} since the form factor is not $\overline{\rm MHV}$. As a result, an easier way to compute this cut is to multiply the appropriate $\overline{\rm MHV}$ component of the tree-level (parity conjugate of the) super-form factor \eqref{eq:FF-Konishi} by the corresponding MHV tree-level $\N\!=\!4$ super-amplitude,
\begin{align}
\A_5^{\rm {MHV}}(\lambda_i, \tilde\lambda_i,\eta_i)\,=\,i\,\frac{\delta^{(8)}\left(\sum_{i=1}^5\lambda_i^{\alpha}\eta_i^A\right)}{\la12\ra\la23\ra\la34\ra\la45\ra\la51\ra}\ ,
\end{align}
and integrate the internal fermionic variables. To this result, we then add the all-plus gluon form factor of \eqref{eq:allplus} multiplied by the corresponding amplitude \eqref{eq:5pt-amp}.
The individual expressions are lengthy and we refrain from presenting them here in full. We discuss the result of this calculation and contrast it with that of the component operator in Section \ref{sec:results}.

\subsubsection{Solving for the three-particle cuts}\label{sec:strategy}

Having collected all the ingredients for the evaluation of the triple cut in the $s_{23}$-channel, we proceed to discuss the methodology for finding the correct two-loop integrand for the desired form factors. Due to the complexity of the terms to be summed in this channel, each depending on high powers of loop momenta, we generate an ansatz with all possible integrand topologies and fix the precise combination by demanding consistency with the cut.
The procedure is as follows, explained here for the component operator $\O_{\mathcal C}$ and equivalent for the supersymmetric operator $\O_{\mathcal S}$:
\begin{itemize}
\item[\bf 1.] We combine the cut integrand expression, consisting of the sum of tree-level form factors \eqref{eq:allplus}--\eqref{eq:ff-fermion} multiplied by the corresponding tree-level amplitudes \eqref{eq:5pt-amp}--\eqref{eq:amp-fermion}, taking into account appropriate multiplicities arising from $R$-symmetry assignment. 

\item[\bf 2.] The integrated form factor does not contain parity-odd terms, but its integrand does. In order to work with a parity even integrand ansatz, we add to the cut expression its parity conjugate (and divide by 2). 
\item[\bf 3.] We construct an ansatz for the integrand in terms of integrals with non-trivial numerators in the following way. All possible two-loop topologies are obtained from the two maximal ones presented in Figure \ref{fig:two-loop-masters}
by pinching propagators; each topology produced in this way must then be cut in the $s_{23}$-channel in all possible ways, thereby generating the ansatz.
\begin{figure}[h]
\centering
\includegraphics[width=0.5\textwidth]{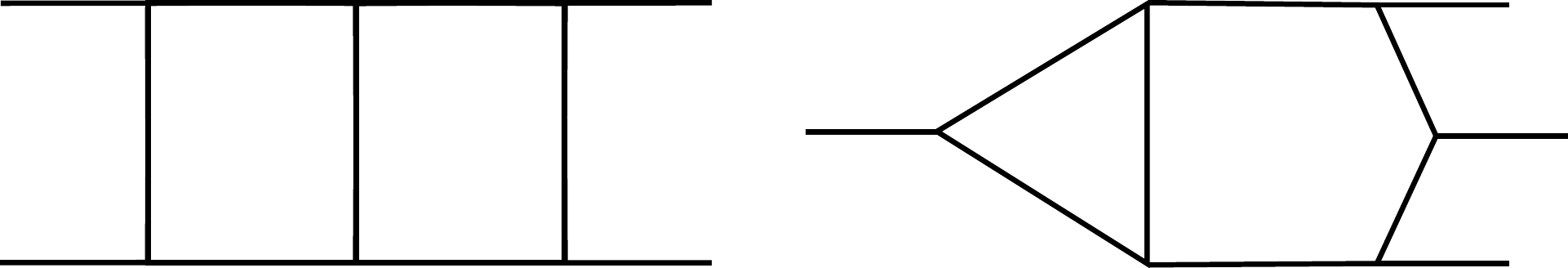}
\caption{\it Maximal two-loop topologies.}
\label{fig:two-loop-masters}
\end{figure}
\item[\bf 4.] Each of these cut topologies can be described using a basis of irreducible scalar products of the two loop momenta and the three external momenta. There are nine irreducible scalar products involving the loop momenta \cite{Lee:2012cn} and three further scalar products involving only the external legs, resulting in twelve irreducible scalar products from which to build numerators.
\item[\bf 5.] 
After choosing a basis of irreducible scalar products for the maximal topologies, we generate all possible numerators, up to a maximum power of loop momenta restricted by a theory-specific power counting. For example, for a Yang-Mills theory, a three-point (minimal) form factor carries three powers of  momenta  and each three-point Yang-Mills vertex carries one power of momentum.
\item[\bf 6.] We then write down a general linear combination of the integral topologies generated above and solve for the coefficients of each integral. Schematically, we have: 
\begin{align}
{\rm Cut\,integrand} \,=\, \sum_{i,j} c_{ij}\, {\rm Numerator}_{ij} \times {\rm \big[Cut\, Topology\big]}_j\,, 
\end{align}
\end{itemize}
where $i$ runs over all possible numerators appearing for a certain topology $j$. 
The result of the computation in this channel consists of hundreds of terms which we need to merge with the integrals obtained in the other cuts (see Table \ref{table:summary2}) to solve for the ambiguities and detect new integrals. In some cases, the comparison is immediate. In others, as discussed next, important subtleties arise.

\subsection{Merging the cuts}
\label{Sec:merging}
In this section, we combine the results of all generalised unitarity cuts of the two-loop form factor to finally obtain its integrand. Having obtained the triple cut in the $s_{23}$-channel we proceed to gather and reconcile the information obtained from different cuts in order to remove any ambiguities in the numerators of integral topologies. 
\begin{figure}[h]
\centering
\includegraphics[width=0.50\textwidth]{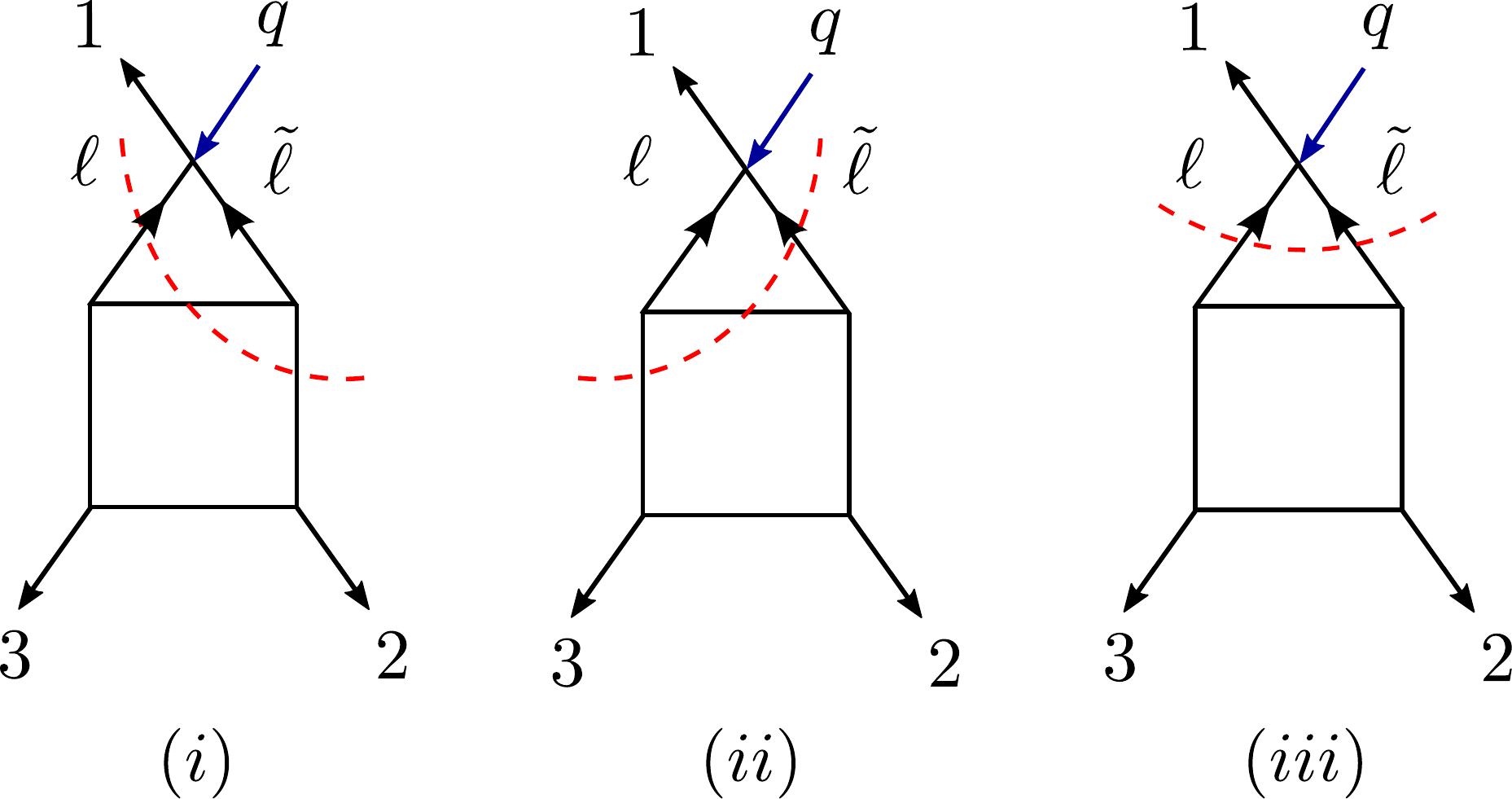}
\caption{\it Three cuts of one of the integral topologies.}
\label{fig:sott-merging}
\end{figure}

We illustrate this procedure using a specific example.
Figure \ref{fig:sott-merging} presents three different cuts of one of the integral topologies contributing to the result for the two-loop form factor. After PV reduction, the three numerators detected by the cuts are:
\begin{align}
N_i \,&=\, -s_{23}\left[s_{23}+4(\ell\cdot p_3)\right]\,,\label{eq:ni}\\
N_{ii} \,&=\, -s_{23}\left[s_{23}+4(\tilde\ell\cdot p_2)\right]\,,\label{eq:nii}\\
N_{iii} \,&=\, s_{23}(s_{2\ell}-s_{3\ell})\,,\label{eq:niii}
\end{align}
and we recall from the discussion in Section \ref{subsec:ff1timesatree} that on the basis of two particle cuts alone we were unable to conclusively tell whether the $s_{2\ell}$ and $s_{3\ell}$ in \eqref{eq:niii} denote the scalar products $2 (p_{2,3} \cdot \ell)$, or the full Mandelstam invariants $(p_{2,3} + \ell)^2$. With additional information from the three-particle cut in the $s_{23}$-channel we are now able to merge the three numerators into an unambiguous expression for the integrand.

The merging between \eqref{eq:ni} and \eqref{eq:nii} is straightforward. We can rewrite the two numerators as
\begin{align}
 N_i \,&=\, -s_{23}\Big[s_{23}+2(\ell+p_3)^2\Big]\,,\qquad
N_{ii} \,=\, -s_{23}\Big[s_{23}+2(\tilde\ell+p_2)^2\Big]\,,\label{eq:nii-bis}
\end{align}
which on the cut, at $\ell^2=0$ and $\tilde\ell^2=0$, respectively reduce to \eqref{eq:ni} and \eqref{eq:nii}. Momentum conservation $\ell+\tilde\ell+p_2+p_3=0$ implies that $(p_3+\ell)^2=(p_2+\tilde\ell)^2$, we see immediately that the two numerators are equivalent.

The merging between these two numerators and \eqref{eq:niii} is more subtle. We rewrite
\begin{align}
\begin{split}
2(\ell+p_3)^2\,&=\,(\ell+p_3)^2 + (\tilde\ell+p_2)^2\\
\,&=\,\ell^2+2(\ell\cdot p_3)+\tilde\ell^2-2(\ell\cdot p_2)-2(p_2 \cdot p_3)\\
\,&=\,\ell^2+\tilde\ell^2+s_{3\ell}\Big|_{\ell^2=0}-s_{2\ell}\Big|_{\ell^2=0}-s_{23}\,,
\end{split}
\end{align}
where in the second line we made use of momentum conservation. As a result, we have 
\begin{align}
N_i\,&=\,-s_{23}\left[s_{23}+2(\ell+ p_3)^2\right]\nonumber\\
\,&=\,-s_{23}(s_{23}+\ell^2+\tilde\ell^2+s_{3\ell}-s_{2\ell}-s_{23})\nonumber\\
\,&=\,N_{iii}-s_{23}(\ell^2+\tilde\ell^2)\,.\label{eq:merging}
\end{align}
The last term in \eqref{eq:merging} constitutes precisely the kind of ambiguity which could not have been detected by any two-particle cut. Using the information obtained from the three-particle cut, we add this term to our numerator, which now becomes:
\begin{align}\label{eq:NumMergedPV}
N\,=\,2s_{23}\left[(\ell\cdot p_2)-(\ell\cdot p_3)\right]-s_{23}(\ell^2+\tilde\ell^2)\,.
\end{align}
We note that the merging procedure could have been carried out using numerators before the PV reduction. We refrain from presenting such discussion here as the numerators involved are more complicated but the outcome is, upon PV reduction, equivalent to~\eqref{eq:NumMergedPV}. 

\begin{figure}[!h]
\centering
\includegraphics[width=0.50\textwidth]{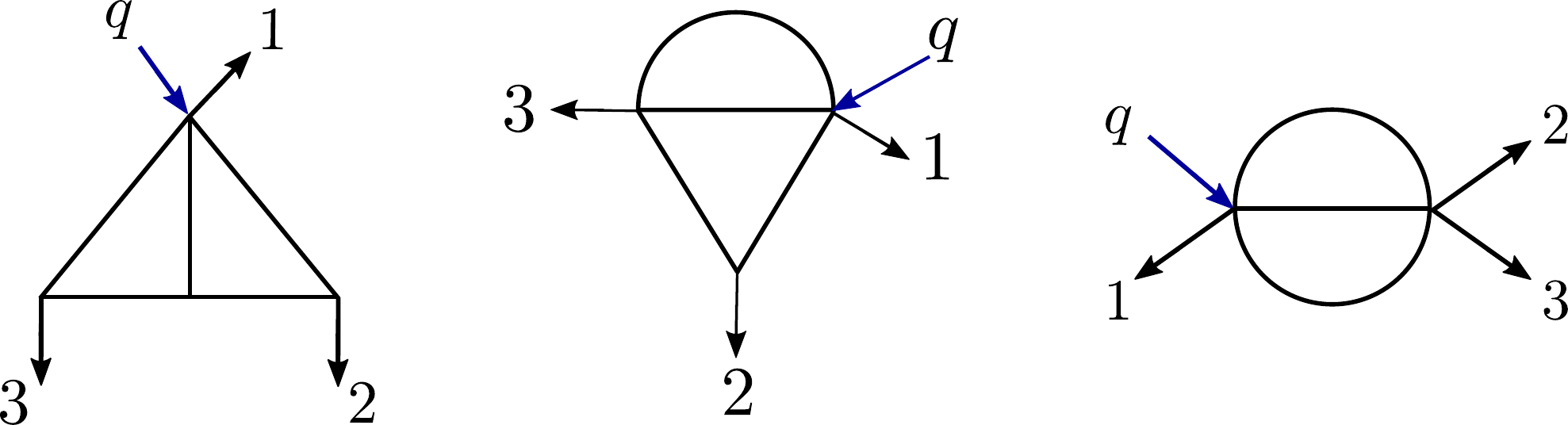}
\caption{\it Examples of topologies with only one valid cut, namely the three-particle cut in the kinematic $s_{23}$-channel.}
\label{fig:s23-cut-only}
\end{figure}

The result of the computation described in Section~\ref{sec:strategy} contains several topologies with {\it only} an $s_{23}$-channel three-particle cut, some of which are presented in Figure~\ref{fig:s23-cut-only}. Since we cannot obtain any other information about numerators of these topologies, we take them directly from the $s_{23}$-channel cut expression, which we then lift off shell. These topologies also do not carry any ambiguities as shrinking of any of the cut propagators would result in a vanishing integral in dimensional regularisation. We are now ready to present the results for the two-loop form factors of $\O_{\mathcal S}$ and $\O_{\mathcal C}$.

\subsection{Final result for the two-loop integrand in $\N\!=\!4$ SYM}\label{sec:results}
We begin by presenting the answer for the two-loop form factor of the supersymmetric operator $\O_{\mathcal S}$ as discussed in Section~\ref{sec:calc-emery}. We then move on to present the result of the component calculation for $\O_{\mathcal C}$ but we note that the sole difference between the two form factors lies in topologies detected only in the $s_{23}$-channel triple cut. In order to avoid redundancy, we will present the component result in terms of a difference from the supersymmetric result. We list integrals constituting the basis in Table~\ref{tab:two-loop basis} and the corresponding numerators in Appendix~\ref{App:Numerators}.

\begin{table}[!h]
\centering
\begin{tabular}{cccc}
$\pic{0.4}{SOTT-nocut}$ & $\pic{0.4}{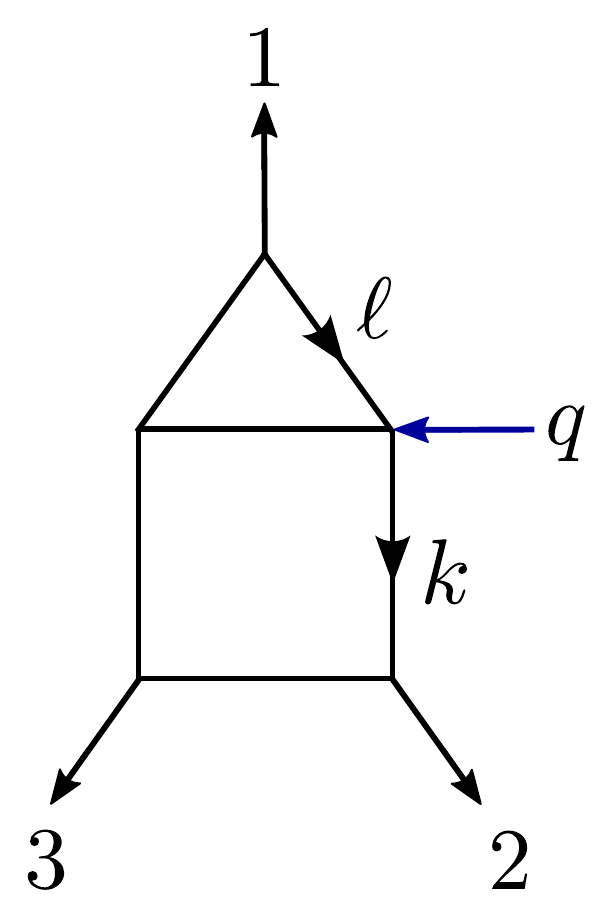}$ & $\pic{0.4}{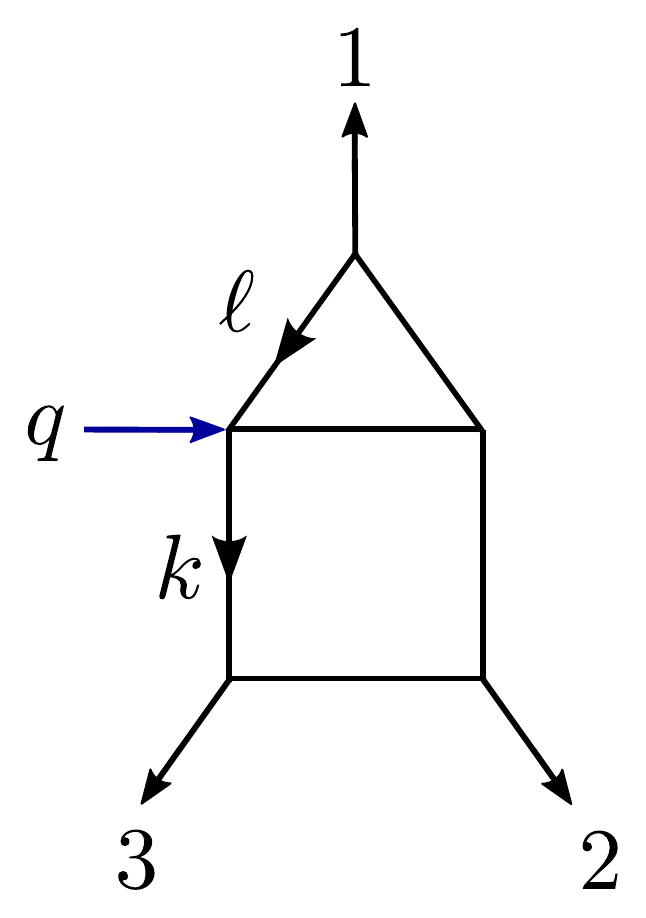}$ & $ \pic{0.4}{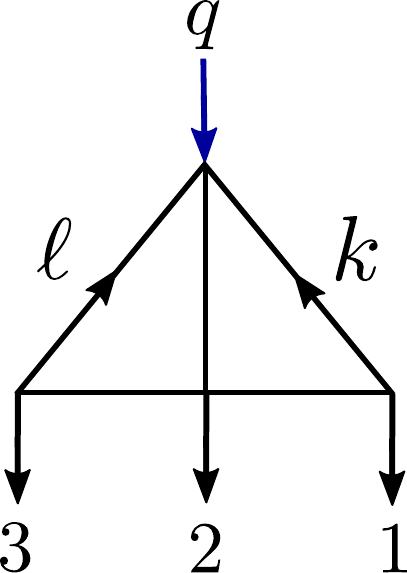} $ \\[30pt]
$I_1$ & $I_2$ &$I_3$ &$I_4$ \\[10pt]
$\pic{0.4}{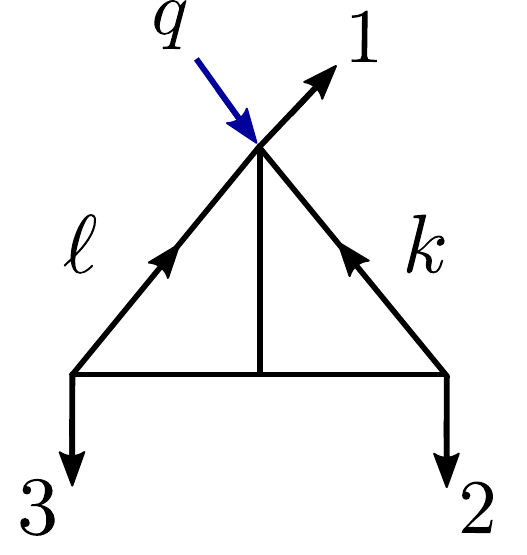} $ & $\pic{0.4}{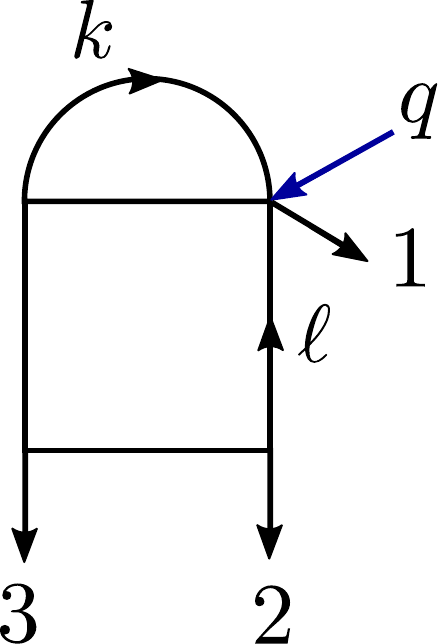}$ & $\pic{0.4}{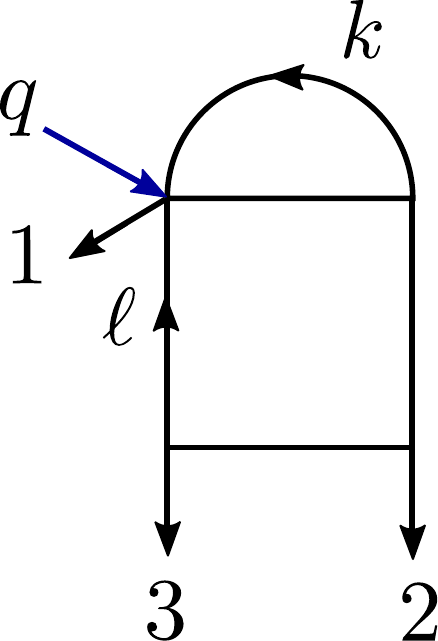}$ & $\pic{0.4}{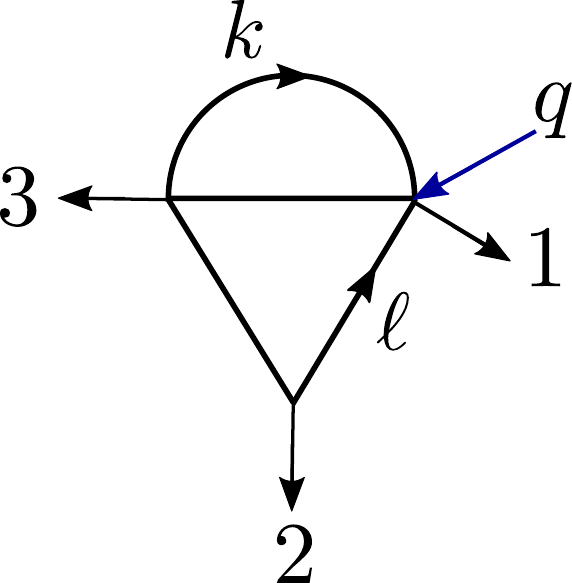}$\\[40pt]
$I_5$ & $I_6$ &$I_7$ &$I_8$ \\[10pt]
$\pic{0.4}{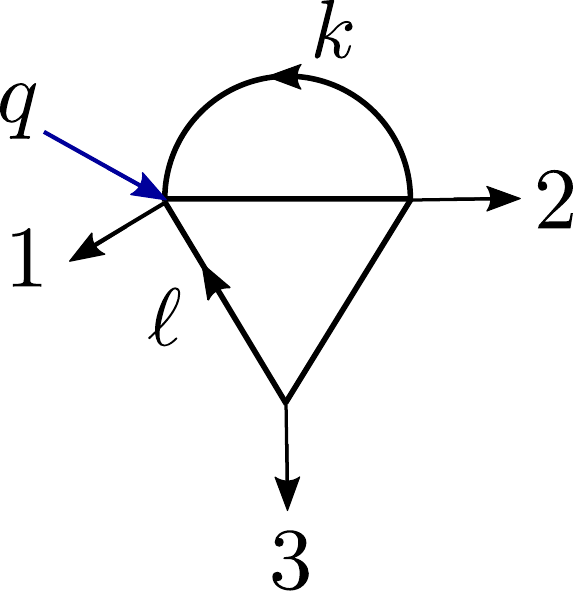} $ & $\pic{0.4}{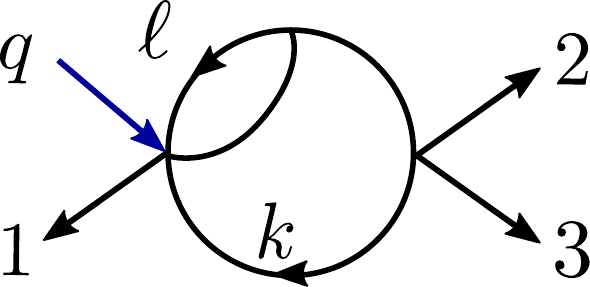} $ & $ \pic{0.4}{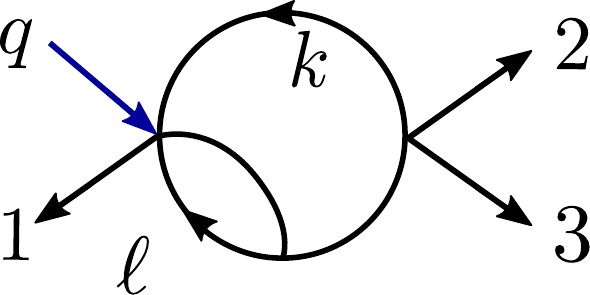} $ & $ \pic{0.4}{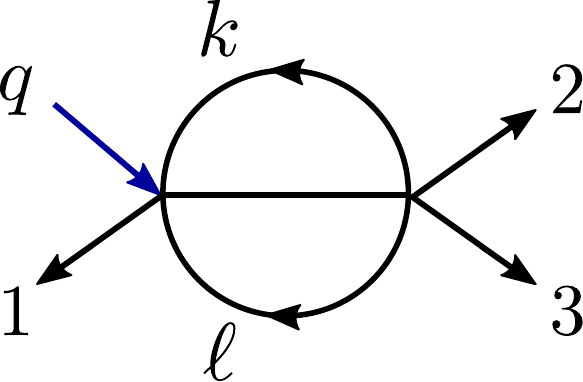}$\\[30pt]
$I_9$ & $I_{10}$ &$I_{11}$ &$I_{12}$ \\[10pt]
\end{tabular}
\caption{\it Integral basis for the two-loop form factor $F_{\O_{\mathcal S},\OC}^{(2)}(1^{+},2^{+},3^{+};q)$ in $\N\!=\!4$ SYM.}
\label{tab:two-loop basis}
\end{table}

\subsubsection{The integrands of the form factors of $\cO_\cS$ and $\cO_\cC$} 
The two-loop integrand of the minimal form factor of the Konishi descendant $\O_{\mathcal S}$ is given by
\begin{align}
F^{(2)}_{\O_{\mathcal S}}\,=\,F^{(0)}_{\O_{\mathcal S}} \,\sum_{i=1}^{12} N_i \times I_i
\ . 
\end{align}
The expressions for the complete numerators are somewhat involved, and we present them in Appendix \ref{app:susyN4}.


In order not to repeat lengthy numerator expressions, we present the result for the two-loop form factor of the component operator $\O_{\mathcal C}$ in terms of a difference when compared to the two-loop form factor of the supersymmetric operator $\O_{\mathcal S}$. Specifically, we have
\begin{align}\label{eq:resulf-difference}
F^{(2)}_{\O_{\mathcal C}}\,=\,F^{(2)}_{\O_{\mathcal S}}+\Delta_{\cN=4}\,,\qquad \Delta_{\cN=4}\,=\, \sum_{i=5}^{12} \tilde N_i \times I_i\,,
\end{align}
{\it i.e.} the difference between the two form factors consists solely of topologies which have only an $s_{23}$ triple cut, denoted by $I_5$ to $I_{12}$ in Table~\ref{tab:two-loop basis}. The numerators are listed in Appendix~\ref{app:compN4}.

\subsection{Components vs. super-cut comparison}\label{sec:compare}
Having obtained and presented the results for the two-loop form factors of supersymmetric operator $\O_{\mathcal S}$ and component operator $\O_{\mathcal C}$ we note the following observations resulting from the comparison of the two results:
\begin{itemize}
\item[\bf 1.] As previously noted, the difference between the two-loop form factors of $\O_{\mathcal S}$ and $\O_{\mathcal C}$ consists of topologies only present in the $s_{23}$-channel triple cut. 
\item[\bf 2.] These topologies have five propagators or fewer and are of sub-maximal transcendental weight. As a result, we observe that the maximally-transcendental part of the form factor is universal for the two operators. 
\item[\bf 3.] Moreover, explicit evaluation of the difference between the two form factors reveals terms of order $1/\eps$ and constant. Therefore, we conclude that the cancellation of infrared poles in the remainder function works exactly in the same way, the difference between the remainders of both operators lying in the $1/\eps$ term which is associated to  renormalisation of the operators.
\end{itemize}
With these observations in mind, we now discuss the remainder function of the two-loop form factor of the supersymmetric operator $\O_{\mathcal S}$.

\subsection{The subminimal form factor $\langle 1^+ 2^+ | \mathcal{O}_{\cS, \cC} | 0\rangle$ at two loops}
\label{subminimal}

\begin{figure}[h]
\centering
\includegraphics[width=0.3\textwidth]{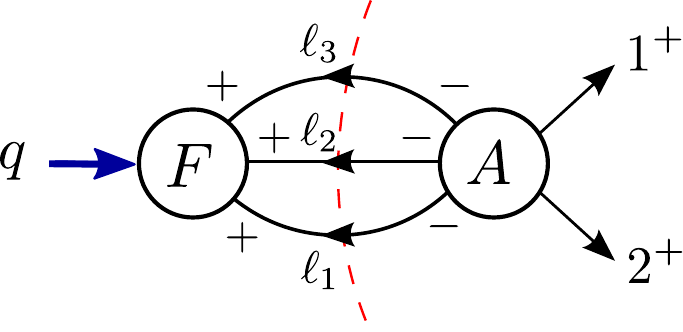}
\caption{\it Triple cut in the $q^2$-channel of the two-loop subminimal form factor $F^{(2)}_{\O_{\mathcal S},\cO_\cC}(1^+,2^+;q)$.}
\label{fig:subminimal}
\end{figure}

In order to discuss,  in the next section,  renormalisation and operator mixing,  we also need
to determine the sub-minimal form factor 
\be
F_{\cO_\cS,\cO_\cC}(1^+, 2^+;q)
\ee
up to two loops. Note that at tree and one-loop level this form factor vanishes.
At two loops, there is only one triple cut contributing which involves the product of the minimal
form factor and a five-point $\overline{\mathrm{MHV}}$ gluon tree-level amplitude, shown in Figure~\ref{fig:subminimal}. Note that, since the form factor is minimal, it can only involve three gluons and hence is identical for both ${\cO_\cS}$ and ${\cO_\cC}$.
\\[5pt]
We find for the cut
\be
F^{(2)}_{\O_{\mathcal S},\cO_\cC}(1^+,2^+;q)\Big|_{3,q^2}\,=\,i^3 \left(- [-\ell_3,-\ell_2][-\ell_2,-\ell_1][-\ell_1,-\ell_3]\right) \frac{(-i) [12]^3}{[2 \ell_1][\ell_1 \ell_2][\ell_2 \ell_3][\ell_3 1]}
\ ,
\ee
with $\ell_3 = -p_1-p_2-\ell_1-\ell_2$.
After some manipulations and taking the loop momenta off shell, we find the two-loop integrand
\be
F^{(0)}_{\cO_\cM}(1^+, 2^+;q) \frac{s_{12} s_{\ell_1 \ell_3}-s_{1 \ell_1} s_{2 \ell_3}+s_{1 \ell_3} s_{2 \ell_1}}{s_{12} s_{2 \ell_1} s_{1 \ell_3} \ell^2_1 \ell^2_2 \ell^2_3} \ ,
\ee
which integrates to our final result
\begin{align}
\begin{split}
\label{subb}
F^{(2)}_{\cO_\cS,\cO_\cC}(1^+, 2^+;q) &=  F^{(0)}_{\cO_\cM}(1^+, 2^+;q) \frac{\epsilon^2 e^{2 \epsilon \gamma_{\mathrm{E}}}}{(1-2 \epsilon)^2} \frac{\Gamma(1+2 \epsilon) \Gamma(-\eps)^3}{\Gamma (2-3 \epsilon)} \frac{(-s_{12})^{1-2 \epsilon}}{s_{12}}
\\
&= F^{(0)}_{\cO_\cM}(1^+, 2^+;q) \left[ \frac{1}{\epsilon}(-s_{12})^{-2 \epsilon} + 7 + \mathcal{O}(\epsilon) \right] \ ,
\end{split}
\end{align}
where 
\beq
F^{(0)}_{\cO_\cM}(1^+, 2^+;q) = \frac{s_{12}^3}{\langle 12 \rangle \langle 21 \rangle}
\ . 
\eeq
This result includes a factor of 2 from the fact that the two orderings of particle 1 and 2 make equal contributions. Interestingly it coincides with the sub-minimal two-loop form factor computed in Section~6 of \cite{Brandhuber:2016fni} up to a factor of $6$ and a spinor bracket.

It is important to note that the result is free of IR divergences, since the tree and one-loop result vanish, and the
$ 1/\eps$ pole of the result has a purely ultraviolet origin.


\section{Remainder functions in $\N\!=\!4$ SYM}\label{Sec:5}

In the previous section we have described the computation of the complete  integrands of the two-loop minimal form factors of the supersymmetric operator $\O_{\mathcal S}$ and of the component operator $\O_{\mathcal C}$ with a final state consisting of three gluons of positive helicity.  In addition, we have also considered the form factor with a  state of two external positive-helicity gluons, which is needed in order to address  mixing and for the study of soft/collinear limits performed in Section \ref{Sec:Discussion}.
In the next step, we have  reduced the corresponding  integrals  to a basis of master integrals using the 
  {\tt Mathematica} package {\tt LiteRed} \cite{Lee:2012cn,Lee:2013mka}. The explicit expressions of all master integrals in terms of (multiple) polylogarithms are provided in   \cite{Gehrmann:1999as,Gehrmann:2000zt}; furthermore,  whenever possible we have simplified
the answer using the symbol of transcendental functions~\cite{Goncharov:2010jf}.

Our next goal  consists in using these results to compute the IR and UV finite remainder functions of the {\it renormalised} operator 
$\O_{\mathcal S}^{\rm ren}$,  whose expression we have to determine by studying mixing. Doing so, we will also  diagonalise the dilatation operator and find
the anomalous dimensions and the appropriate diagonal operator. With this information at hand, we compute the two-loop remainder functions of the form factor of the renormalised operator $\O_{\mathcal S}^{\rm ren}$. 
For completeness, we also present the remainder function of the two-loop form factor of the bare operators $\O_{\mathcal S}$ and $\O_{\mathcal C}$.

\subsection{Disentangling operator mixing and the dilatation operator} 
\label{section:mixing}
We have already briefly mentioned mixing in Section \ref{section:mixing-first}. Expanding on that discussion, we note that the operator $\cO_{\cS}$  can mix with three other operators -- limiting ourselves to external states containing only up to three gluons of positive helicity, the only operators that can mix at two loops are $\cO_{\cS}$  and the operator $\cO_{\cM}\sim {\rm Tr} \big[D_\mu F_{\nu \rho}\, D^\mu F^{\nu \rho}\big]\sim q^2 {\rm Tr} \big(F^2\big)$ whose precise definition is given in \eqref{eq:operator2}. In order to simplify the discussion we can in fact take $\cO_{\cM}\sim q^2 \cL_\text{on-shell}$, where  we recall that the on-shell Lagrangian is a protected operator (and hence its form factors are UV finite).%
\footnote{The tree-level definition \eqref{eq:operator2} is unaltered by this choice. At one loop there is no difference in the UV divergences of form factors of ${\rm Tr}(F^2)$ and $\cL_\text{on-shell}$, while any difference at two loops between the corresponding form factors will not be relevant at the loop order we are working.}
In summary, we need to solve mixing in a two-dimensional space, similarly to what was done in the $SU(2|3)$ sector in \cite{Brandhuber:2016fni}. \\[5pt]
We define the renormalised operators as
\begin{align}\label{MixMatrix}
\begin{pmatrix}
\O_\cS^{\rm ren} \\[10pt]
\O_\cM^{\rm ren} 
\end{pmatrix}\,=\,\begin{pmatrix}
\cZ_S^{\phantom{F}S} & \cZ_S^{\phantom{F}M} \\[10pt]
 \cZ_M^{\phantom{B}S} & \cZ_M^{\phantom{B}M}
\end{pmatrix} \begin{pmatrix}
\O_\cS \\[10pt]
\O_\cM
\end{pmatrix} \ , 
\end{align}
 where $\O_\cS$ and $\O_\cM$ are the bare operators  used to compute form factors in earlier sections. The matrix of renormalisation constants $\cZ$, or mixing matrix, is  determined by requiring the UV-finiteness of the form factors of the renormalised operators $\O_\cS^{\rm ren}$ and $\O_\cM^{\rm ren}$ with the external states $\langle  1^+ 2^+  3^+|$ and $\langle  1^+ 2^+ |$. 

The UV divergences of the  form factors of these two operators with three and two positive-helicity gluons have the following structure: 
\begin{align}
\label{eq:FF-divergence}
\begin{split}
\left. F^{(1)}_{\cO_\cS}(1^+, 2^+, 3^+;q) \right|_{\rm UV} &=  a (\mu_R) \,  {b_1^{(1)} \over \epsilon} \, F^{(0)}_{\cO_\cS}(1^+, 2^+, 3^+;q)\ , 
\\
\left.F^{(2)}_{\cO_\cS}(1^+, 2^+, 3^+;q)\right|_{\rm UV}   &= a^2(\mu_R)
\Big[{b^{(2)}_1\over \epsilon} +  {{b^{(2)}_2\over \epsilon^2}}\Big]   F^{(0)}_{\cO_\cS}(1^+, 2^+, 3^+;q)  +  a^2(\mu_R) {\hat{b}^{(2)}_1 \over\epsilon} 
\, F^{(0)}_{\cO_\cM}(1^+, 2^+, 3^+;q) , \\
\left. F^{(2)}_{\cO_\cS}(1^+, 2^+;q) \right|_{\rm UV} &= {a^2(\mu_R)\over g}  \,  {k \over \epsilon}  F^{(0)}_{\cO_\cM}(1^+, 2^+;q) \ , 
\end{split}
\end{align}
where $F^{(0)}_{\cO_\cS}(1^+, 2^+, 3^+;q)$ and $F^{(0)}_{\cO_\cM}(1^+, 2^+, 3^+;q)$ are given in    \eqref{FF-OS-n3} and \eqref{eq:operator2}, respectively; furthermore, 
from  the one- and two-loop computations of the preceding sections, we can infer the values for the coefficients in \eqref{eq:FF-divergence},
\begin{align}
\begin{split}
\label{explicit}
b^{(1)}_1 &= \, -6\, , \qquad \\
b^{(2)}_1 &= \, 12\, , \qquad b^{(2)}_2  \ = \, 18\, ,\qquad \hat{b}^{(2)}_1 \ = \ 1 \ , 
\\
k & = 1 \ .
\end{split}
\end{align}
 Here 
\beq
\label{Hooft-run}
a (\mu_R) \ :=\ {g^2 N  e^{- \epsilon \gamma_{\rm E}}\over  (4 \pi )^{2-\epsilon }}\left( {\mu_R \over \mu}\right)^{- 2 \epsilon} 
\ , 
\eeq 
is the running 't Hooft coupling,  and 
  $\mu_R$ is the renormalisation  scale.  Note that the form factors of $\cO_\cM \sim q^2 \cL_\text{on-shell}$ are UV finite, which is why they do not make an appearance  in the previous list.\\[5pt]
A comment is in order here. From \eqref{eq:operator2} we know that
  \beq
  \label{1/uvw**}
  F^{(0)}_{\cO_\cM}(1^+, 2^+, 3^+;q) \ = \  { F^{(0)}_{\cO_\cS,\cO_\cC}(1^+, 2^+, 3^+;q)\over u v w }\ . 
  \eeq
The presence of the $1/ (u v w)$ factor gives a distinctive, useful  signature of mixing in all quantities we compute in this paper. \\[5pt]
Next we   introduce the    renormalisation constants  that are relevant for our problem:
\begin{align}
\begin{split}
\cZ_S^{\phantom{F}S}  &=1\, + \, a(\mu_R) \, 
{\tilde{b}_1^{(1)}\over \epsilon}\,  + \, a^2(\mu_R)\Big( {\tilde{b}^{(2)}_1\over \epsilon}\, + \, 
{\tilde{b}^{(2)}_2\over \epsilon^2}\Big) \, + \, \cdots \  ,  
\\
\cZ_S^{\phantom{F}M}& = {a^2(\mu_R)\over g} {\tilde{B}\over \epsilon} \, + \cdots \ ,
\end{split}
\end{align}
where the ellipses denote terms of higher order in $a(\mu_R)$.
In addition  we can solve the mixing with the further simple choices: 
\begin{align}
\cZ_M^{\phantom{F}S}  &=0 \, + \, \cdots\ , \\
\cZ_M^{\phantom{F}M} & = 1 \, + \, \cdots \ , 
\end{align}
where the dots stand for terms that can be discarded at two loops. 
Requiring the finiteness of the form factors of the renormalised operators leads to the  conditions: 
\begin{align}
\begin{split}
\tilde{b}_1^{(1)} & = \ - b_1^{(1)}\ , 
\\
\tilde{b}^{(2)}_1& = - b^{(2)}_1\ , \\
 \tilde{b}^{(2)}_2& = - \big[ b^{(2)}_2 - \big( b^{(1)}_1 \big)^2\big]
\ , 
\\
\tilde{B} & = \ - \hat{b}^{(2)}_1 \ = \ - k
\ .
\end{split}
\end{align}
Note the appearance of a consistency condition  $\hat{b}^{(2)}_1  =   k$ which is indeed satisfied given our results \eqref{explicit}. In conclusion, we arrive at the following expansion for the renormalisation constants:
\begin{align}
\begin{split}
\label{explicit2}
\cZ_S^{\phantom{F}S}  &=1\, + \, a (\mu_R)
{6\over \epsilon}\,  + a^2(\mu_R) \Big( - {12 \over \epsilon}\, + \, 
{18\over \epsilon^2}\Big) \, + \, \cdots \  ,  
\\
\cZ_S^{\phantom{F}M}& = - {a^2(\mu_R)\over g} {1\over \epsilon} \, + \cdots \ , \\
\cZ_M^{\phantom{F}S}  &=0 \, + \, \cdots\ , \\
\cZ_M^{\phantom{F}M} & = 1 \, + \, \cdots \ , 
\end{split}
\end{align}
from which one can determine the renormalised operators using \eqref{MixMatrix}.

Next we derive the form of the  dilatation operator $\mathfrak{D}:= \uno + \delta \mathfrak{D}$. Its quantum corrections are encoded in the matrix $\delta \mathfrak{D}$, which  is related to the mixing matrix $\cZ$ as
\begin{align}
\label{eq:delta-D}
\delta\mathfrak{D}\,=\, \lim_{\eps\rightarrow 0}\Big[-\mu_R\frac{\partial}{\partial \mu_R}\log(\cZ)\Big] \ . 
\end{align}
In order to compute $\delta\mathfrak{D}$, we first compute the matrix $\log \cZ$. Up to two loops we find 
\beq
\label{ourZ}
\log \cZ \ = \ 
\begin{pmatrix}
a(\mu_R) \, \dfrac{6}{\epsilon}\,  -\,  a^2(\mu_R) \, \dfrac{12}{\epsilon} \ \  \ \  & -\dfrac{a^2(\mu_R)}{g} \, \dfrac{1}{\epsilon} \\[10pt]
0 & 0
\end{pmatrix}\ , 
\eeq
where we note the cancellation of all $1 / \eps^2$ poles. Finally,  using \eqref{eq:delta-D} we arrive at 
\beq
\delta\mathfrak{D} \ = \ 
2\times \begin{pmatrix}
6 \, a - 24 a^2  \ \  \ \  & -2 \, \dfrac{a^2}{g} \,  \\[10pt]
0 & 0
\end{pmatrix}\ , 
\eeq
where $a$ is the 't Hooft coupling defined in \eqref{eq:tHooft}. 
Indicating by $\tilde\O_\cS$ and $\tilde\O_\cM$ the eigenvectors of $\delta\mathfrak{D}$, we find that that the corresponding eigenvalues are, up to two loops, 
\beq
\gamma_{\tilde\O_\cS}\ = \ 12 \, a \, -\, 48\, a^2   \ , \qquad 
\gamma_{\tilde\O_\cM} \ = 0 
\ . 
\eeq
Note that $\gamma_{\tilde\O_\cS}$ precisely coincides with the anomalous dimension of the  Konishi multiplet at this loop order. This is an important consistency check of our calculation. 
It might also be of interest to  compute the eigenvector    corresponding to $\gamma_{\tilde\O_\cS}$. The result
of this is 
\begin{align}
\label{eq:diag-op}
\tilde\O_\cS^{\rm ren}  \ = \ \O_{\cS}^{\rm ren}  \ - \   {a\over 3\, g} \, \O_{\cM}^{\rm ren}
\ . 
\end{align}
In the next section we will compute various remainders, and  in particular the remainder of the renormalised operator 
$\O_\cS^{\rm ren}$. For convenience, in the following we  choose  the renormalisation scale to be 
\beq
\mu_R^2 \ =\ q^2\ .
\eeq

\subsection{Definition of the BDS form factor remainder} 
The remainder function for form factors  in $\cN\!=\!4$ SYM \cite{Brandhuber:2012vm} is defined in the same way as for scattering amplitudes, namely through the subtraction of the BDS ansatz \cite{Anastasiou:2003kj,Bern:2005iz}. 
For a generic operator $\O$, the form factor remainder function at two loops is defined as 
\begin{align}\label{eq:remainder}
\cR^{(2)}_{\O} \coloneqq \ \mathcal{F}_{\O}^{(2)}(\epsilon )\, - \, {1\over 2} \big( \mathcal{F}_{\O}^{(1)} (\epsilon) \big)^2 - f^{(2)} (\epsilon)\ \mathcal{F}_{\O}^{(1)} ( 2 \epsilon )
+ \O (\epsilon )\,, 
\end{align}
where $\mathcal{F}^{(L)}_\mathcal{O}= F^{(L)}_\mathcal{O}/F^{(0)}_\mathcal{O}$. 
The function $f^{(2)} (\eps) = -2( \zeta_2 + \eps \, \zeta_3 + \eps^2 \, \zeta_4)$ is determined from the iteration of the splitting amplitudes \cite{Anastasiou:2003kj,Bern:2005iz} and hence it is the same for form factors, as was explicitly shown in \cite{Brandhuber:2012vm}. 
Note that we define the remainders (bare and renormalised)  by taking out a factor of 
\beq
a\big[4 \pi e^{- \gamma_{\rm E}}   \big]^{\epsilon}
\eeq
per loop, where $a$ is the 't Hooft coupling, defined in \eqref{eq:tHooft}.

In dimensional regularisation, the definition \eqref{eq:remainder} allows for the cancellation of all infrared  poles as well as the $1/\eps^2$ pole of ultraviolet origin. Computing remainders of renormalised operators, also $1/ \eps$ poles of UV origin cancel.
We note that in theories with non-trivial beta functions, the BDS remainder \eqref{eq:remainder} is not appropriate and in the companion paper \cite{Part2} we will switch to the more general  remainder introduced by Catani \cite{Catani:1998bh}.

\subsection{The remainder of $\cO_\cS^{\rm ren}$} 
\label{sec:susy-op}

Our result for the remainder of the form factor of $\O_{\mathcal S}^{\rm ren}$ has the following properties:
\begin{itemize}
	\item[{\bf 1.}]  All poles $1/\eps^k$ vanish 
	as expected -- there are no UV poles since we are using renormalised operators, and there are no IR poles since we are computing the BDS remainder, which is taking care of all infrared divergences. 

			\item[{\bf 2.}] The finite part of the remainder function is surprisingly simple for an operator as intricate as $\O_{\mathcal S}^{\rm ren}$: it is comprised of classical polylogarithms only and classical zeta functions. It can be split into slices of fixed transcendentality ranging from zero to four. Moreover, each slice features universal building blocks which have appeared already for operators in other sectors.
\end{itemize}
In the following, we present and discuss each transcendentality slice of the remainder function in turn.

\noindent{\bf Transcendentality four:} We find that the maximally transcendental slice of the remainder function is the same as that of the BPS operator $\Tr(\phi^3)$ \cite{Brandhuber:2014ica},
\begin{align}\label{eq:remainderBPS}
\cR^{(2)}_{\O_{\mathcal S}^{\rm ren};4}\ =\	\cR^{(2)}_{\rm BPS} \ = \ 
& -\frac{3}{2}\, \text{Li}_4(u)+\frac{3}{4}\,\text{Li}_4\left(-\frac{u v}{w}\right) 
-\frac{3}{2}\log(w) \, \text{Li}_3 \left(-\frac{u}{v} \right)+\frac{1}{16} {\log}^2(u)\log^2(v) \nonumber \\
&+{\log^2 (u) \over 32} \Big[ \log^2 (u) - 4 \log(v) \log(w) \Big]+{\zeta_2 \over 8 }\log(u) \Big[ 5\log(u)- 2\log (v)\Big] \nonumber \\
&+{\zeta_3 \over 2} \log(u) + \frac{7}{16}\, \zeta_4 + {\rm perms}\, (u,v,w) \, . 
\end{align}

\noindent{\bf Transcendentality three:} The transcendentality-three piece has a feature which was also observed in the $SL(2)$ sector in \cite{Loebbert:2016xkw}: it contains terms with kinematic-dependent prefactors taken from the list 
\beq
\label{eq:listT3}
\left\{{u\over v}, \, {v\over u}, \,{v\over w}, \, {w\over v}, \, {u\over w}, \, {w\over u}\right\}
\eeq
in addition to terms without any kinematic-dependent prefactor -- which we refer to as ``pure".
The pure part of the degree-three slice is
\begin{align}
\label{5.10}
\cR^{(2)}_{{\O_{\mathcal S}^{\rm ren}};3}\Big|_{\rm pure}\ =\ 
&\text{Li}_3(u)+ \text{Li}_3(1-u)- {1\over 4} \log^2(u) \log \left({v w\over (1-u)^2} \right)
+{1\over 3} \log (u) \log (v) \log (w)\nonumber \\
&+\zeta_2 \log (u) 
+{13\over 3}\zeta_3 \, + \,\text{perms}\, (u,v,w)\, .
\end{align}
Interestingly, this result can be related to another known quantity, the remainder function of the operator $\Tr(X[Y,Z])$ calculated in \cite{Brandhuber:2016fni}:
\begin{align}
\cR^{(2)}_{{\O_{\mathcal S}^{\rm ren}};3}\Big|_{\rm pure} = {1\over 2}\Big(\cR^{(2)}_{\text{non-BPS};3}+
4 \zeta_2\log(uvw)-24\, \zeta_3 \Big)\,, 
\end{align}
where $\cR^{(2)}_{\text{non-BPS};3}$ is given in (4.11) of \cite{Brandhuber:2016fni}.
The term with coefficient $u/w$ in the ``non-pure" part of the transcendentality-three piece is
\begin{align}
\label{tre}
\cR^{(2)}_{{\O_{\mathcal S}^{\rm ren}};3}\Big|_{u/w}=& \Big[-\text{Li}_3\left(-{u\over w}\right)+\log(u)\text{Li}_2\left({v\over 1-u}\right)-{1\over 2}\log(1-u)\log(u)\log\left({ w^2\over 1-u }\right)
\nonumber \\
&+{1\over 2}\text{Li}_3\left(-{uv\over w}\right)+{1\over 2}\log(u)\log(v)\log(w)+{1\over 12}\log^3(w) +(u\leftrightarrow v) \Big] \nonumber \\
&+\text{Li}_3(1-v)-\text{Li}_3(u)+{1\over 2}\log^2(v)\log\left({1-v\over u}\right)\ 
- \ 
\zeta_2 \log\left( {u v\over w}\right) \,.
\end{align}
The coefficients of the other factors in the list \eqref{eq:listT3} are obtained by taking the appropriate permutation of the function above.
We also anticipate that there is an intriguing relation between \eqref{5.10} and the quantity obtained after summing \eqref{tre} over permutations of $(u,v,w)$, as we discuss in the next section.

\noindent{\bf Transcendentality two:} The degree-two part also contains terms with kinematic-dependent prefactors taken from the list
\beq
\label{eq:listT2}
\left\{{u^2\over v^2}, \, {v^2\over u^2}, \, {u^2\over w^2}, \, {v^2\over w^2}, \, {w^2\over u^2}, \, {w^2\over v^2}\right\}\ .
\eeq
The pure part reads
\begin{align}
\cR^{(2)}_{{\O_{\mathcal S}^{\rm ren}};2}\Big|_{\rm pure}&= -\text{Li}_2(1-u)-\log^2(u)+{1\over 2}\log(u)\log(v)-{13\over2}\zeta_2 \, + \,\text{perms}\, (u,v,w)\,,
\end{align}
while the coefficient of the $u^2/w^2$ part is given by
\begin{align}\label{du}
\cR^{(2)}_{{\O_{\mathcal S}^{\rm ren}};2}\Big|_{u^2/w^2}&= \text{Li}_2(1 - u) + \text{Li}_2(1 - v) + \log(u) \log(v)-\zeta_2\,.
\end{align}
Again, the coefficients of the other terms in \eqref{eq:listT2} are obtained through permutations of the function above.\\[10pt]
\noindent{\bf Transcendentality one and zero:} The transcendentality-one slice is simply given by
\begin{align}\label{eins}
\cR^{(2)}_{{\O_{\mathcal S}^{\rm ren}};1}&= \left(-4 + \frac{v }{w}+ \frac{u^2 }{2 v w} \right) \log(u)\, + \,\text{perms}\, (u,v,w)\, .
\end{align}
Finally, the degree-zero part of the remainder is 
\begin{align}\label{smile}
\cR^{(2)}_{{\O_{\mathcal S}^{\rm ren}};0}&=7 \left(12+\frac{1}{uvw} \right)\,.
\end{align}

\subsection{The remainder of the bare $\O_{\cS}$ operator}

In this section we quote for completeness the remainder function of the bare operator $\O_{\cS}$. 
A short calculation making use of the mixing matrix \eqref{MixMatrix} and \eqref{explicit2}, as well as the definition of BDS remainder given in
\eqref{eq:remainder} and of the running coupling constant \eqref{Hooft-run} shows that the non-renormalised remainder still contains a $1/\eps$ pole of UV origin,
\begin{align}
\label{eq:RenBareOS}
 \cR_{\cO_{\cS}}^{(2)} \ = \  \cR_{\cO_{\cS}^{\rm ren}}^{(2)}+ \left( - {q^2\over \mu^2}\right)^{- 2 \epsilon}\Big[ {1\over \epsilon} \Big( 12 - 6 \zeta_2 + {1\over u v w } \Big) - 6 \zeta_3 \Big] 
 \ . 
\end{align}
Note that the $1/ (u v w)$ pole is due to the  mixing between $\O_{\cS}$ and $\O_{\cM}$, as alluded to in \eqref{1/uvw**}, while the  $\zeta_2$ and $\zeta_3$ terms arise from the last term in the BDS remainder \eqref{eq:remainder}.

\subsection{The remainder of the bare $\cO_\cC$ operator} 
We now discuss the remainder of the two-loop form factor of the bare component operator $\O_{\mathcal C}$. 
It has   the following properties:
\begin{itemize}
\item[{\bf 1.}] 
Like the remainder of $\mathcal{O}_\mathcal{S}$, it has a $1/\eps$ pole arising from a UV divergence,
\beq 
\label{1/uvw-bis}
\left.\cR^{(2)}_{\O_\cC}\right|_{1\over \eps}\ = \ 
9-6\, \zeta_2 + {1\over u v w}
\ .
\eeq
Again, the term $1/(uvw)$ indicates mixing with $\cO_\cM$.
\item[{\bf 2.}] We recall from Section \ref{sec:compare} that the difference between the form factors of operators $\O_{\mathcal S}$ and $\O_{\mathcal C}$, denoted as $\Delta_{\cN=4}$ in \eqref{eq:resulf-difference} contained only terms of order $1/\eps$ and a constant. As a result, also for $\cR_{\O_{\mathcal C}}^{(2)}$ all poles in $1/\eps^k$ vanish for $k>1$, as expected.

\item[{\bf 3.}] Even more strikingly, we find that the remainder function of the operator $\O_{\mathcal C}$ is almost identical to that of operator $\O_{\mathcal S}$ given in  \eqref{eq:RenBareOS},
\begin{align}
\label{eq:RenBareOC}
 \cR_{\cO_{\cC}}^{(2)} \ = \  \cR_{\cO_{\cS}^{\rm ren}}^{(2)}+ \left( - {q^2\over \mu^2}\right)^{- 2 \epsilon}\Big[ {1\over \epsilon} \Big( 9- 6 \zeta_2 + {1\over u v w } \Big) - 6 \zeta_3 \Big]  +\log(u v w)  - \frac{51}{2}\ .
\end{align}
In particular, this implies that
\begin{align}
\cR^{(2)}_{\O_{\mathcal C};i}\,=\,\cR^{(2)}_{\O_{\mathcal S};i}\,,\qquad i=4,3,2\,.
\end{align}
\end{itemize}


\section{Consistency checks of the result and discussion}
\label{Sec:Discussion}
In this final section we comment on some nontrivial consistency checks of the result and make some final  observations on  the results we have presented. 

\subsection{Soft and collinear limits of the bare two-loop form factor}

We can obtain some nontrivial consistency checks on our calculations by considering soft and collinear limits of the results. For clarity, we find it convenient to present our discussion at the level of the bare quantities. 

The first observation is that, at tree level and at one loop, soft (and collinear) limits vanish because of the explicit form of the tree-level form factors \eqref{FF-OS-n3} and \eqref{eq:one-loop-result}. This is consistent with factorisation theorems, since the minimal form factors cannot factorise on anything at this loop order. 

The situation is more interesting at two loops, since at this order $F^{(2)}_{\O_{\mathcal S}}(1^+,2^+,3^+;q)$
can factorise onto the subminimal form factor $F^{(2)}_{\cO_\cS}(1^+, 2^+;q)$ computed in Section \ref{subminimal}. This form factor  is for the first time non-vanishing at two loops; its expression  is given in \eqref{subb}, and contains only two terms, which we will now identify in the factorisation. 
Beginning with the   soft limit  $p_2 \to 0$, at two loops we expect 
\beq
F^{(2)}_{\O_{\mathcal S}}(1^+,2^+,3^+;q) \to \text{Soft}^{\rm tree} (1, 2^+, 3) \ F^{(2)}_{\O_{\mathcal S}}(1^+,3^+;q) \ ,  
\eeq
with 
\beq
\text{Soft}^{\rm tree} (a, s^+, b) \ = \ {\lan a b\ran\over \lan as\ran \lan sb\ran}
\ .
\eeq
In the collinear limit  $p_1 || p_2$ we expect
\beq
F^{(2)}_{\O_{\mathcal S}}(1^+,2^+,3^+;q) \to \text{Split}^{\rm tree}_{-} (1^+,  2^+) \ F^{(2)}_{\O_{\mathcal S}}\big((1+2)^+,3^+;q\big) \ ,  
\eeq
with  
\beq
\text{Split}^{\rm tree}_{-} (a^+, b^+) \ = \ {1\over \sqrt{z (1-z)} \lan ab\ran}
\ , 
\eeq
where in the collinear limit $p_1\to z P$, $p_2\to (1-z) P$ with $P=p_1 + p_2$. 

Due to the vanishing of the tree-level prefactor in the soft/collinear limits, we only need to consider terms in the result with rational factors that could lead to additional poles in the limit such as $1/ (u v  w)$. We now organise the discussion by degree of transcendentality. 

At transcendentality degree four, we have only pure terms without any rational factors.%
\footnote{The soft/collinear limits of the maximally transcendental terms were already studied in \cite{Brandhuber:2014ica}.
} 
A particular feature of the remainder described in the previous section is that
 ``non-pure'' terms with rational coefficients of the type
$v/u$, $v^2/u^2$ and $v w/u^2$ come at transcendentality three, two and one,  respectively.
At first glance they are problematic as they could potentially lead to unphysical simple or even double poles in collinear/soft limits when one or two of the three kinematic ratios $u$, $v$ and $w$ tend to zero. This may occur  in the collinear limit $p_1 || p_2$,  where   $u \to 0$, or in the soft limit $p_2 \to 0$ where we have both $u \to 0$ and $v \to 0$. 

Let us begin by looking at the ``non-pure" transcendentality-three terms given by \eqref{tre} (plus permutations of  $(u,v,w)$)
with rational coefficients such as  $v/u$.
To study the collinear limit $u \to 0$ (with  $v \neq 0, 1$) we simply expand \eqref{tre} around $u=0$. Keeping
only the terms diverging in the limit we find
\begin{align}
\begin{split}
{u\over w} \cR^{(2)}_{{\O_{\mathcal S}};3}\Big|_{u/w}\, + \, {\rm perms} (u,v,w)  
  \ \underset{u\to 0}{\to}  \ 
& \log(u)\frac{v^2(\log(v) \log(1-v)-\zeta_2)+(2 v-1) \mathrm{Li}_2(v)}{v(1-v)} \\
&\!\!\!\!\!\! -
\color{black} \hf \color{black}\log(u)^2 \frac{v^2 \log(v)+(1-v)^2 \log(1-v)}{v(1-v)} + \mathrm{finite} \ ,
\end{split}
\end{align}
which displays only logarithmic divergences. Importantly, all potential simple poles have cancelled out, and since the overall 
tree-level form factor vanishes in this limit, these contributions to the form factor vanish in the limit too. 

Similarly, for the soft limit $p_2 \to 0$ we need to expand around $u=v=0$ with the result
\be
{u\over w} \cR^{(2)}_{{\O_{\mathcal S}};3}\Big|_{u/w}\, + \, {\rm perms} (u,v,w)  
  \ \underset{(u,v)\to (0,0)}{\to}  \ 
2\, +\, 2 \zeta_2 -\log(u)+{\log(u)^2 \over 2} -\log(v)+{\log(v)^2 \over 2} \, + \, \mathrm{finite}\ .
\ee
Again there are only logarithmic divergences and the dangerous poles have cancelled.

Next let us consider the transcendentality-two terms given by \eqref{du} (plus permutations of  $(u,v,w)$)
which contain potentially even more problematic double poles. Following the same procedure as for the transcendentality-three terms one finds now not only logarithmic singularities -- the simple poles do not cancel. Naively one would expect that terms of different degree of transcendentality separately have the correct kinematic limits, and this would be a serious problem. 
However it magically turns out that we have to add the transcendentality-one terms \eqref{eins} in order to cancel the dangerous poles. Doing so, in the collinear limit $u \to 0$ we find only logarithmic terms 
\begin{align}
\begin{split}
{u^2\over w^2} \cR^{(2)}_{{\O_{\mathcal S}};2}\Big|_{u^2/w^2}\, & + \, 
\cR^{(2)}_{{\O_{\mathcal S}};1}\, + \, 
{\rm perms} (u,v,w)  
\ 
\underset{u\to 0}{\to}  \ 
\\
&\log(u) \frac{v(1-v)(1-10v(1-v))+v^4 \log(v)+(1-v)^4 \log(1-v)}{v^2(1-v)^2} + \mathrm{finite} \ ,
\end{split}
\end{align}
while in the soft limit $p_2 \to 0$ we expand around $u = v = 0$,
\be
{u^2\over w^2} \cR^{(2)}_{{\O_{\mathcal S}};2}\Big|_{u^2/w^2}\,  + \, 
\cR^{(2)}_{{\O_{\mathcal S}};1}\, + \, 
{\rm perms} (u,v,w)  
 \ \underset{(u,v)\to (0,0)}{\to}  \ 
-\frac{1}{2}\left[ 1+ 15 \log(u v)\right] + \mathrm{finite} \ .
\ee
Hence we find that the transcendentality-two and one terms of the remainder conspire in a way to cancel all
unphysical poles,  leaving only logarithmic terms  which vanish   in soft/collinear limits due to the presence  of the tree-level prefactor.


Finally we come to the transcendentality-zero term in  \eqref{smile}, which turns out to be  particularly interesting.  In the soft/collinear  limits the rational  term  $7/ ( u v w )$   survives and combines with 
the UV divergent term of the form factor, that is cancelled by the ${\O_{\mathcal M}}$ counterterm. The only relevant terms of the bare form factor contributing in the soft/collinear limits are
\be
-\frac{[12][23][31]}{u v w} \left[ {(-q^2)^{-2 \epsilon} \over \epsilon}+7 \right] \ ,
\ee
which reproduces exactly the expected soft/collinear factorisation -- for instance,  in the soft limit $p_2 \to  0$ we find
\be
\frac{\langle 13 \rangle}{\langle 12 \rangle\langle 23 \rangle} \frac{s_{13}^3}{\langle 13 \rangle \langle 31 \rangle} \left[ {(-s_{13})^{-2 \epsilon} \over \epsilon}+7 \right] =  \text{Soft}^{\rm tree} (1, 2^+, 3) \ F^{(2)}_{\O_{\mathcal S}}(1^+,3^+;q) \ ,
\ee
where the expression for the  sub-minimal form factor $F^{(2)}_{\O_{\mathcal S}}$ can be found in \eqref{subb}.
This provides a strong consistency check of our results and highlights   an intricate conspiracy among the  peculiar rational factors appearing in the remainder function. 
We also note that the discussion  for both operators considered in this paper, namely $\cO_\cC$ and $\cO_\cS$, is identical since their remainders only differ by terms without rational factors.

\subsection{Further observations on the result}

{\bf 1.} In \cite{Loebbert:2016xkw}, the authors discuss the idea of assigning a degree of transcendentality to harmonic numbers, already explored in {\it e.g.} \cite{Fleischer:1998nb} and propose the concept of ``hidden maximal transcendentality" of the remainder function. For our purposes, we are particularly interested in assigning transcendentality to ratios of Mandelstam invariants which multiply the ``non-pure" pieces of the remainder, presented in \eqref{tre} and \eqref{du}. It turns out that we can think of ratios of invariants such as $(1-v)/w$ as having transcendentality degree one, due to the expansion
\begin{align}
\lim_{m\rightarrow\infty}\sum_{k=1}^m\frac{1}{k}\left(\frac{1-v}{w}\right)^k\,=\,-\log\left(1-\frac{1-v}{w}\right)\,.
\end{align}
In order to see the hidden maximal transcendentality manifest itself in the (part of) our result we rewrite the ratios of Mandelstam invariants multiplying the transcendentality-three piece in \eqref{tre} using the fact that $u+v+w\!=\!1$, for example
\begin{align}
\frac{u}{w}\,=\,\frac{1-v-w}{w}\,=\,\frac{1-v}{w}-1\,.
\end{align}
Upon such trivial rewriting, it turns out that the pure transcendentality-three part of the remainder (almost) cancels out, namely
\begin{align}
\cR^{(2)}_{{\O_{\mathcal S}};3}\Big|_{u/w} \, + \, \textrm{perms}\, (u,v,w) \ = \ 
\cR^{(2)}_{{\O_{\mathcal S}};3}\Big|_{\rm pure}\, \color{black} - \color{black}\, 4\zeta_2\log(u v w) \color{black} +\color{black} 6\zeta_3
\,,
\end{align}
leaving ``non-pure" terms, now multiplied by ratios such as $(1-v)/w$ -- resulting in uniform transcendentality four. 


{\bf 2.} Finally, we note that the constant part of the remainder in \eqref{smile}, when multiplied by $-4/7$ gives the value of the two-loop Konishi anomalous dimension, {\it i.e.} $-48$. The same feature was first noted in \cite{Loebbert:2015ova} for remainders of operators in the $SU(2)$ sector.

\newpage

\section*{Acknowledgements}
	
We would like to thank Zvi Bern, John Joseph Carrasco and Henrik Johansson for very helpful discussions, Claude Duhr for sharing a package for handling polylogarithms, and Sophia Borowka, Lance Dixon, Claude Duhr, Paul Heslop, Florian Loebbert, Jan Plefka, Emery Sokatchev and Donovan Young for conversations. 
The work of AB and GT was supported by the Science and Technology Facilities Council (STFC) Consolidated Grant ST/L000415/1 
\textit{``String theory, gauge theory \& duality"}. The work of MK is supported by an STFC quota studentship. BP is funded by the ERC Starting Grant 637019 ``\emph{MathAm}''. AB and GT would like to thank the KITP at the University of California, Santa Barbara, where their research was supported by the National Science Foundation under Grant No.~NSF PHY-1125915.
GT is grateful to the Alexander von Humboldt Foundation for support through a Friedrich Wilhelm Bessel Research Award, and to the Institute for Physics and IRIS Adlershof at Humboldt University, Berlin, for their warm hospitality.
This research was supported in part by the Munich Institute for Astro and Particle Physics (MIAPP) of the DFG cluster of excellence ``Origin and Structure of the Universe".

\newpage	
\appendix

\section{One-loop integral functions}\label{App:Integrals} 
 
Throughout the paper, we use the following conventions for the one-loop massless scalar integrals in dimensional regularisation (upper/lower-case letters correspond to massive/massless momenta) \cite{Bern:1994cg}:
\begin{table}[h!]
\centering
\begin{tabular}{cl}
$\pic{0.6}{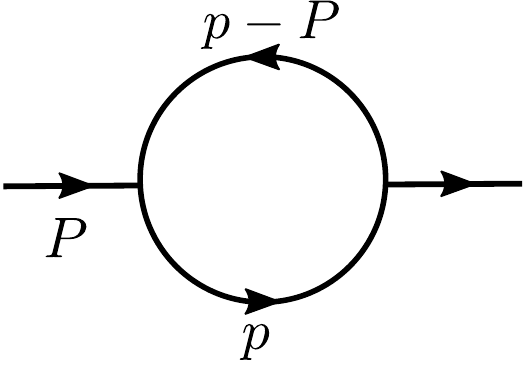}$& $ \,=\, \displaystyle\int \dfrac{d^{4-2\epsilon}p}{(2\pi)^{4-2\epsilon}}\dfrac{1}{p^2(p-P)^2} \,=\, i\, \frac{c_{\Gamma}}{\epsilon(1-2\epsilon)}\left(-{P^2\over \mu^2}\right)^{-\epsilon}\ ,$\\[10pt]
$\pic{0.7}{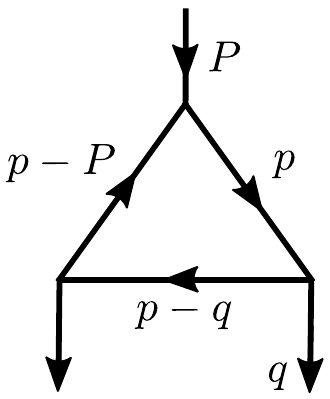}$ &$\,=\, \displaystyle \int \dfrac{d^{4-2\epsilon}p}{(2\pi)^{4-2\epsilon}}\dfrac{1}{p^2(p-q)^2(p-P)^2}= -i\, \frac{c_\Gamma}{\epsilon^2}
\dfrac{\ \ \left(-P^2 / \mu^2 \right)^{-\epsilon}}{(- P^2)}\ ,$ \\[10pt]
$\pic{0.7}{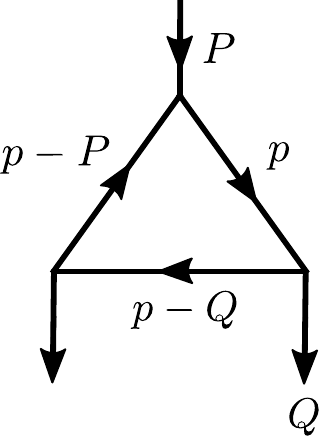}$ &$\,=\, \displaystyle \int \dfrac{d^{4-2\epsilon}p}{(2\pi)^{4-2\epsilon}}\dfrac{1}{p^2(p-Q)^2(p-P)^2}\,=\, -i\,\frac{c_\Gamma}{\epsilon^2}\frac{(-P^2/ \mu^2)^{-\epsilon}-(-Q^2/\mu^2)^{-\epsilon}}{(-P^2)-(-Q^2)}\ ,$
\\[10pt]
$\pic{0.8}{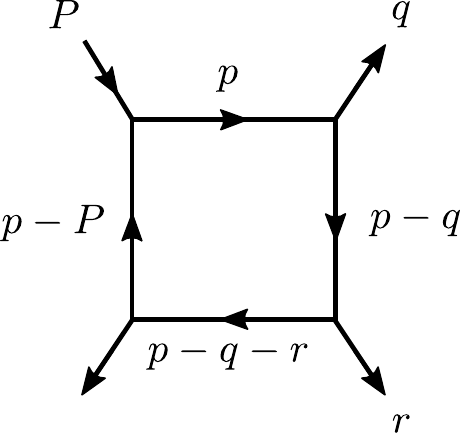} $ & $
\,=\, \displaystyle \int\! \dfrac{d^{4-2\eps} p}{(2\pi)^{4-2\eps}}\,\dfrac{1}{p^2 (p-q)^2(p-q-r)^2(p-P)^2}$ \\
& $ \displaystyle \,=\, -i\,\frac{2c_{\Gamma}}{st}\Big\{-\frac{1}{\eps^2}\Big[\Big(-{s\over \mu^2}\Big)^{-\eps}+\Big(-{t\over \mu^2}\Big)^{-\eps} -\Big(-{P^2\over \mu^2}\Big)^{-\eps}\Big] $
\\[15pt]
& $\displaystyle \,+ \text{Li}_2\Big(1-\frac{P^2}{s}\Big) + \text{Li}_2\Big(1-\frac{P^2}{t}\Big) +\frac{1}{2}\log^2\Big(\frac{s}{t}\Big) + \frac{\pi^2}{6}\Big\}\ .$
\end{tabular}
\end{table}

\noindent 
where
\begin{align*}
c_{\Gamma} \,=\, \frac{1}{(4\pi)^{2-\epsilon}} \frac{\Gamma(1+\epsilon)\Gamma(1-\epsilon)^2}{\Gamma(1-2\epsilon)}\,.
\end{align*}
	
\newpage

\section{Numerators}\label{App:Numerators}
In this appendix we present the numerators of the integral topologies which constitute the two loop integrands for form factors of $\O_{\mathcal S}$ and $\O_{\mathcal C}$ in $\N\!=\!4$ SYM. The integral topologies, denoted as $I_i$, $i=1,\ldots,12$ are presented in Table \ref{tab:two-loop basis}.

\subsection{Two-loop integrand for the $\cO_\cS$ form factor in $\N\!=\!4$ SYM}
\label{app:susyN4}

The integrand of the two-loop minimal form factor of the Konishi descendant operator $\O_{\mathcal S}$ is given by
\begin{align}\nonumber
F^{(2)}_{\O_{\mathcal S}}\,=\,F^{(0)}_{\O_{\mathcal S}}\,\sum_{i=1}^{12} N_i \times I_i \ , 
\end{align}
where\footnote{Note that the $N_1$ quoted here is before the PV reduction, in contrast to \eqref{eq:NumMergedPV}. PV reduction procedure relates the two, but it affects the numerators $N_6$ and $N_7$ accordingly.} 
\begin{align}
\begin{split}
\label{eq:numerators}
N_1\,=\,&\hf\frac{s_{23}}{s_{12}s_{13}}\left[2s_{12}s_{23}s_{13}-2p_1\cdot(p_3+\ell)s_{23}(s_{12}-s_{13})+(s_{12}+s_{13})^2(p_3+\ell)^2\right]\,,\\[5pt]
N_2\,=\,&\frac{\Tr(1\,q\,k\, q\, \ell\,k\, q\,1\,3\,2)}{s_{12}s_{23}s_{13}}\,, \\[5pt]
N_3\,=\,&N_2\,\Big|_{p_2\leftrightarrow p_3}\,,\\[5pt]
N_4\,=\,&\frac{s_{123}}{s_{12}s_{23}s_{13}}\Tr(1q\ell kq3)\,,\\[5pt]
N_5\,=\, &\frac{1}{2}\Big[-3 (s_{2\ell}+s_{23}+ s_{1k})-\frac{s_{23}^3+2 s_{23} s_{3 k} s_{1\ell}+s_{23} s_{3 k} s_{2\ell}+ 2 s_{23}^2 (s_{1k}+s_{2\ell})}{2 s_{12} s_{13}}\\[5pt]
&-\frac{s_{23} \left(s_{1k}+s_{2\ell}+2s_{3 k}+4 s_{1\ell}+ 2 s_{23}\right)+2s_{1k} s_{2\ell}+s_{13} \left(s_{3 k}+s_{1\ell}-3 s_{2\ell}+s_{23}\right)+s_{3 k} \left(s_{1\ell}+s_{2\ell}\right)}{s_{12}}\\[5pt]
&+\frac{s_{12} s_{3 k}-s_{1k} s_{2\ell}}{s_{23}}+\frac{s_{12} s_{3 k}(s_{12}-s_{1\ell})}{s_{13} s_{23}}\Big]+(p_2\leftrightarrow p_3,k\leftrightarrow \ell)\\[5pt]
N_6\,=\,&s_{23}\left(\frac{s_{1\ell}}{s_{12}}-\frac{s_{1\ell}}{s_{13}}+\frac{s_{13}}{2 s_{12}}-\frac{s_{12}}{s_{13}}-\frac{1}{2}\right)\,, \\[5pt]
N_7\,=\,& N_6\,\Big|_{p_2\leftrightarrow p_3}\,,\\[5pt]
N_8\,=\,&-2+\frac{s_{23} (s_{1\ell}-s_{23})}{2 s_{12} s_{13}}+\frac{s_{12} s_{1\ell}}{2 s_{13} s_{23}} +\frac{s_{1\ell}-2s_{23}-s_{13}}{2 s_{12}}+\frac{2s_{1\ell}-s_{23}+2s_{12}}{2s_{13}}+\frac{s_{1\ell}-s_{12}-s_{13}}{2s_{23}}\,,
\end{split}
\end{align}
\newpage
\begin{align*}
\phantom{N_8\,=\,}&\phantom{-2+\frac{s_{23} (s_{1\ell}-s_{23})}{2 s_{12} s_{13}}+\frac{s_{12} s_{1\ell}}{2 s_{13} s_{23}} +\frac{s_{1\ell}-2s_{23}-s_{13}}{2 s_{12}}+\frac{2s_{1\ell}-s_{23}+2s_{12}}{2s_{13}}+\frac{s_{1\ell}-s_{12}-s_{13}}{2s_{23}}\,,}\\[-1.2cm]
N_9\,=\, &N_8\,\Big|_{p_2\leftrightarrow p_3}\,, \\[5pt]
N_{10}\,=\,&-\frac{\left(s_{12}+s_{13}\right){}^2}{s_{12} s_{13}}\,,\\[5pt]
N_{11}\,=\,&N_{10}\,,\\[5pt]
N_{12}\,=\,&\frac{s_{12}+s_{23}+s_{13}}{2 s_{12} s_{13}}\,.
\end{align*}

\subsection{Two-loop integrand for the $\cO_\cC$ form factor in $\N\!=\!4$ SYM}
\label{app:compN4}
The two-loop integrand of the form factor of the component operator $\O_{\mathcal C}$ can be conveniently expressed 
in terms that of the supersymmetric operator $\O_{\mathcal S}$ plus an offset term: 
\begin{align}\nonumber
F^{(2)}_{\O_{\mathcal C}}\,=\,F^{(2)}_{\O_{\mathcal S}}+\Delta_{\cN=4}\,,\qquad \Delta_{\cN=4}\,=\, \sum_{i=5}^{12} \tilde N_i \times I_i\,,
\end{align}
\begin{align}
\label{eq:ff-cpt}
\begin{split}
\tilde N_5\,=\,&\frac{s_{3 k} s_{2\ell}}{s_{23}}-\frac{s_{3 k} s_{1\ell}}{s_{13}}-\frac{s_{1k} s_{3 k} s_{2\ell}}{s_{12} s_{23}}+\frac{s_{3 k}^2}{2 s_{23}}+\frac{5 s_{3 k}}{2}-\frac{3 s_{1k} s_{3 k}}{2 s_{12}}-\frac{3 s_{23} s_{1k}}{2 s_{12}}+s_{23}+(p_2\leftrightarrow p_3,k\leftrightarrow \ell)\,,\\[5pt]
\tilde N_6\,=\,&\frac{s_{2 k} s_{1\ell}}{2 s_{13}}-\frac{s_{3 k} s_{1\ell}}{2 s_{12}}-\frac{s_{23} s_{1k}}{2 s_{13}}+\frac{s_{2 k}}{2}+\frac{s_{3 k}}{2}+\frac{s_{12} \left(s_{2 k}+s_{3 k}\right)}{2 s_{13}}\,,\\[5pt]
\tilde N_7\,=\,&\tilde N_6\,\Big|_{p_2\leftrightarrow p_3}\,, \\[5pt]
\tilde N_8\,=\,&4+\frac{s_{2 k} s_{1\ell}}{s_{12} s_{23}}+\frac{4 s_{2 k}+3 s_{3 k}+6 s_{3\ell}}{2 s_{23}}+\frac{s_{2 k} s_{1\ell}+s_{12} \left(s_{2 k}+s_{3 k}+s_{3\ell}\right)}{s_{13} s_{23}}-\frac{s_{1k}}{s_{13}}-\frac{3 s_{1\ell}}{s_{12}}+\frac{3 s_{12}}{2 s_{13}}\ ,\\[5pt]
\tilde N_9\,=\, &\tilde N_8\,\Big|_{p_2\leftrightarrow p_3}\,, \\[5pt]
\tilde N_{10}\,=\,& -\frac{s_{1k}}{2 s_{12}}+\frac{s_{2 k}}{s_{23}}+\frac{s_{13} s_{2 k}}{2 s_{12} s_{23}}+\frac{s_{12} s_{2 k}}{2 s_{13} s_{23}}+(p_2\leftrightarrow p_3)\,,\\[5pt]
\tilde N_{11}\,=\,& \tilde N_{10}\,, \\[5pt]
\tilde N_{12}\,=\, &\frac{3 s_{12}-s_{1k}}{s_{13} s_{23}}+\frac{3 s_{13}-s_{1\ell}}{s_{12} s_{23}}+\frac{8}{s_{23}}\,.
\end{split}
\end{align}

	\pagebreak
	\bibliographystyle{utphys}
	\bibliography{remainder}

\providecommand{\href}[2]{#2}\begingroup\raggedright\begin{thebibliography}{10}

\bibitem{Wilczek:1977zn}
F.~Wilczek, ``{Decays of Heavy Vector Mesons Into Higgs Particles},''
\href{http://dx.doi.org/10.1103/PhysRevLett.39.1304}{{\em Phys. Rev. Lett.}
  {\bfseries 39} (1977) 1304}.

\bibitem{Shifman:1979eb}
M.~A. Shifman, A.~I. Vainshtein, M.~B. Voloshin, and V.~I. Zakharov,
  ``{Low-Energy Theorems for Higgs Boson Couplings to Photons},'' {\em Sov. J.
  Nucl. Phys.} {\bfseries 30} (1979) 711--716.
[Yad. Fiz.30,1368(1979)].

\bibitem{Dawson:1990zj}
S.~Dawson, ``{Radiative corrections to Higgs boson production},''
\href{http://dx.doi.org/10.1016/0550-3213(91)90061-2}{{\em Nucl. Phys.}
  {\bfseries B359} (1991) 283--300}.

\bibitem{Buchmuller:1985jz}
W.~Buchmuller and D.~Wyler, ``{Effective Lagrangian Analysis of New
  Interactions and Flavor Conservation},''
\href{http://dx.doi.org/10.1016/0550-3213(86)90262-2}{{\em Nucl. Phys.}
  {\bfseries B268} (1986) 621--653}.

\bibitem{Neill:2009tn}
D.~Neill, ``{Two-Loop Matching onto Dimension Eight Operators in the Higgs-Glue
  Sector},''
\href{http://arxiv.org/abs/0908.1573}{{\ttfamily arXiv:0908.1573 [hep-ph]}}.

\bibitem{Neill:2009mz}
D.~Neill, ``{Analytic Virtual Corrections for Higgs Transverse Momentum
  Spectrum at $O(\alpha_s^2/m_t^3)$ via Unitarity Methods},''
\href{http://arxiv.org/abs/0911.2707}{{\ttfamily arXiv:0911.2707 [hep-ph]}}.

\bibitem{Harlander:2013oja}
R.~V. Harlander and T.~Neumann, ``{Probing the nature of the Higgs-gluon
  coupling},'' \href{http://dx.doi.org/10.1103/PhysRevD.88.074015}{{\em Phys.
  Rev.} {\bfseries D88} (2013) 074015},
\href{http://arxiv.org/abs/1308.2225}{{\ttfamily arXiv:1308.2225 [hep-ph]}}.

\bibitem{Dawson:2014ora}
S.~Dawson, I.~M. Lewis, and M.~Zeng, ``{Effective field theory for Higgs boson
  plus jet production},''
  \href{http://dx.doi.org/10.1103/PhysRevD.90.093007}{{\em Phys. Rev.}
  {\bfseries D90} no.~9, (2014) 093007},
\href{http://arxiv.org/abs/1409.6299}{{\ttfamily arXiv:1409.6299 [hep-ph]}}.

\bibitem{Gracey:2002he}
J.~A. Gracey, ``{Classification and one loop renormalization of dimension-six
  and dimension-eight operators in quantum gluodynamics},''
  \href{http://dx.doi.org/10.1016/S0550-3213(02)00334-6,
  10.1016/j.nuclphysb.2004.06.053}{{\em Nucl. Phys.} {\bfseries B634} (2002)
  192--208}, \href{http://arxiv.org/abs/hep-ph/0204266}{{\ttfamily
  arXiv:hep-ph/0204266 [hep-ph]}}.
[Erratum: Nucl. Phys.B696,295(2004)].

\bibitem{Benincasa:2011pg}
P.~Benincasa and E.~Conde, ``{Exploring the S-Matrix of Massless Particles},''
  \href{http://dx.doi.org/10.1103/PhysRevD.86.025007}{{\em Phys. Rev.}
  {\bfseries D86} (2012) 025007},
\href{http://arxiv.org/abs/1108.3078}{{\ttfamily arXiv:1108.3078 [hep-th]}}.

\bibitem{Dixon:1993xd}
L.~J. Dixon and Y.~Shadmi, ``{Testing gluon selfinteractions in three jet
  events at hadron colliders},''
  \href{http://dx.doi.org/10.1016/0550-3213(94)90563-0,
  10.1016/0550-3213(95)00450-7}{{\em Nucl. Phys.} {\bfseries B423} (1994)
  3--32}, \href{http://arxiv.org/abs/hep-ph/9312363}{{\ttfamily
  arXiv:hep-ph/9312363 [hep-ph]}}.
[Erratum: Nucl. Phys.B452,724(1995)].

\bibitem{Dixon:2004za}
L.~J. Dixon, E.~W.~N. Glover, and V.~V. Khoze, ``{MHV rules for Higgs plus
  multi-gluon amplitudes},''
  \href{http://dx.doi.org/10.1088/1126-6708/2004/12/015}{{\em JHEP} {\bfseries
  12} (2004) 015},
\href{http://arxiv.org/abs/hep-th/0411092}{{\ttfamily arXiv:hep-th/0411092
  [hep-th]}}.

\bibitem{Cohen:2010mi}
T.~Cohen, H.~Elvang, and M.~Kiermaier, ``{On-shell constructibility of tree
  amplitudes in general field theories},''
  \href{http://dx.doi.org/10.1007/JHEP04(2011)053}{{\em JHEP} {\bfseries 04}
  (2011) 053},
\href{http://arxiv.org/abs/1010.0257}{{\ttfamily arXiv:1010.0257 [hep-th]}}.

\bibitem{Broedel:2012rc}
J.~Broedel and L.~J. Dixon, ``{Color-kinematics duality and double-copy
  construction for amplitudes from higher-dimension operators},''
  \href{http://dx.doi.org/10.1007/JHEP10(2012)091}{{\em JHEP} {\bfseries 10}
  (2012) 091},
\href{http://arxiv.org/abs/1208.0876}{{\ttfamily arXiv:1208.0876 [hep-th]}}.

\bibitem{Eden:2011yp}
B.~Eden, P.~Heslop, G.~P. Korchemsky, and E.~Sokatchev, ``{The
  super-correlator/super-amplitude duality: Part I},''
  \href{http://dx.doi.org/10.1016/j.nuclphysb.2012.12.015}{{\em Nucl. Phys.}
  {\bfseries B869} (2013) 329--377},
\href{http://arxiv.org/abs/1103.3714}{{\ttfamily arXiv:1103.3714 [hep-th]}}.

\bibitem{Brandhuber:2011tv}
A.~Brandhuber, O.~Gurdogan, R.~Mooney, G.~Travaglini, and G.~Yang, ``{Harmony
  of Super Form Factors},''
  \href{http://dx.doi.org/10.1007/JHEP10(2011)046}{{\em JHEP} {\bfseries 10}
  (2011) 046},
\href{http://arxiv.org/abs/1107.5067}{{\ttfamily arXiv:1107.5067 [hep-th]}}.

\bibitem{Witten:2003nn}
E.~Witten, ``{Perturbative gauge theory as a string theory in twistor space},''
  \href{http://dx.doi.org/10.1007/s00220-004-1187-3}{{\em Commun. Math. Phys.}
  {\bfseries 252} (2004) 189--258},
\href{http://arxiv.org/abs/hep-th/0312171}{{\ttfamily arXiv:hep-th/0312171
  [hep-th]}}.

\bibitem{Brandhuber:2012vm}
A.~Brandhuber, G.~Travaglini, and G.~Yang, ``{Analytic two-loop form factors in
  $\mathcal{N}\!=\!4$ SYM},''
  \href{http://dx.doi.org/10.1007/JHEP05(2012)082}{{\em JHEP} {\bfseries 05}
  (2012) 082},
\href{http://arxiv.org/abs/1201.4170}{{\ttfamily arXiv:1201.4170 [hep-th]}}.

\bibitem{Brandhuber:2010ad}
A.~Brandhuber, B.~Spence, G.~Travaglini, and G.~Yang, ``{Form Factors in N=4
  Super Yang-Mills and Periodic Wilson Loops},''
  \href{http://dx.doi.org/10.1007/JHEP01(2011)134}{{\em JHEP} {\bfseries 01}
  (2011) 134},
\href{http://arxiv.org/abs/1011.1899}{{\ttfamily arXiv:1011.1899 [hep-th]}}.

\bibitem{Gehrmann:2011aa}
T.~Gehrmann, M.~Jaquier, E.~W.~N. Glover, and A.~Koukoutsakis, ``{Two-Loop QCD
  Corrections to the Helicity Amplitudes for $H \to$ 3 partons},''
  \href{http://dx.doi.org/10.1007/JHEP02(2012)056}{{\em JHEP} {\bfseries 02}
  (2012) 056},
\href{http://arxiv.org/abs/1112.3554}{{\ttfamily arXiv:1112.3554 [hep-ph]}}.

\bibitem{Penante:2014sza}
B.~Penante, B.~Spence, G.~Travaglini, and C.~Wen, ``{On super form factors of
  half-BPS operators in N=4 super Yang-Mills},''
  \href{http://dx.doi.org/10.1007/JHEP04(2014)083}{{\em JHEP} {\bfseries 1404}
  (2014) 083},
\href{http://arxiv.org/abs/1402.1300}{{\ttfamily arXiv:1402.1300 [hep-th]}}.

\bibitem{Brandhuber:2014ica}
A.~Brandhuber, B.~Penante, G.~Travaglini, and C.~Wen, ``{The last of the simple
  remainders},'' \href{http://dx.doi.org/10.1007/JHEP08(2014)100}{{\em JHEP}
  {\bfseries 1408} (2014) 100},
\href{http://arxiv.org/abs/1406.1443}{{\ttfamily arXiv:1406.1443 [hep-th]}}.

\bibitem{Koster:2016loo}
L.~Koster, V.~Mitev, M.~Staudacher, and M.~Wilhelm, ``{All tree-level MHV form
  factors in $ \mathcal{N} $ = 4 SYM from twistor space},''
  \href{http://dx.doi.org/10.1007/JHEP06(2016)162}{{\em JHEP} {\bfseries 06}
  (2016) 162},
\href{http://arxiv.org/abs/1604.00012}{{\ttfamily arXiv:1604.00012 [hep-th]}}.

\bibitem{Chicherin:2016qsf}
D.~Chicherin and E.~Sokatchev, ``{Composite operators and form factors in
  ${\mathcal N}\!=\!4$ SYM},''
  \href{http://dx.doi.org/10.1088/1751-8121/aa72fe}{{\em J. Phys.} {\bfseries
  A50} no.~27, (2017) 275402},
\href{http://arxiv.org/abs/1605.01386}{{\ttfamily arXiv:1605.01386 [hep-th]}}.

\bibitem{Part2}
A.~Brandhuber, M.~Kostaci\'{n}ska, B.~Penante, and G.~Travaglini, ``{${\rm
  Tr}\, F^3$ supersymmetric form factors and maximal transcendentality Part II:
  $0<\mathcal{N}<4$ super Yang-Mills},''
\href{http://arxiv.org/abs/1804.05828}{{\ttfamily arXiv:1804.05828 [hep-th]}}.

\bibitem{Brandhuber:2017bkg}
A.~Brandhuber, M.~Kostaci\'{n}ska, B.~Penante, and G.~Travaglini, ``{Higgs
  amplitudes from $\mathcal{N}=4$ super Yang-Mills theory},''
  \href{http://dx.doi.org/10.1103/PhysRevLett.119.161601}{{\em Phys. Rev.
  Lett.} {\bfseries 119} no.~16, (2017) 161601},
\href{http://arxiv.org/abs/1707.09897}{{\ttfamily arXiv:1707.09897 [hep-th]}}.

\bibitem{Kotikov:2002ab}
A.~V. Kotikov and L.~N. Lipatov, ``{DGLAP and BFKL equations in the
  $\mathcal{N}\!=\!4$ supersymmetric gauge theory},''
  \href{http://dx.doi.org/10.1016/S0550-3213(03)00264-5,
  10.1016/j.nuclphysb.2004.02.032}{{\em Nucl. Phys.} {\bfseries B661} (2003)
  19--61}, \href{http://arxiv.org/abs/hep-ph/0208220}{{\ttfamily
  arXiv:hep-ph/0208220 [hep-ph]}}.
[Erratum: Nucl. Phys.B685,405(2004)].

\bibitem{Kotikov:2004er}
A.~V. Kotikov, L.~N. Lipatov, A.~I. Onishchenko, and V.~N. Velizhanin, ``{Three
  loop universal anomalous dimension of the Wilson operators in $N=4$ SUSY
  Yang-Mills model},'' \href{http://dx.doi.org/10.1016/j.physletb.2004.05.078,
  10.1016/j.physletb.2005.11.002}{{\em Phys. Lett.} {\bfseries B595} (2004)
  521--529}, \href{http://arxiv.org/abs/hep-th/0404092}{{\ttfamily
  arXiv:hep-th/0404092 [hep-th]}}.
[Erratum: Phys. Lett.B632,754(2006)].

\bibitem{Moch:2004pa}
S.~Moch, J.~A.~M. Vermaseren, and A.~Vogt, ``{The Three loop splitting
  functions in QCD: The Nonsinglet case},''
  \href{http://dx.doi.org/10.1016/j.nuclphysb.2004.03.030}{{\em Nucl. Phys.}
  {\bfseries B688} (2004) 101--134},
\href{http://arxiv.org/abs/hep-ph/0403192}{{\ttfamily arXiv:hep-ph/0403192
  [hep-ph]}}.

\bibitem{Vogt:2004mw}
A.~Vogt, S.~Moch, and J.~A.~M. Vermaseren, ``{The Three-loop splitting
  functions in QCD: The Singlet case},''
  \href{http://dx.doi.org/10.1016/j.nuclphysb.2004.04.024}{{\em Nucl. Phys.}
  {\bfseries B691} (2004) 129--181},
\href{http://arxiv.org/abs/hep-ph/0404111}{{\ttfamily arXiv:hep-ph/0404111
  [hep-ph]}}.

\bibitem{Bern:1994cg}
Z.~Bern, L.~J. Dixon, D.~C. Dunbar, and D.~A. Kosower, ``{Fusing gauge theory
  tree amplitudes into loop amplitudes},''
  \href{http://dx.doi.org/10.1016/0550-3213(94)00488-Z}{{\em Nucl. Phys.}
  {\bfseries B435} (1995) 59--101},
\href{http://arxiv.org/abs/hep-ph/9409265}{{\ttfamily arXiv:hep-ph/9409265
  [hep-ph]}}.

\bibitem{Bern:1993mq}
Z.~Bern, L.~J. Dixon, and D.~A. Kosower, ``{One loop corrections to five gluon
  amplitudes},'' \href{http://dx.doi.org/10.1103/PhysRevLett.70.2677}{{\em
  Phys. Rev. Lett.} {\bfseries 70} (1993) 2677--2680},
\href{http://arxiv.org/abs/hep-ph/9302280}{{\ttfamily arXiv:hep-ph/9302280
  [hep-ph]}}.

\bibitem{Bedford:2004nh}
J.~Bedford, A.~Brandhuber, B.~J. Spence, and G.~Travaglini,
  ``{Non-supersymmetric loop amplitudes and MHV vertices},''
  \href{http://dx.doi.org/10.1016/j.nuclphysb.2005.01.032}{{\em Nucl. Phys.}
  {\bfseries B712} (2005) 59--85},
\href{http://arxiv.org/abs/hep-th/0412108}{{\ttfamily arXiv:hep-th/0412108
  [hep-th]}}.

\bibitem{Banerjee:2016kri}
P.~Banerjee, P.~K. Dhani, M.~Mahakhud, V.~Ravindran, and S.~Seth, ``{Finite
  remainders of the Konishi at two loops in $ \mathcal{N}=4 $ SYM},''
  \href{http://dx.doi.org/10.1007/JHEP05(2017)085}{{\em JHEP} {\bfseries 05}
  (2017) 085},
\href{http://arxiv.org/abs/1612.00885}{{\ttfamily arXiv:1612.00885 [hep-th]}}.

\bibitem{Loebbert:2015ova}
F.~Loebbert, D.~Nandan, C.~Sieg, M.~Wilhelm, and G.~Yang, ``{On-Shell Methods
  for the Two-Loop Dilatation Operator and Finite Remainders},''
  \href{http://dx.doi.org/10.1007/JHEP10(2015)012}{{\em JHEP} {\bfseries 10}
  (2015) 012},
\href{http://arxiv.org/abs/1504.06323}{{\ttfamily arXiv:1504.06323 [hep-th]}}.

\bibitem{Brandhuber:2016fni}
A.~Brandhuber, M.~Kostaci\'{n}ska, B.~Penante, G.~Travaglini, and D.~Young,
  ``{The $SU(2|3)$ dynamic two-loop form factors},''
  \href{http://dx.doi.org/10.1007/JHEP08(2016)134}{{\em JHEP} {\bfseries 08}
  (2016) 134},
\href{http://arxiv.org/abs/1606.08682}{{\ttfamily arXiv:1606.08682 [hep-th]}}.

\bibitem{Loebbert:2016xkw}
F.~Loebbert, C.~Sieg, M.~Wilhelm, and G.~Yang, ``{Two-Loop $SL(2)$ Form Factors
  and Maximal Transcendentality},''
  \href{http://dx.doi.org/10.1007/JHEP12(2016)090}{{\em JHEP} {\bfseries 12}
  (2016) 090},
\href{http://arxiv.org/abs/1610.06567}{{\ttfamily arXiv:1610.06567 [hep-th]}}.

\bibitem{Li:2014afw}
Y.~Li, A.~von Manteuffel, R.~M. Schabinger, and H.~X. Zhu, ``{Soft-virtual
  corrections to Higgs production at N$^3$LO},''
  \href{http://dx.doi.org/10.1103/PhysRevD.91.036008}{{\em Phys. Rev.}
  {\bfseries D91} (2015) 036008},
\href{http://arxiv.org/abs/1412.2771}{{\ttfamily arXiv:1412.2771 [hep-ph]}}.

\bibitem{Li:2016ctv}
Y.~Li and H.~X. Zhu, ``{Bootstrapping Rapidity Anomalous Dimensions for
  Transverse-Momentum Resummation},''
  \href{http://dx.doi.org/10.1103/PhysRevLett.118.022004}{{\em Phys. Rev.
  Lett.} {\bfseries 118} no.~2, (2017) 022004},
\href{http://arxiv.org/abs/1604.01404}{{\ttfamily arXiv:1604.01404 [hep-ph]}}.

\bibitem{Dixon:2017nat}
L.~J. Dixon, ``{The Principle of Maximal Transcendentality and the Four-Loop
  Collinear Anomalous Dimension},''
  \href{http://dx.doi.org/10.1007/JHEP01(2018)075}{{\em JHEP} {\bfseries 01}
  (2018) 075},
\href{http://arxiv.org/abs/1712.07274}{{\ttfamily arXiv:1712.07274 [hep-th]}}.

\bibitem{Beisert:2003ys}
N.~Beisert, ``{The $su(2|3)$ dynamic spin chain},''
  \href{http://dx.doi.org/10.1016/j.nuclphysb.2003.12.032}{{\em Nucl. Phys.}
  {\bfseries B682} (2004) 487--520},
\href{http://arxiv.org/abs/hep-th/0310252}{{\ttfamily arXiv:hep-th/0310252
  [hep-th]}}.

\bibitem{talk12}
A.~Brandhuber, M.~Kostaci\'{n}ska, B.~Penante, and G.~Travaglini, ``\text{Talks
  at the XXIII IFT Christmas workshop (Madrid, 2017) and the 2018 Bethe }
  \text{Forum (Bonn)},'' 2017.
\newblock \url{https://workshops.ift.uam-csic.es/Xmas17/program} and
  \url{https://indico.desy.de/indico/event/18613/}.

\bibitem{0309040}
C.~Anastasiou, Z.~Bern, L.~J. Dixon, and D.~A. Kosower, ``{Planar amplitudes in
  maximally supersymmetric Yang-Mills theory},''
  \href{http://dx.doi.org/10.1103/PhysRevLett.91.251602}{{\em Phys. Rev. Lett.}
  {\bfseries 91} (2003) 251602},
\href{http://arxiv.org/abs/hep-th/0309040}{{\ttfamily arXiv:hep-th/0309040
  [hep-th]}}.

\bibitem{Bern:2006vw}
Z.~Bern, M.~Czakon, D.~A. Kosower, R.~Roiban, and V.~A. Smirnov, ``{Two-loop
  iteration of five-point N=4 super-Yang-Mills amplitudes},''
  \href{http://dx.doi.org/10.1103/PhysRevLett.97.181601}{{\em Phys. Rev. Lett.}
  {\bfseries 97} (2006) 181601},
\href{http://arxiv.org/abs/hep-th/0604074}{{\ttfamily arXiv:hep-th/0604074
  [hep-th]}}.

\bibitem{Bern:2008ap}
Z.~Bern, L.~J. Dixon, D.~A. Kosower, R.~Roiban, M.~Spradlin, C.~Vergu, and
  A.~Volovich, ``{The Two-Loop Six-Gluon MHV Amplitude in Maximally
  Supersymmetric Yang-Mills Theory},''
  \href{http://dx.doi.org/10.1103/PhysRevD.78.045007}{{\em Phys. Rev.}
  {\bfseries D78} (2008) 045007},
\href{http://arxiv.org/abs/0803.1465}{{\ttfamily arXiv:0803.1465 [hep-th]}}.

\bibitem{Nandan:2014oga}
D.~Nandan, C.~Sieg, M.~Wilhelm, and G.~Yang, ``{Cutting through form factors
  and cross sections of non-protected operators in N=4 SYM},''
\href{http://arxiv.org/abs/1410.8485}{{\ttfamily arXiv:1410.8485 [hep-th]}}.

\bibitem{Nair:1988bq}
V.~P. Nair, ``{A Current Algebra for Some Gauge Theory Amplitudes},''
\href{http://dx.doi.org/10.1016/0370-2693(88)91471-2}{{\em Phys. Lett.}
  {\bfseries B214} (1988) 215--218}.

\bibitem{Lee:2012cn}
R.~N. Lee, ``{Presenting LiteRed: a tool for the Loop InTEgrals REDuction},''
\href{http://arxiv.org/abs/1212.2685}{{\ttfamily arXiv:1212.2685 [hep-ph]}}.

\bibitem{Lee:2013mka}
R.~N. Lee, ``{LiteRed 1.4: a powerful tool for reduction of multiloop
  integrals},'' \href{http://dx.doi.org/10.1088/1742-6596/523/1/012059}{{\em J.
  Phys. Conf. Ser.} {\bfseries 523} (2014) 012059},
\href{http://arxiv.org/abs/1310.1145}{{\ttfamily arXiv:1310.1145 [hep-ph]}}.

\bibitem{Mangano:1990by}
M.~L. Mangano and S.~J. Parke, ``{Multiparton amplitudes in gauge theories},''
  \href{http://dx.doi.org/10.1016/0370-1573(91)90091-Y}{{\em Phys. Rept.}
  {\bfseries 200} (1991) 301--367},
\href{http://arxiv.org/abs/hep-th/0509223}{{\ttfamily arXiv:hep-th/0509223
  [hep-th]}}.

\bibitem{Cachazo:2004kj}
F.~Cachazo, P.~Svrcek, and E.~Witten, ``\text{MHV} vertices and tree amplitudes
  in gauge theory,'' {\em JHEP} {\bfseries 09} (2004) 006,
\href{http://arxiv.org/abs/hep-th/0403047}{{\ttfamily hep-th/0403047}}.

\bibitem{Gehrmann:1999as}
T.~Gehrmann and E.~Remiddi, ``{Differential equations for two loop four point
  functions},'' \href{http://dx.doi.org/10.1016/S0550-3213(00)00223-6}{{\em
  Nucl. Phys.} {\bfseries B580} (2000) 485--518},
\href{http://arxiv.org/abs/hep-ph/9912329}{{\ttfamily arXiv:hep-ph/9912329
  [hep-ph]}}.

\bibitem{Gehrmann:2000zt}
T.~Gehrmann and E.~Remiddi, ``{Two loop master integrals for $\gamma^\star
  \rightarrow$ 3 jets: The planar topologies},''
  \href{http://dx.doi.org/10.1016/S0550-3213(01)00057-8}{{\em Nucl. Phys.}
  {\bfseries B601} (2001) 248--286},
\href{http://arxiv.org/abs/hep-ph/0008287}{{\ttfamily arXiv:hep-ph/0008287
  [hep-ph]}}.

\bibitem{Goncharov:2010jf}
A.~B. Goncharov, M.~Spradlin, C.~Vergu, and A.~Volovich, ``{Classical
  Polylogarithms for Amplitudes and Wilson Loops},''
  \href{http://dx.doi.org/10.1103/PhysRevLett.105.151605}{{\em Phys.Rev.Lett.}
  {\bfseries 105} (2010) 151605},
\href{http://arxiv.org/abs/1006.5703}{{\ttfamily arXiv:1006.5703 [hep-th]}}.

\bibitem{Anastasiou:2003kj}
C.~Anastasiou, Z.~Bern, L.~J. Dixon, and D.~A. Kosower, ``{Planar amplitudes in
  maximally supersymmetric Yang-Mills theory},''
  \href{http://dx.doi.org/10.1103/PhysRevLett.91.251602}{{\em Phys. Rev. Lett.}
  {\bfseries 91} (2003) 251602},
\href{http://arxiv.org/abs/hep-th/0309040}{{\ttfamily arXiv:hep-th/0309040
  [hep-th]}}.

\bibitem{Bern:2005iz}
Z.~Bern, L.~J. Dixon, and V.~A. Smirnov, ``{Iteration of planar amplitudes in
  maximally supersymmetric Yang-Mills theory at three loops and beyond},''
  \href{http://dx.doi.org/10.1103/PhysRevD.72.085001}{{\em Phys. Rev.}
  {\bfseries D72} (2005) 085001},
\href{http://arxiv.org/abs/hep-th/0505205}{{\ttfamily arXiv:hep-th/0505205
  [hep-th]}}.

\bibitem{Catani:1998bh}
S.~Catani, ``{The Singular behavior of QCD amplitudes at two loop order},''
  \href{http://dx.doi.org/10.1016/S0370-2693(98)00332-3}{{\em Phys. Lett.}
  {\bfseries B427} (1998) 161--171},
\href{http://arxiv.org/abs/hep-ph/9802439}{{\ttfamily arXiv:hep-ph/9802439
  [hep-ph]}}.

\bibitem{Fleischer:1998nb}
J.~Fleischer, A.~V. Kotikov, and O.~L. Veretin, ``{Analytic two loop results
  for selfenergy type and vertex type diagrams with one nonzero mass},''
  \href{http://dx.doi.org/10.1016/S0550-3213(99)00078-4}{{\em Nucl. Phys.}
  {\bfseries B547} (1999) 343--374},
\href{http://arxiv.org/abs/hep-ph/9808242}{{\ttfamily arXiv:hep-ph/9808242
  [hep-ph]}}.

\end{thebibliography}\endgroup

\end{document}